\documentclass[showpacs,preprintnumbers,amsmath,amssymb,11pt]{article}
\usepackage[utf8]{inputenc}
\usepackage[english]{babel}
\usepackage{amsmath}
\usepackage{graphicx}
\usepackage[colorinlistoftodos]{todonotes}
\usepackage{caption}
\usepackage{tabularx}
\usepackage{bbm}
\usepackage[normalem]{ulem}
\usepackage{listings}
\usepackage{amsthm}
\usepackage{algorithm}
\usepackage{algpseudocode}
\usepackage{mathtools}
\usepackage{amsmath}
\usepackage[toc,page]{appendix}
\usepackage{booktabs}
\usepackage{xcolor}
\usepackage{amsfonts}
\usepackage{epstopdf}
\usepackage{adjustbox}
\usepackage{csquotes}
\usepackage{enumitem} 
\usepackage{makecell}
\usepackage{subfigure}
\usepackage{tikz}
\usepackage{amssymb}
\usetikzlibrary{decorations.pathreplacing}

\usetikzlibrary{arrows}
\usetikzlibrary{shapes}

\usepackage[numbers]{natbib}

\makeatletter
\def\BState{\State\hskip-\ALG@thistlm}
\makeatother

\definecolor{darkgreen}{rgb}{0,0.5,0}
\definecolor{darkred}{rgb}{0.5,0,0}

\usepackage{amsmath,amsfonts,amssymb,amsthm,epsfig,epstopdf,titling,url,array}

\theoremstyle{plain}
\newtheorem{thm}{Theorem}[section]

\theoremstyle{plain}

\theoremstyle{remark}
\newtheorem{rem}{Remark}

\newcommand{\E}{\mathbb{E}}

\newcommand{\R}{\mathbb{R}}
\newcommand{\Q}{\mathbb{Q}}

\newcommand{\Levy}{L\'{e}vy}

\def\ito{It\={o}}

\newcommand\coolleftbrace[2]{%
	#1\left\{\vphantom{\begin{matrix} #2 \end{matrix}}\right.}

\title{A Heath-Jarrow-Morton framework for energy markets: a pragmatic approach}

\author{
	Matteo Gardini\thanks{Eni Plenitude, Via Ripamonti 85, 20136, Milan, Italy, email matteo.gardini@eniplenitude.com, \\ Department of Mathematics, University of Genoa, Via Dodecaneso 45, 16146, Genoa, Italy, email gardini@dima.unige.it} 	\and
	Edoardo Santilli\thanks{Eni Plenitude, Via Ripamonti 85, 20136, Milan, Italy, email edoardo.santilli@eniplenitude.com}}

\date{\today}
\begin{document}
\maketitle

\begin{abstract}
In this article we discuss the application of the Heath-Jarrow-Morton framework \citet{HJM1992} to energy markets. The goal of the article is to give a detailed overview of the topic, focusing on practical aspects rather than on theory, which has been widely studied in literature. This work aims to be a guide for practitioners and for all those who deal with the practical issues of this approach to energy market. In particular, we focus on the markets' structure, model calibration by dimension reduction with Principal Component Analysis (PCA), Monte Carlo simulations and derivatives pricing. As application, we focus on European power and gas markets: we calibrate the model on historical futures quotations, we perform futures and spot simulations and we analyze the results.\\

\vspace{0.2cm}
\end{abstract}

\section{Introduction}
Electricity markets across the world differ for many factors, both fundamental, such as demand and generation mix, and regulatory. In the most modern countries one goal in deregulating energy markets is to allow them to
respond to supply and demand variation in a more efficient way. Focusing on US electricity market, \citet{PARK2006} has shown that as a result of deregulation,
more competitive and interrelated environments are developing
in the electricity and natural gas markets. In a previous paper \citet{Emery2002} have discovered that the daily settlement prices of
New York Mercantile Exchange’s (NYMEX’s) California–Oregon Border
(COB) and Palo Verde (PV) electricity futures contracts are cointegrated
with the prices of its natural-gas futures contracts. Such a result has been confirmed by \citet{Mjelde2009} which have shown how electricity prices mainly response to shocks in coal market, whereas \citet{Bachmeier2006} deeply investigated the level of market integration between crude oil, coal, and natural gas prices. Moving to European countries several authors discussed how electricity prices react to variations in fuel prices. Testing for market integration between natural gas, electricity and oil prices in the UK in the period in which the natural gas market was deregulated but not yet linked to the continental European gas market, \citet{Asche2006} have highlighted evidences for an integrated energy market. \citet{Panagiotidis2007} analyzed the relationship
between UK wholesale gas prices and Brent oil price finding co-integration over the entire sample period(1996–2003). Likewise, using daily price data for Brent crude oil, NBP UK natural gas and EEX electricity \citet{Bencivenga2010} have shown that gas, oil and electricity markets are integrated. On the other hand
as a result of a robust multivariate long-run dynamics analysis \citet{Bosco2010} have revealed the presence of four highly integrated central European electricity markets (France, Germany, the Netherlands and Austria). The trend shared by these four electricity markets appears to be common also to gas prices, but not to oil prices. 
\par
\begin{figure}
	\centering
	\subfigure[Future prices, delivery year 2023]{\includegraphics[width=1\textwidth]{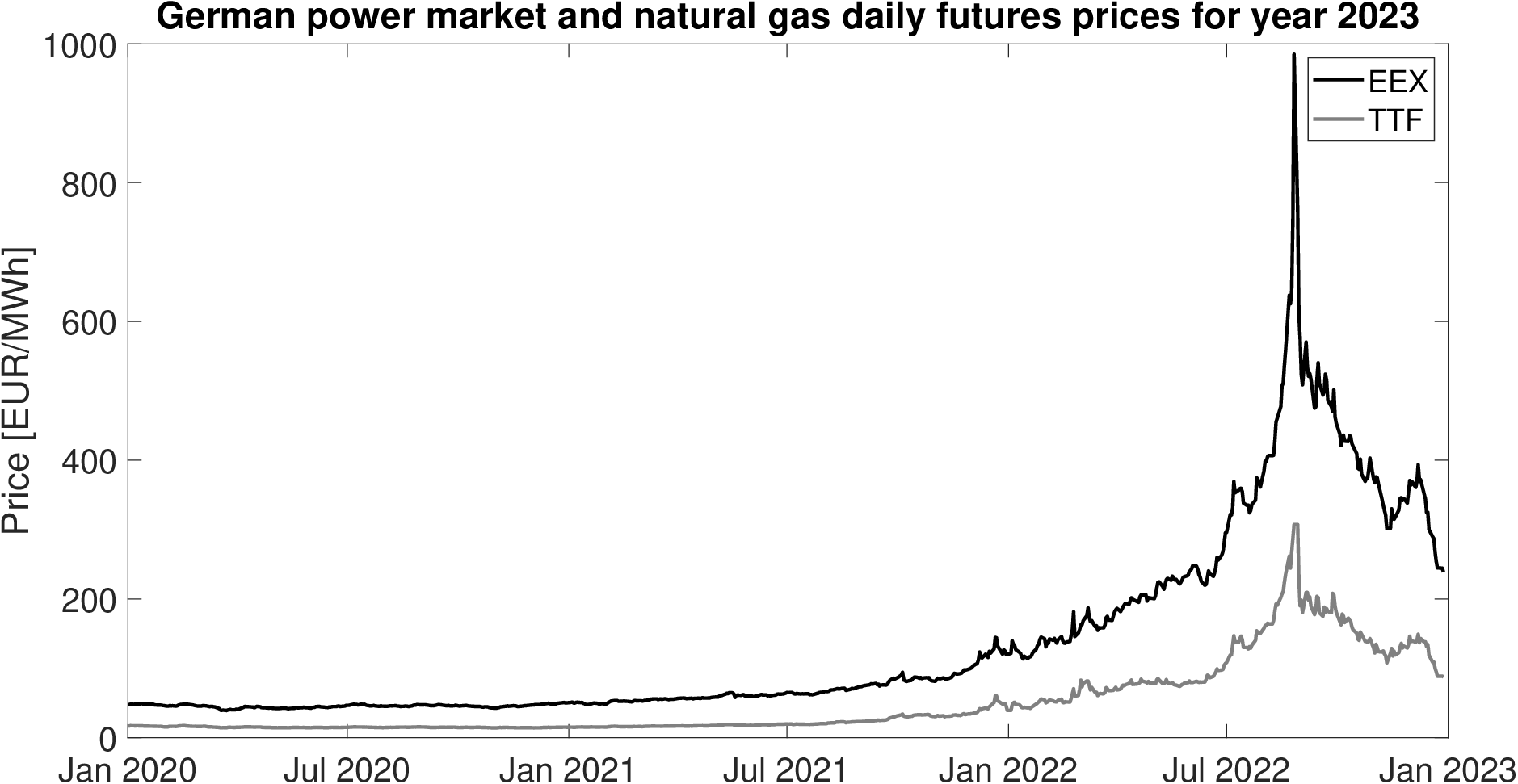}\label{fig:historical_spot_prices_ttf_eex}}
	\subfigure[Spot prices]{\includegraphics[width=1\textwidth]{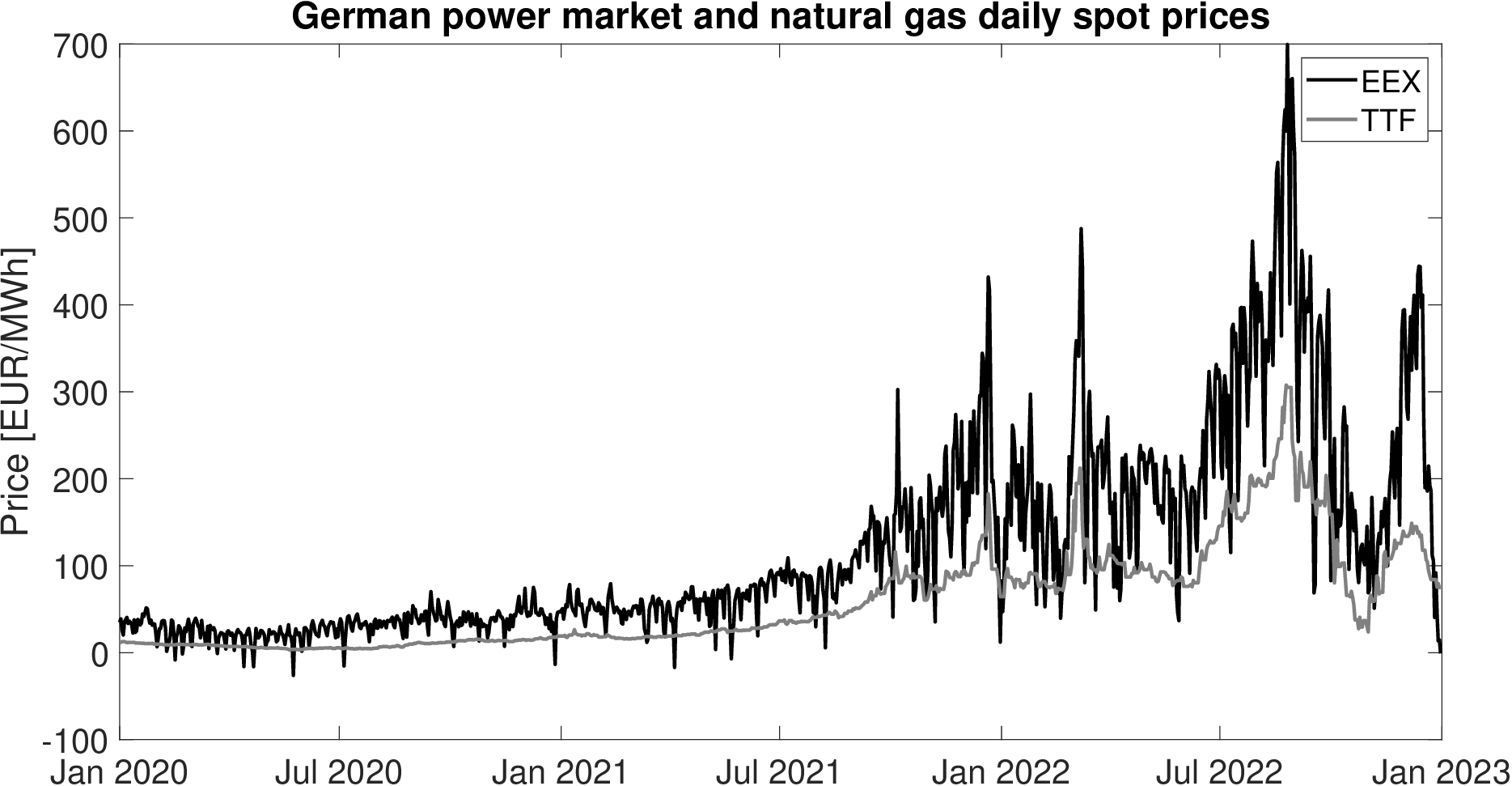}\label{fig:historical_spot_prices_ttf_eex}}
	\caption{German power and natural TTF gas daily prices from1/1/2020 to the 31/12/2021.}\label{fig:ttf_eex_markets}
\end{figure}

 The recent invasion of Ukraine by Russia, and the fear of a possible shortage in gas supply for Europe, led to an increase in gas and electricity prices\footnote{Such an increase has begun before the invasion, after the Covid-19 pandemic and many complex factors contributed to it. Nevertheless, the effect of the war on European energy commodities prices has been evident.} which has never seen before as shown in Figure \ref{fig:ttf_eex_markets}. This is easy to explain from an economical point of view, since in Europe natural gas is used to produce the $19.2\%$ of electricity and natural gas power plants usually play the role of marginal technology in the electricity supply curve.

\noindent At this point it should be evident that integration between energy markets must be taken into account if one is interested in energy commodities modeling, risk management or in derivatives pricing. 
\par Over the years many approaches have been proposed to model energy markets in a univariate setting. Pioneering papers in this field dates back at the beginning of the century: \citet{SchwSchm00}, \citet{LS02} and \citet{SCHWARTZ97} focused mainly purely Gaussian framework, whereas \citet{CarteaFigueroa} proposed a mean-reverting model with jumps and a deterministic seasonality component for electricity spot prices. \citet{saifert2006} compared different modeling approaches in power markets, whereas a good summary of energy markets modeling is contained in \citet{Benth2008b}. Many non-Gaussian models based on \Levy\ processes which have been proposed for equities, such as the variance gamma (\citet{MadanCarrChang1998}), the jump-diffusion model (\citet{Merton76}) and the normal-inverse Gaussian process (\citet{BN98}): in particular a Two-Factor version of this process have been recently applied to the energy context by \citet{Piccirilli2021}. Furthermore, many stochastic volatility models, like the ones proposed by \citet{Heston93} and \citet{Bates1996}, can be adapted to model commodity prices behavior. All these powerful tools can be used to properly consider many stylized facts such as jumps in price trajectories, skewness and fat-tails in log-returns distribution and volatility smiles. A review of financial modeling with jump processes can be found in \citet{CT2003}.\\

\par Financial modeling in univariate setting has been deeply investigated, but challenging issues arise when we scale to a multi-commodities market. Within this context, the former modeling techniques become harder to apply in practice and literature is not as rich as in the one dimensional framework. \citet{cs15_2} have shown how some standard models such as the ones proposed by \citet{BLS1973}, \citet{SchwSchm00} and \citet{CarteaFigueroa} can be extended to a multivariate context by adding dependent jumps which are modeled using self-decomposable laws, whereas \citet{KIESEL2016} introduced a structural model to properly consider the market coupling effect in electricity spot markets, in the spirit of what has been proposed by \citet{Carmona2014}.  
\par  A widely recognized approach for energy markets modeling has been proposed by \citet{BenthBenth07} which adapted the framework introduced by \citet{HJM1992} to energy futures market. This modeling technique, together with the calibration of the underlying model, has been studied by many authors such as \citet{Sclavounos2009}, \citet{Hinderks2020}, \citet{Benth2017}, \citet{Broszkiewicz2005}, \citet{Edoli2013} and \citet{Feron2020}. Despite their mathematical completeness and accurateness, these articles seems hard to be used in practice since many practicalities are not covered. The data preparation, a clear explanation on how the model should be implemented in practice, a \enquote{practitioner} interpretation of the results, the management of typical issues arising during the implementation are often missing. This works aims at filling this gap, by collecting all the results presented so far concerning the application of the Heat-Jarrow-Morton (HJM) framework to energy markets. By focusing on European power and gas futures markets, we present a very general approach for data analysis and preparation, for model calibration and, finally, for simulation. Furthermore, we discuss some model limitations and we suggest some possible extensions, trying to preserve both the numerical and the mathematical tractability of the framework. \\

\par The article is organized as follows: Section \ref{sec:market_structure_and_analysis}, focuses on a slice of European power and gas futures markets, and shows how the HJM approach might be the appropriate modeling framework. Section \ref{sec:model} introduces the model and briefly explain its behavior with the support of some \enquote{toy-examples}. In this section we also show how the PCA can be used in order to calibrate the market parameters.
Section \ref{sec:calibration} considers real-market data, deals with different approaches to data preparation and shows how to properly calibrate the model on power and gas European futures prices. In Section \ref{sec:numerical_results} we show simulation results of the whole market considered in Section \ref{sec:calibration}: we analyze the outcomes and we briefly discuss model's strengths and limitations. Section \ref{sec:conclusions} concludes the work and discusses some possible model extensions.

\section{Market structure and analysis}
\label{sec:market_structure_and_analysis}
In this section we focus on the European power and gas futures markets. The results we obtain are valid within this framework, but the approach can be easily adapted to any market. In particular, we consider data from the European Energy Exchange (EEX) for the power markets, whereas data regarding the natural gas markets comes from the Intercontinental Exchange (ICE).

\par Recently, the power and gas futures markets in Europe has experimented an era of expansion and increasing interest. \textcolor{black}{Rules and specifications for trading energy contract on gas and electricity vary among the different power exchange, but the fact that there contracts are settled financially against a reference price, let us consider them as side bets on the physical system. The basic exchange traded contracts are written on the weighted average of the hourly system price over a specified delivery period. During the delivery period the contract is settled against the system price, hence these type of contract are in fact \textit{swap contracts}, exchanging a fixed price against a floating spot price. However, these contracts are commonly called \emph{futures} or \emph{forwards} even if the denomination might be misleading. See the book of \citet{Benth2008b} for an overview on these markets. Another important fact is that these contracts are not traded over the delivery period and the usual financial term time to maturity is often replaced by \emph{time to delivery}.} 
\par Within this market many contracts are present but the most traded ones are those with monthly, quarterly and yearly delivery. For example, a power futures (or swap) 2024 calendar is a contract between two counterparts to buy or sell a specific volume of energy in MWh at fixed price, decided at time $t$, for all the hours of the year. In this case \textcolor{black}{the delivery period lies between} $\tau^{s} = 1/1/2024$ and $\tau^{e}=31/12/2024$. Moreover, at time $\tau^{s}$ the product expires and hence $\tau^{s}$ plays the role of maturity\footnote{Actually the product with delivery starting at $\tau^{s}$ is traded until the \textcolor{black}{business} day before the beginning of the delivery period, but in order to simplify the notation we consider $\tau^{s}$ as the maturity.}. Futures contracts with monthly or quarterly delivery are defined similarly. We will denote the price of such a contract at time $t$ by $F(t,\tau^{s},\tau^{e})$. Concerning the gas market, the situation is slightly different, but for the aim of this paper we can consider it to be similar to power one (see again \citet[Chapter~1]{Benth2008b}). In particular, we can assume that the same contract  $F(t,\tau^{s},\tau^{e})$ written on natural gas, will delivery the gas you need to produce 1 MWh of electricity for the whole delivery period $\left[\tau^{s},\tau^{e}\right]$. \textcolor{black}{From a modeling point of view it is worth it to introduce what we call \emph{fixed-delivery} products, which are denoted hereafter by $F_{d}(t,T)$. This is a contract with maturity $T$ with delivery period given by $d$}. These latter products are not quoted in the market but can be easily obtaining by rearranging $F(t,\tau^{s},\tau^{e})$ as we will show in detail in Section \ref{sec:monthly_flat_fwd_curve}.  \textcolor{black}{For example, assume that today is a day in December 2023 and that we observe the future price $F(t,\tau^{s},\tau^{e})$ with delivery period the whole year 2024. The corresponding fixed delivery contract is given by $F_{Y_1}\left(t,T\right)$ where $T=\tau_{s}$ and $Y_1$ means that it delivers in the following year. Clearly, when January 2024 arrives the contract with delivery for the whole 2024 is no more traded and hence by $F_{Y_1}\left(t,T\right)$ we refer to the product which delivers over the subsequent year, namely 2025. The same \emph{rolling mechanism} applies to product with finer delivery (namely quartes and months).   }

\par
\begin{figure}
	\centering
	\subfigure{\includegraphics[width=0.48\textwidth]{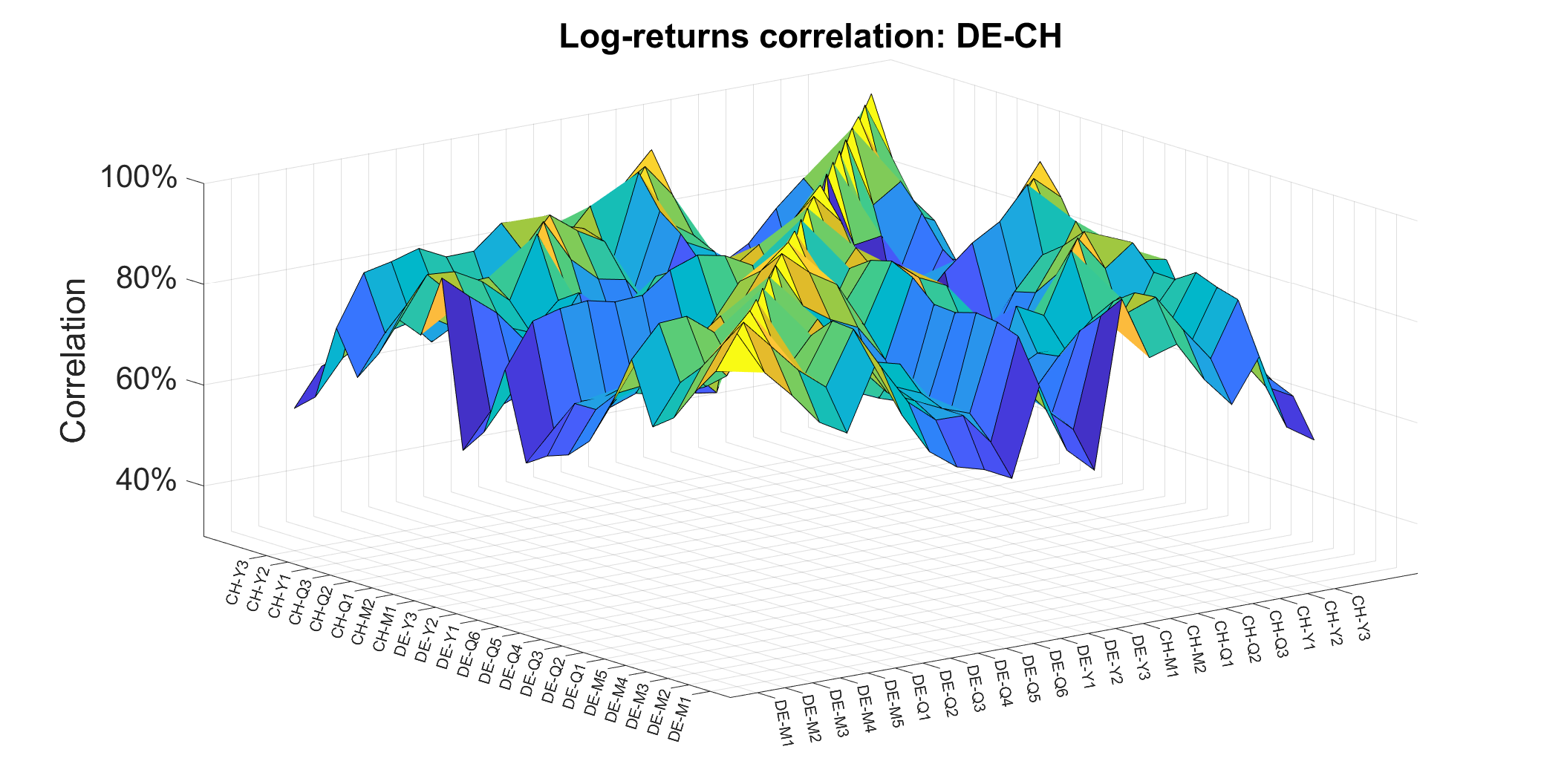}} 
	\subfigure{\includegraphics[width=0.48\textwidth]{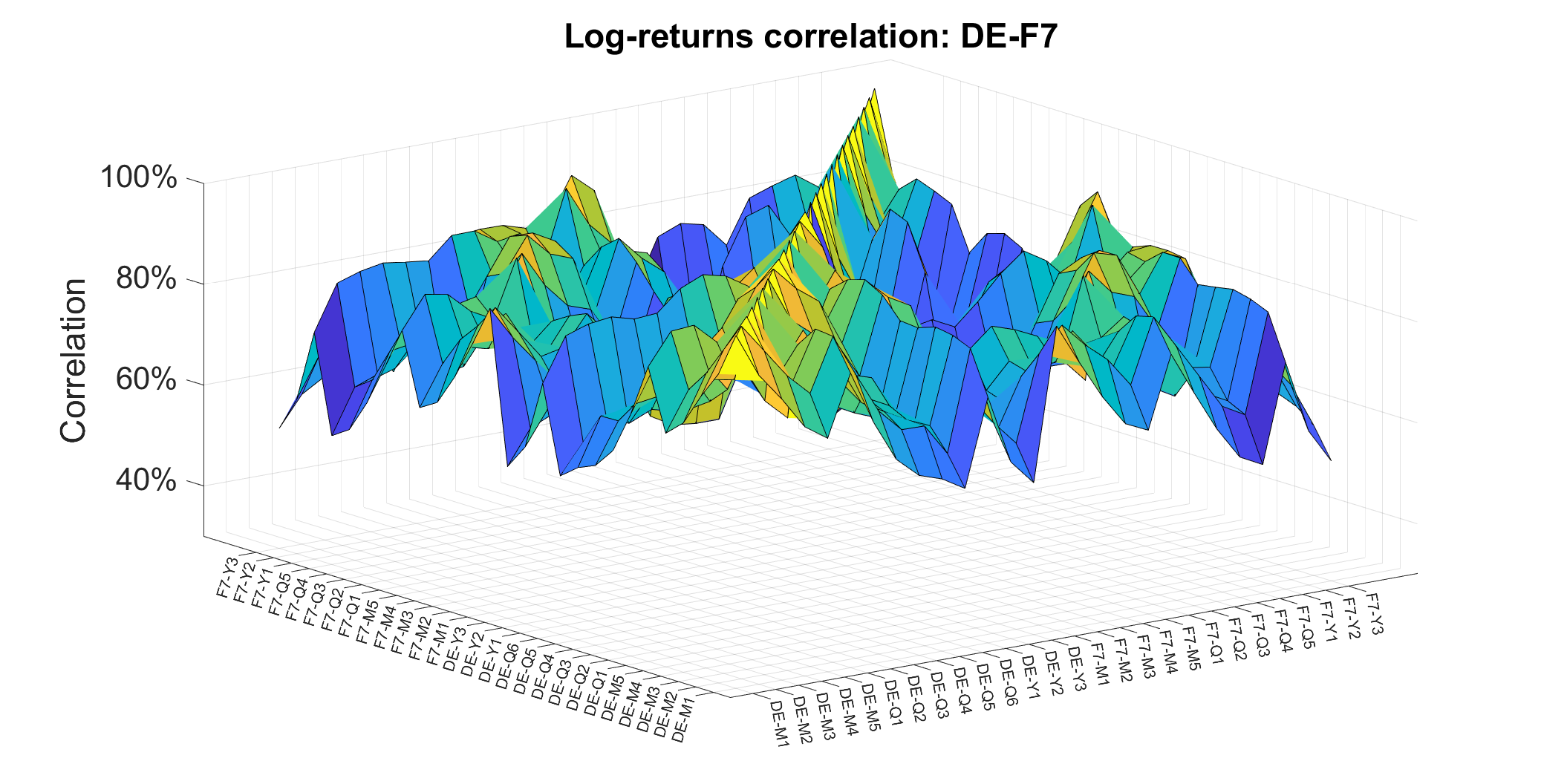}} 
	\subfigure{\includegraphics[width=0.48\textwidth]{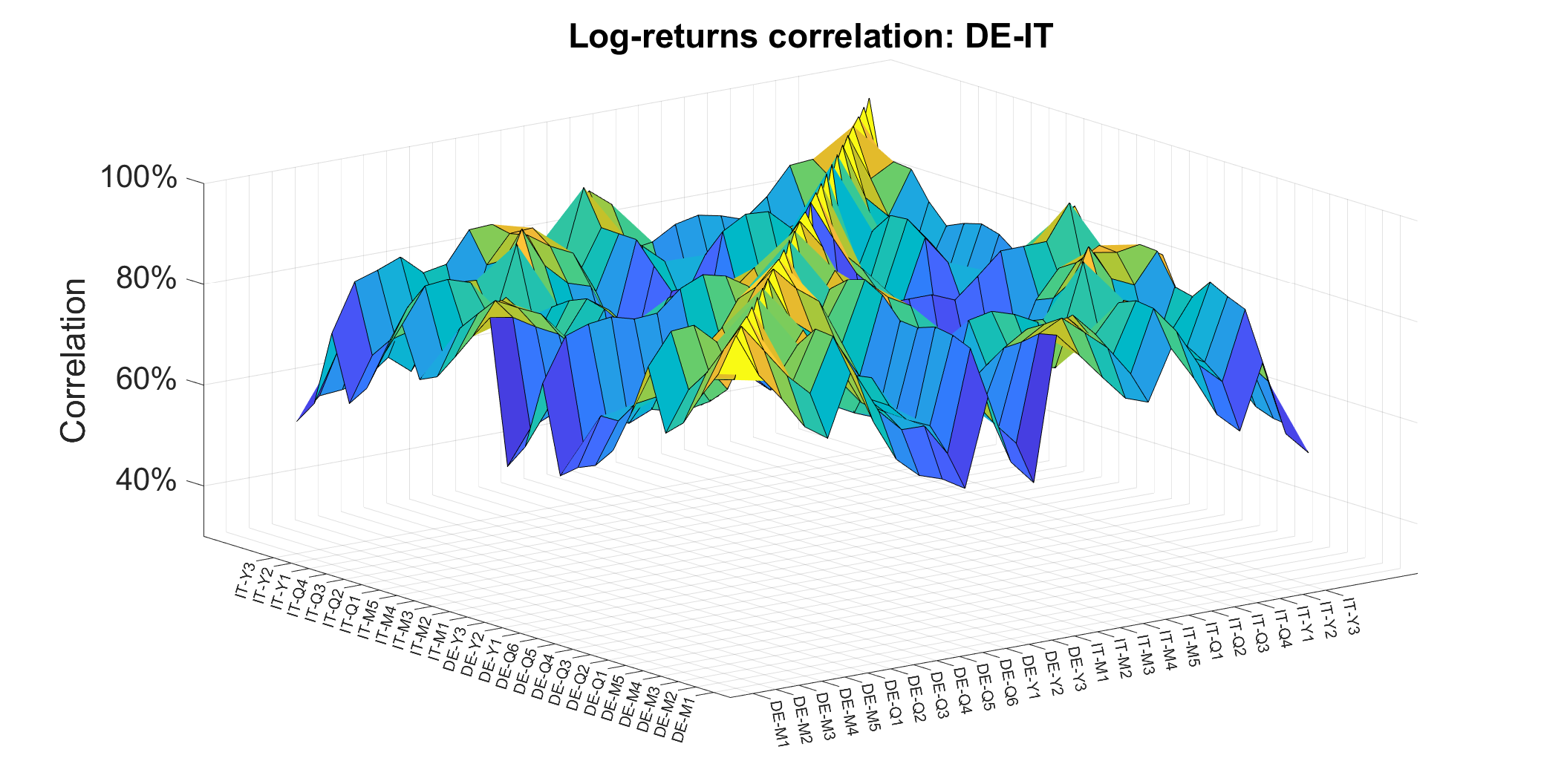}}
	\subfigure{\includegraphics[width=0.48\textwidth]{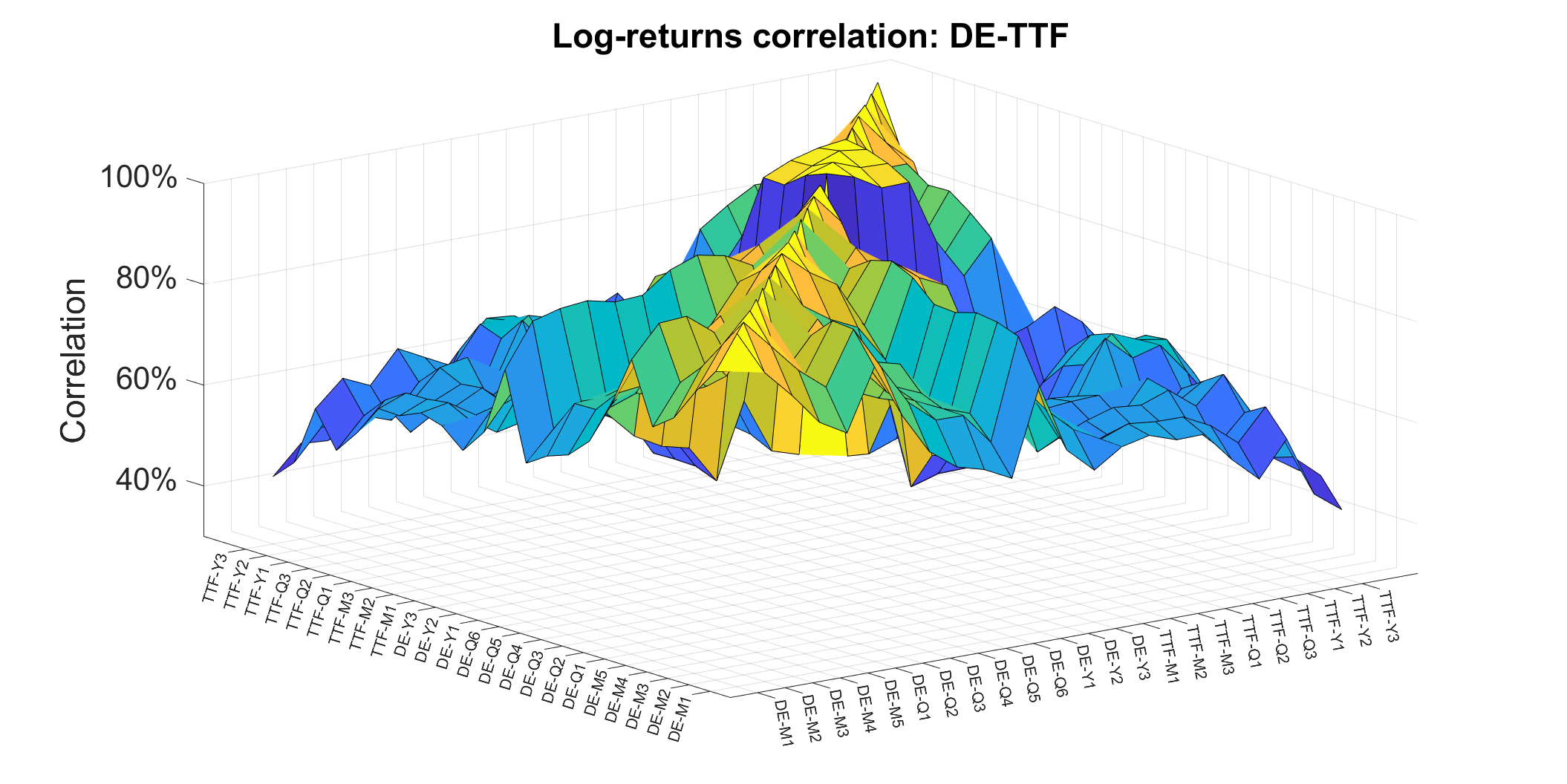}}
	\subfigure{\includegraphics[width=0.48\textwidth]{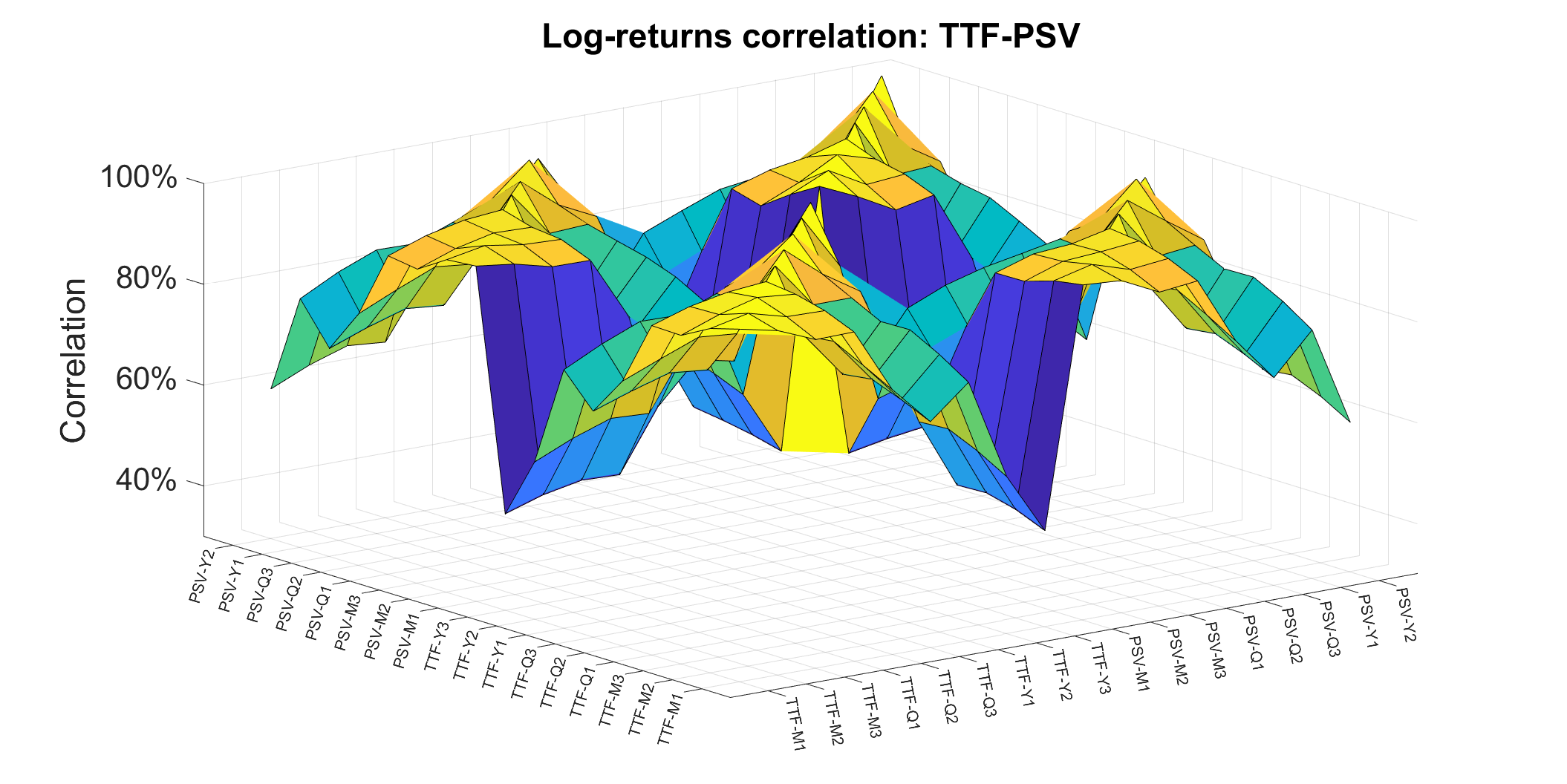}}
	\subfigure{\includegraphics[width=0.48\textwidth]{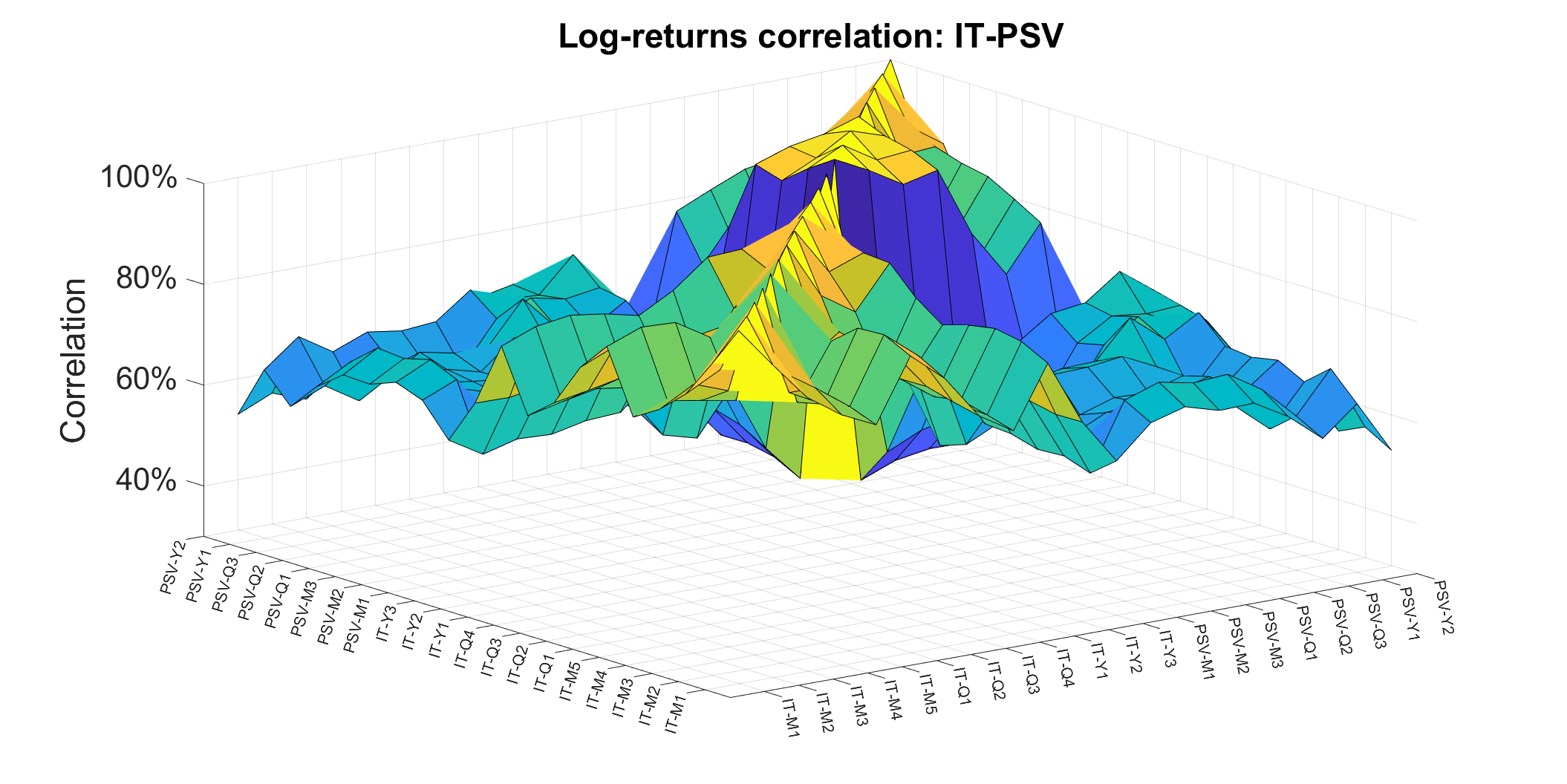}}
	\caption{Log-returns correlation surfaces for different commodities. }
	\label{fig:correlation_surfaces}
\end{figure}

 Without loss of generality, in this paper we focus on four power markets and on two natural gas markets. In particular we consider the German (DE), Italian (IT), French (F7) and Swiss (CH) power futures markets and the TTF and PSV which are the Dutch and the Italian hubs for the natural gas, respectively. Of course, the analysis can be extended to an arbitrary number of markets. \\
The final goal is to model the dynamics of the whole forward curve for the aforementioned markets under a HJM framework. We can assume that each of traded contract acts as a source of uncertainty (which we call a \enquote{random factor}) for the determination of the forward curve dynamic. On the other hand, as we stated before, all these markets are co-integrated and hence the hope is that we can use a small number of random factors to successfully model the forward curves dynamic. In oder to verify this assumption, we consider daily futures prices between $1/1/2020$ and  $31/12/2022$ and we compute the correlation of the log-returns, following the approach proposed by \citet{Sclavounos2009}. In Figure \ref{fig:correlation_surfaces} we plot the correlation surfaces between daily log-returns of several futures products with fixed delivery 
: we observe that the linear correlation coefficient is significant and hence this bodes well that only few \emph{stochastic factors} seem to drive the whole structure of the market.

\par
\begin{figure}
	\centering
	\subfigure{\includegraphics[width=0.48\textwidth]{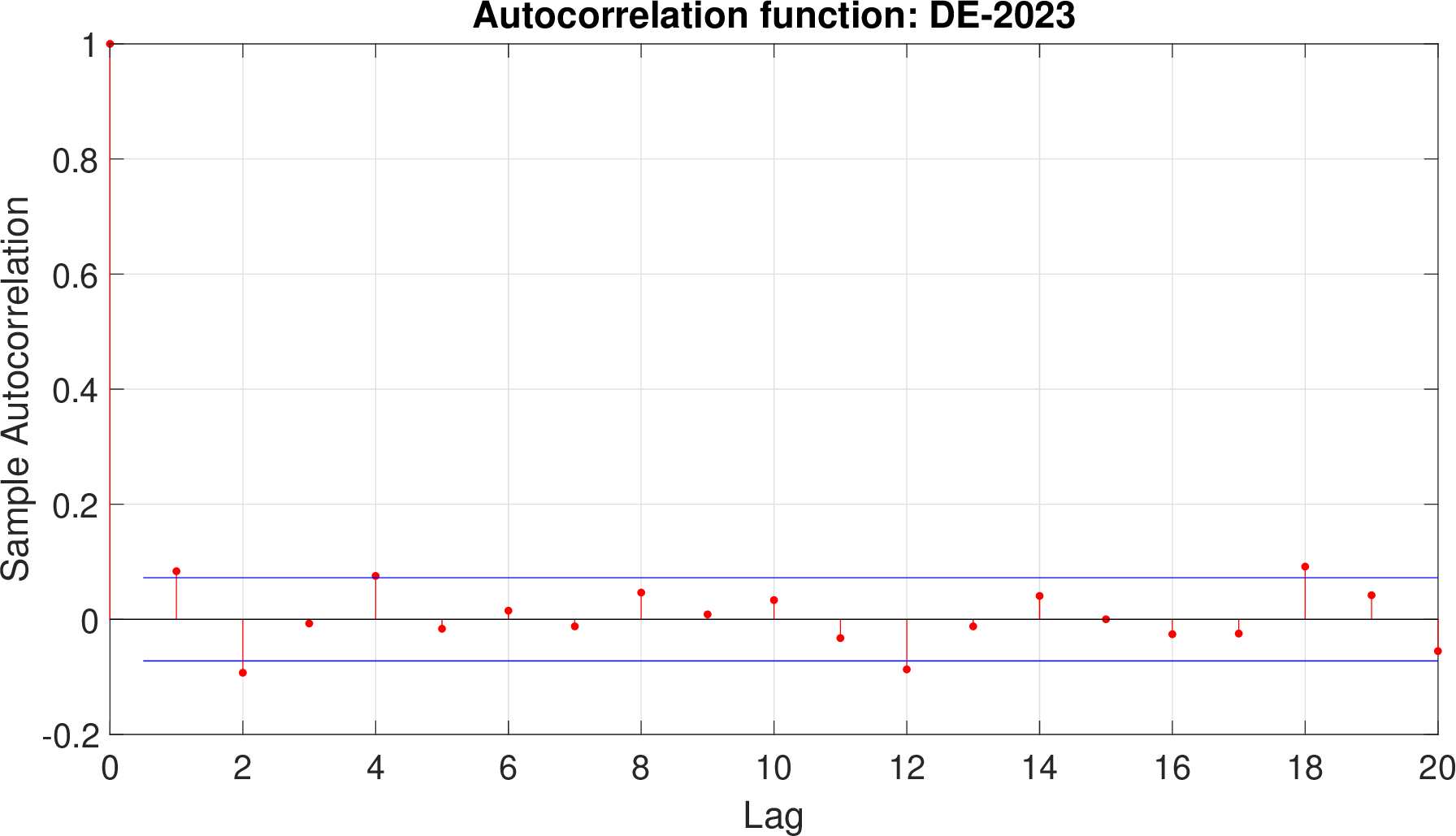}} 
	\subfigure{\includegraphics[width=0.48\textwidth]{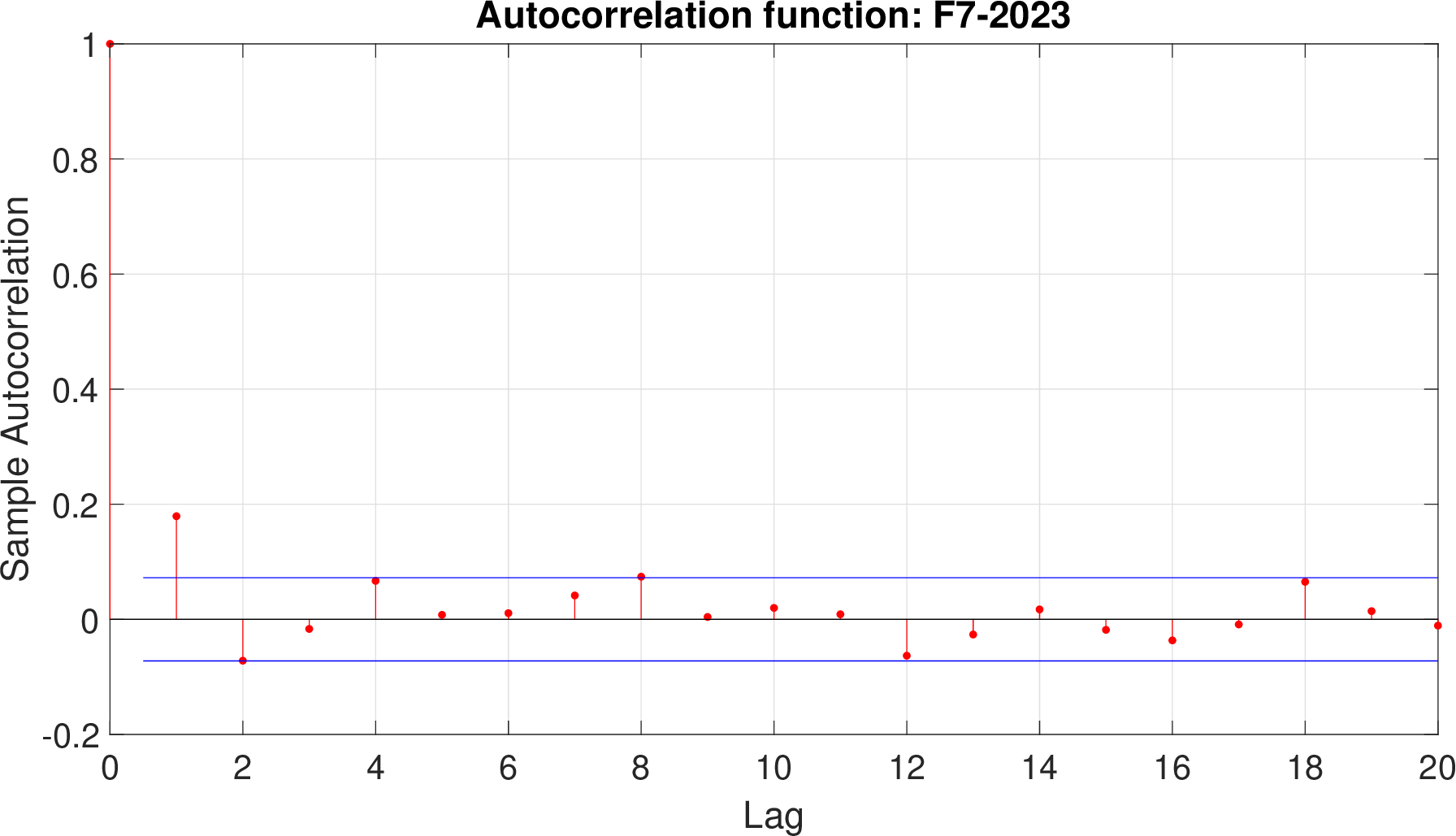}} 
	\subfigure{\includegraphics[width=0.48\textwidth]{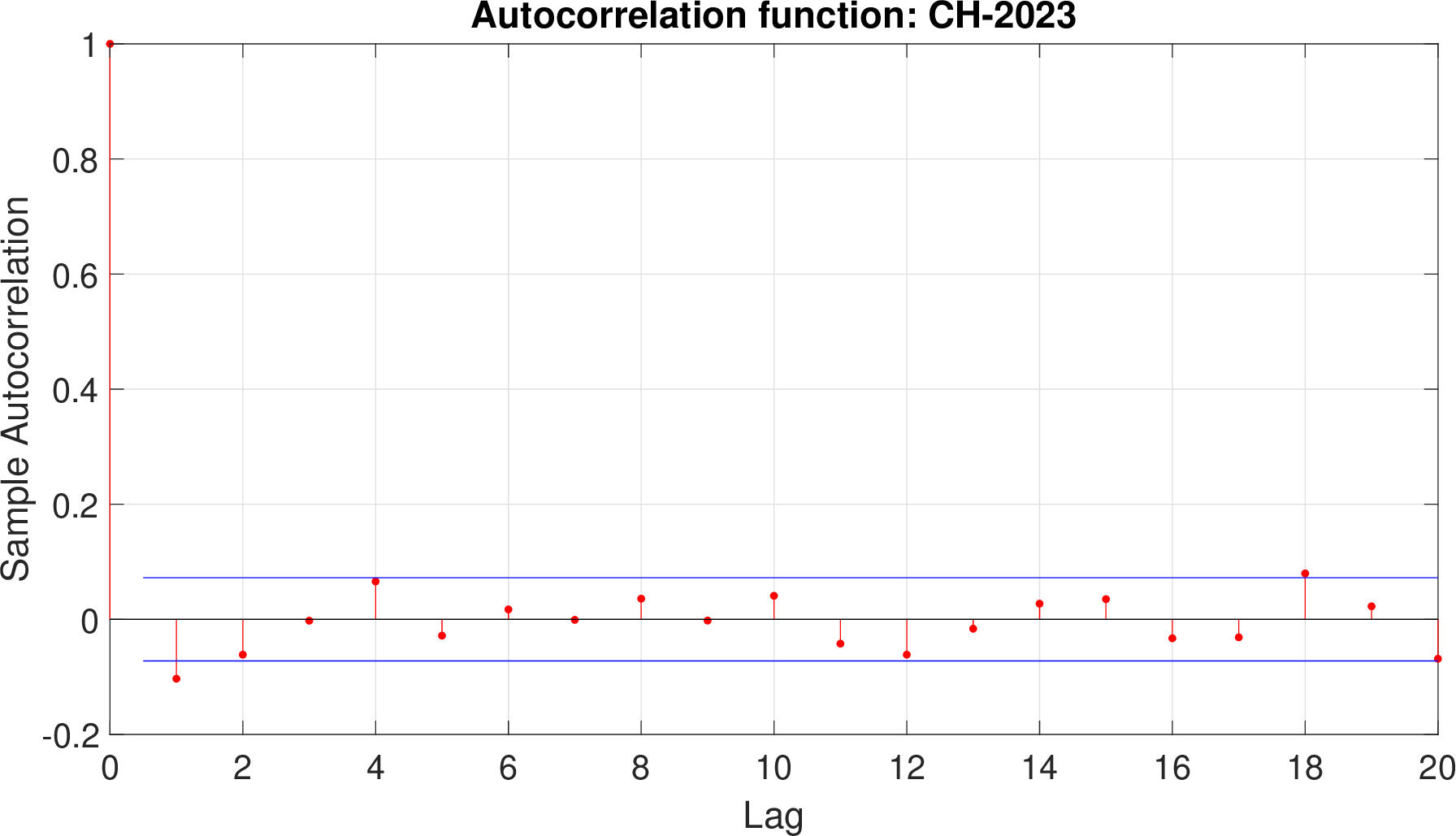}}
	\subfigure{\includegraphics[width=0.48\textwidth]{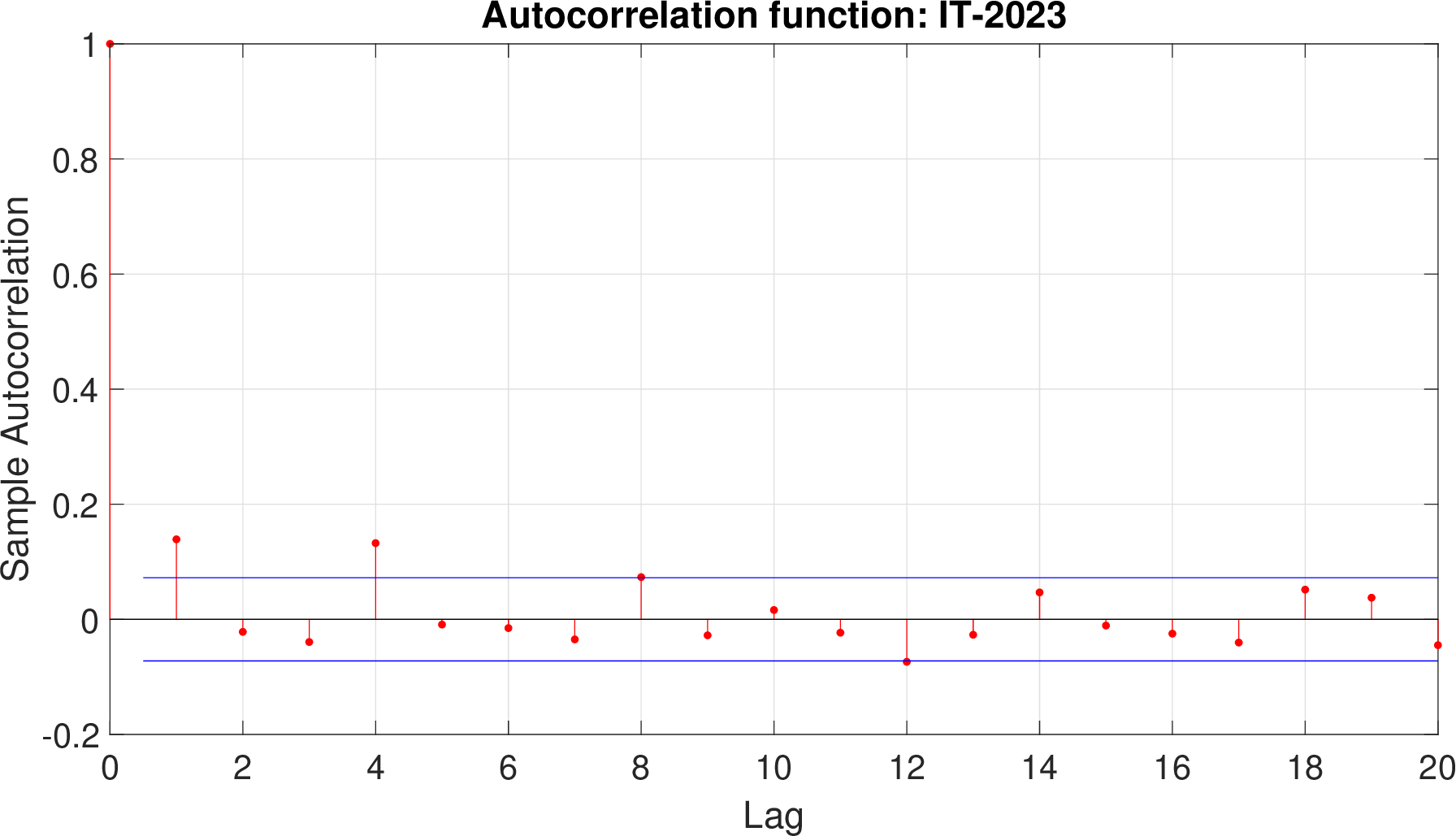}}
	\subfigure{\includegraphics[width=0.48\textwidth]{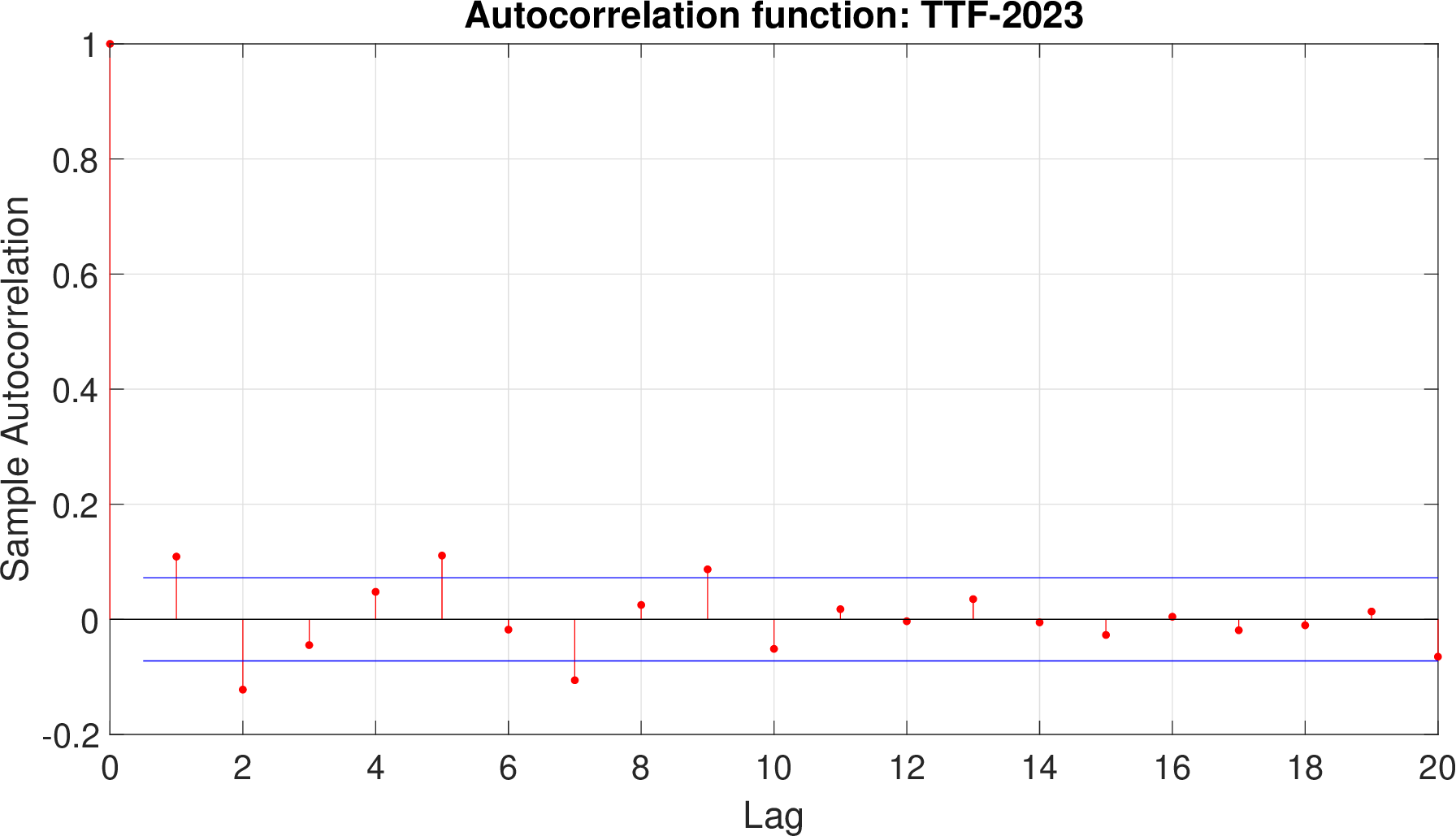}}
	\subfigure{\includegraphics[width=0.48\textwidth]{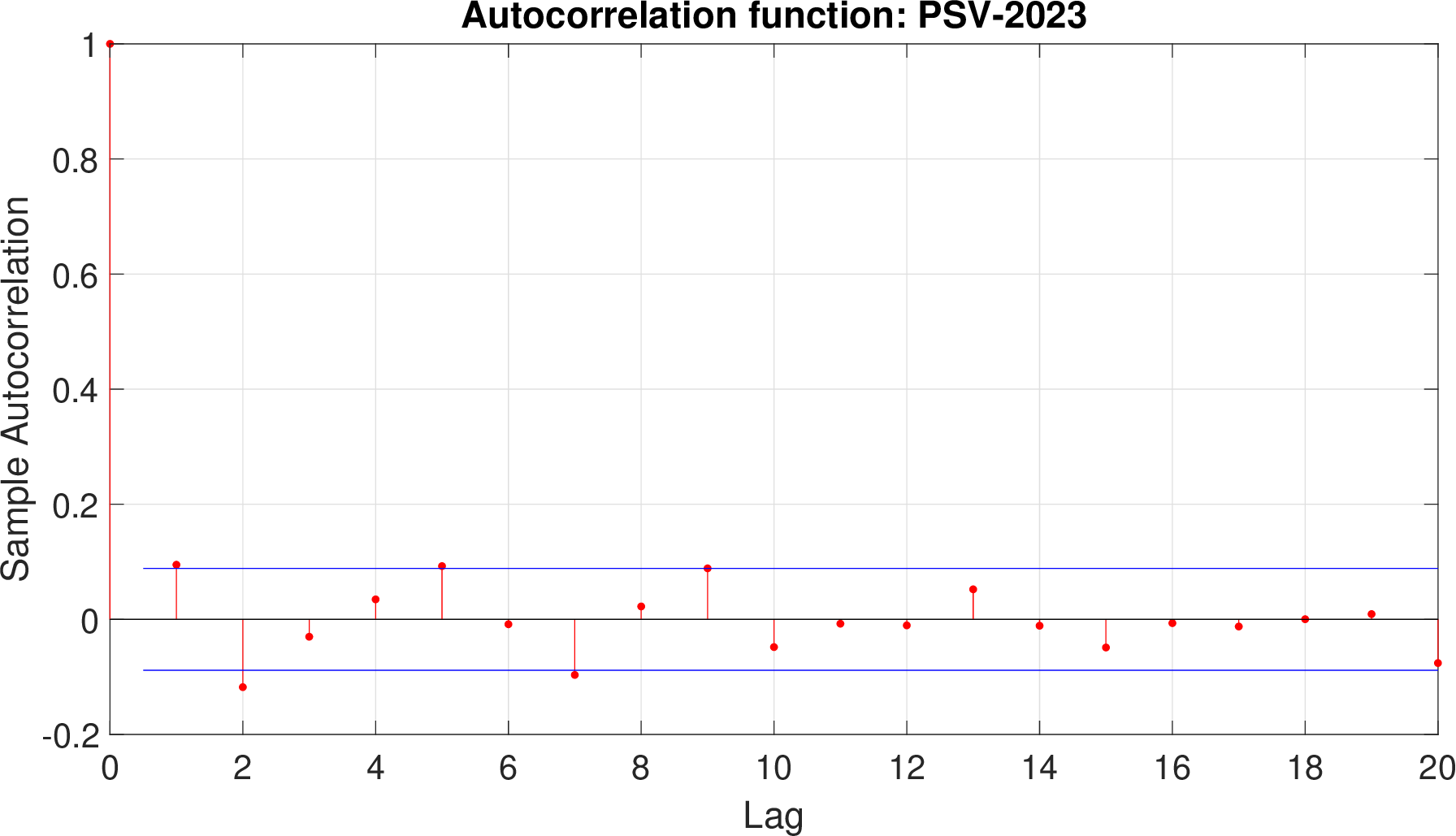}}
	\caption{Sample ACF computed on log-returns.}
	\label{fig:ACF_historical_log_ret}
\end{figure}

 Most financial models relies upon stochastic processes in continuous time and, mainly, on \Levy\ processes: the well known Brownian motion is just the simplest of them.  Working in a \Levy\ framework leads to the development models characterized by a very reach structure which can be used to efficiently include many stylized facts. The interested reader can refer to \citet{Sato} and \citet{applebaum2009} for an overview on \Levy\ processes and to \citet{CT2003} for applications to financial markets. By definition, \Levy\ processes has independent increments. For this reason, before using \Levy\ processes for modeling purposes, one has to check that the increments are independent. Following the approach proposed in \citet{Brigo2007}, we compute the auto-correlation function (ACF) on six different time series, one for each market, on the calendar product with delivery the year 2023. We consider the series of daily log-returns $x_{1},x_{2},\dots,x_{n}$, with
\begin{equation*}
	x_{i} = \ln \frac{P_{t_{i+1}}}{P_{t_{i}}},
\end{equation*}
where $P_{t_{i}}$ denotes the price of a (general) risky asset at time $t_{i}$ and we compute the ACF with lag $k$ as:
\begin{equation*}
	ACF(k) = \frac{1}{(n-k)\hat{v}} \sum_{i=1}^{n-k}\left(x_{i} - \hat{m}\right)\left(x_{i+k} - \hat{m}\right),\quad k=1,\dots,20,
\end{equation*}
where $\hat{m}$ and $\hat{v}$ are the sample mean and variance. Roughly speaking we can consider the ACF as an estimate of the correlation between the random variables $X(t_{i})$ and $X(t_{i+k})$. 
ACF for the six selected products above are shown in the charts in Figure \ref{fig:ACF_historical_log_ret}. For the all of them we do not observe any significant lags in the historical return time series, which means the independence assumption is acceptable in this case. Changing the delivery period of the product we get similar results. Therefore \Levy\ processes can be used in order to properly model the futures prices.

\par
\begin{figure}
	\centering
	\subfigure{\includegraphics[width=0.48\textwidth]{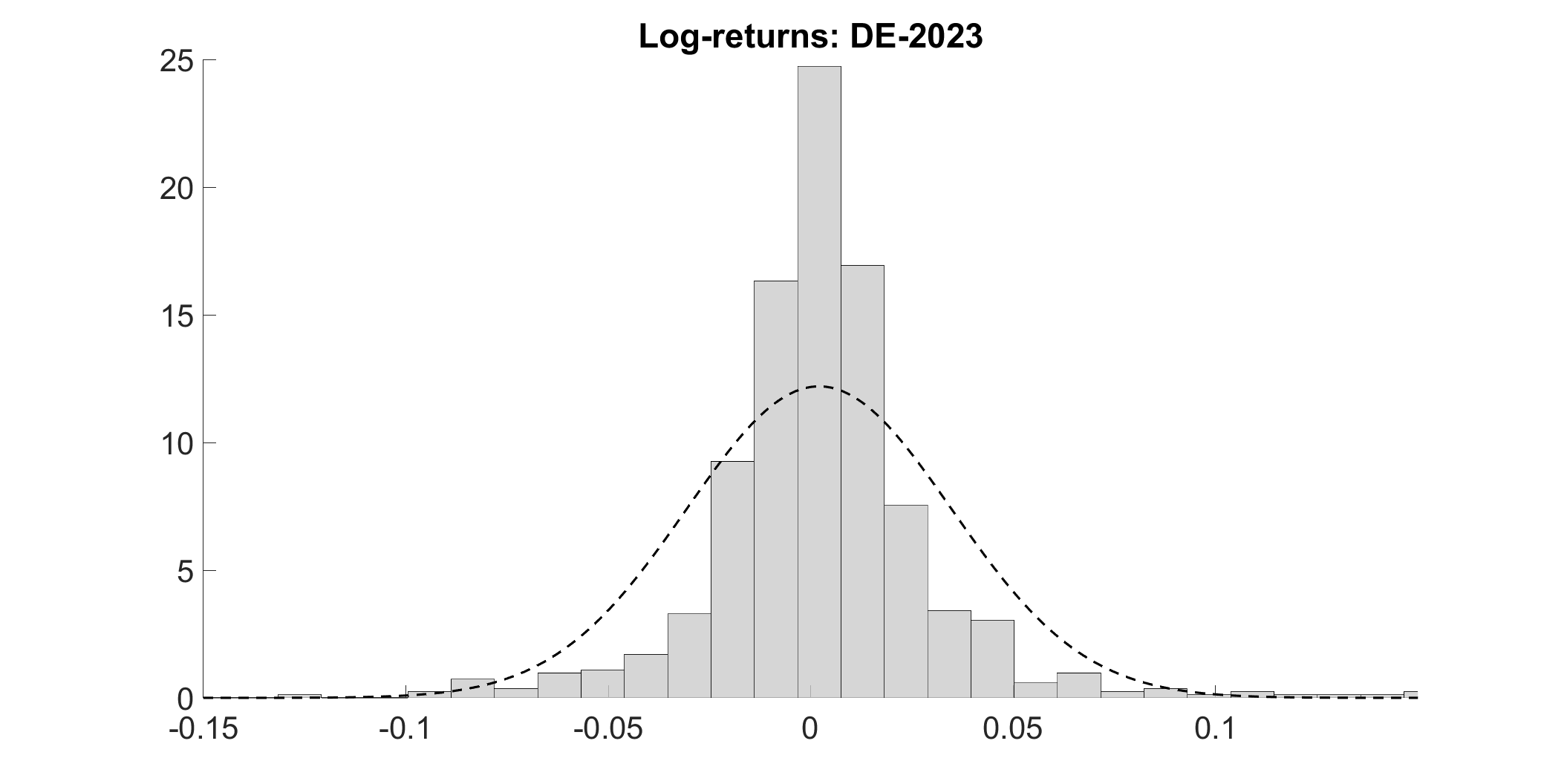}} 
	\subfigure{\includegraphics[width=0.48\textwidth]{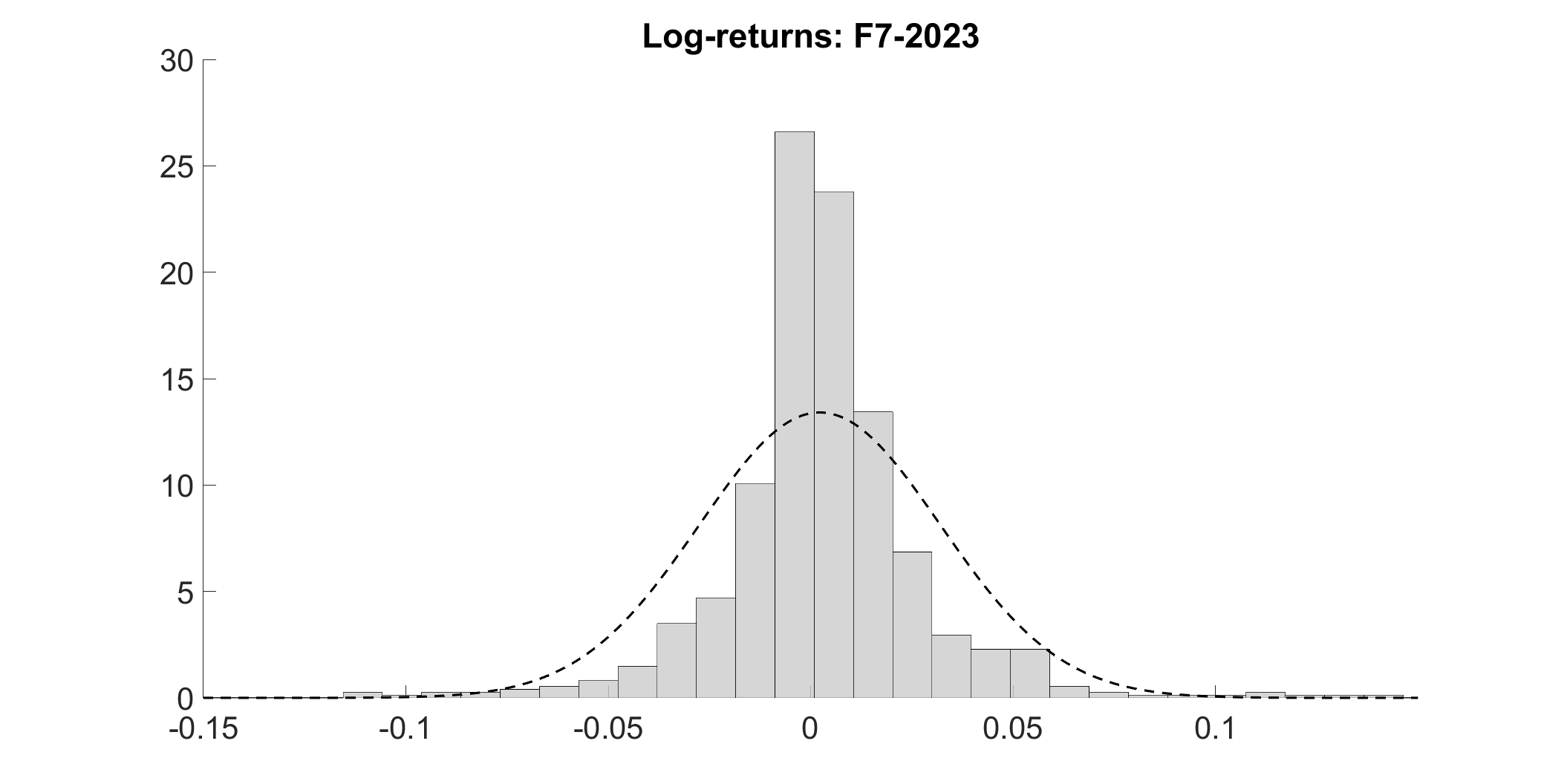}} 
	\subfigure{\includegraphics[width=0.48\textwidth]{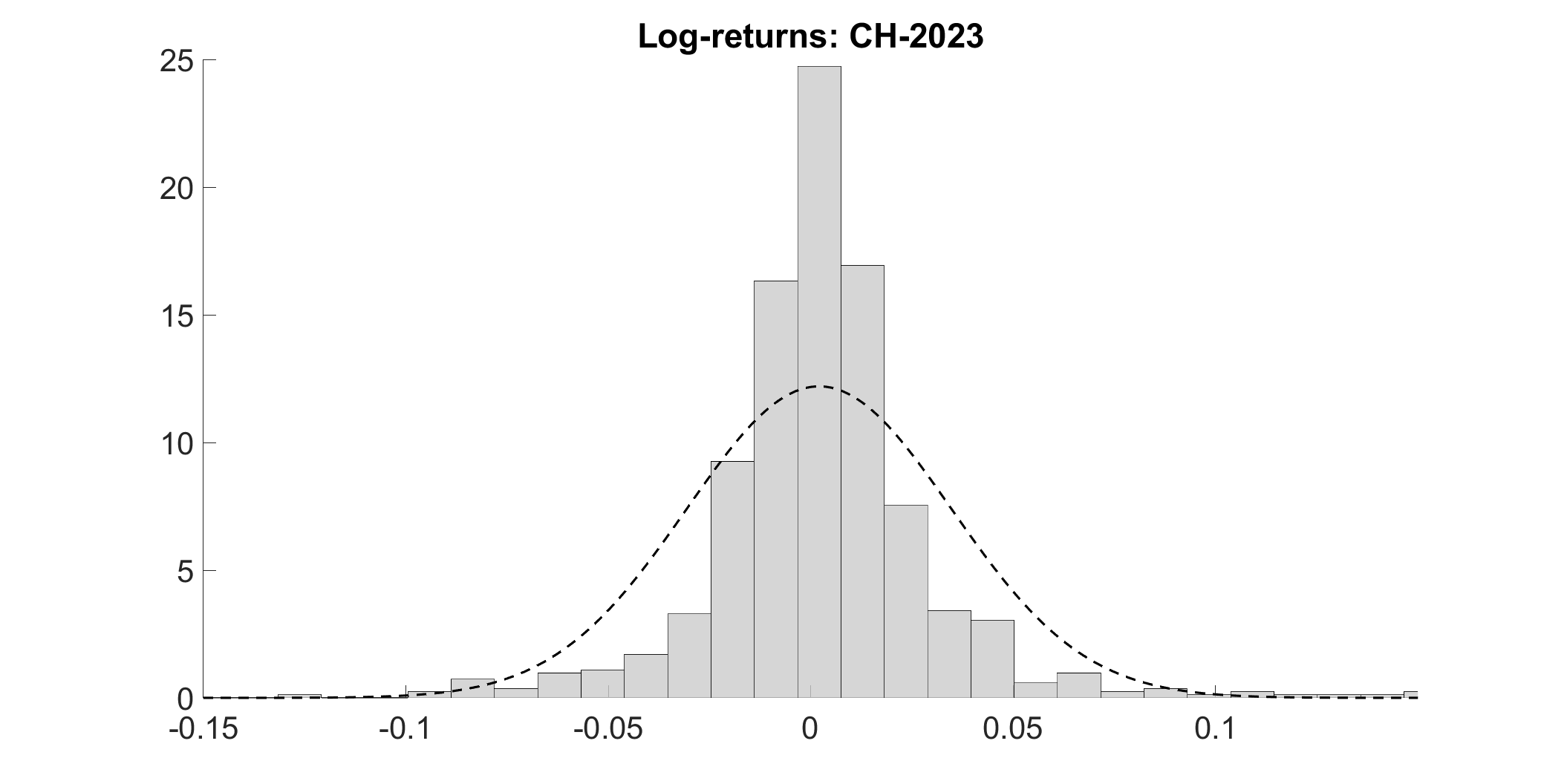}}
	\subfigure{\includegraphics[width=0.48\textwidth]{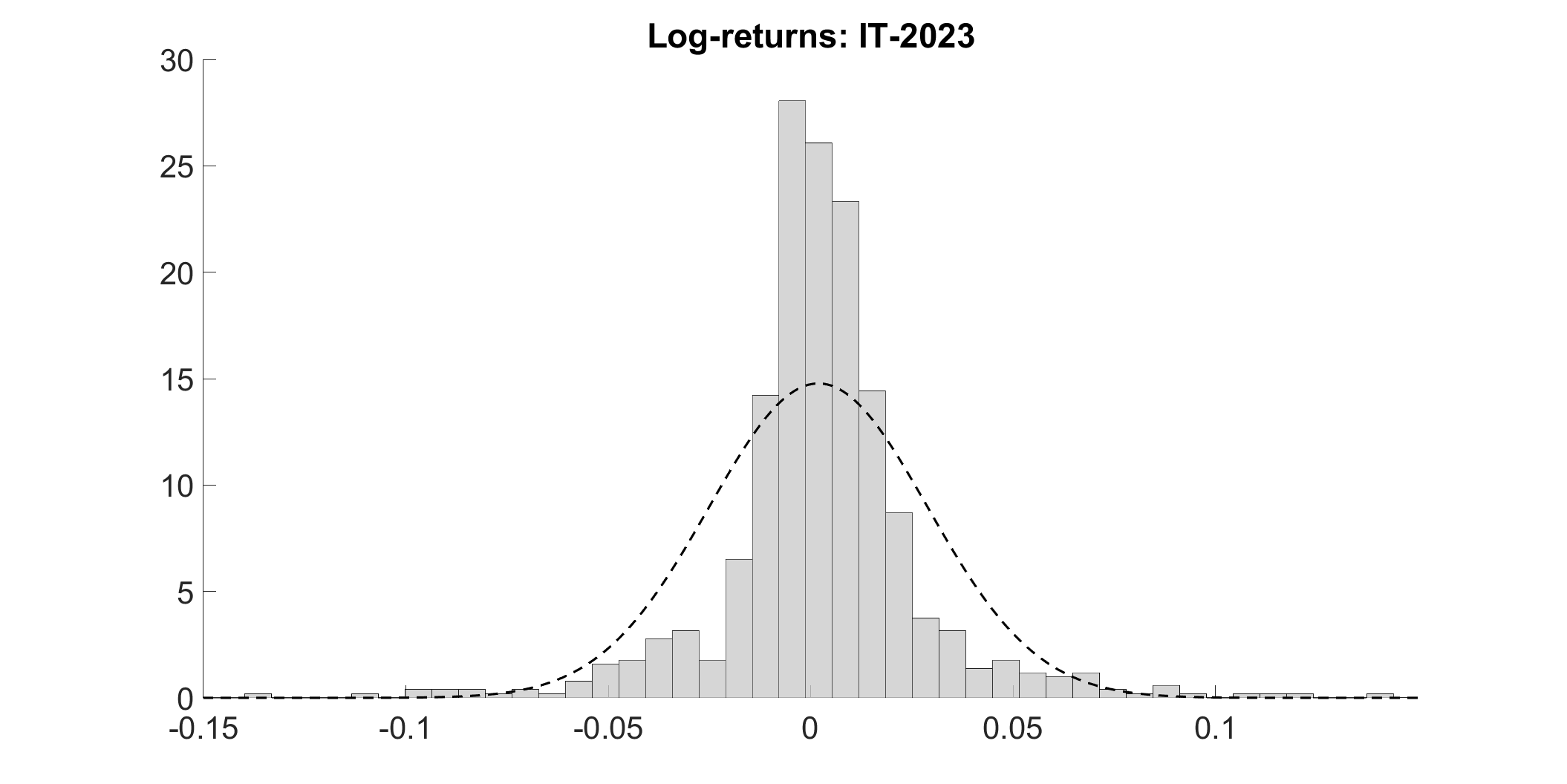}}
	\subfigure{\includegraphics[width=0.48\textwidth]{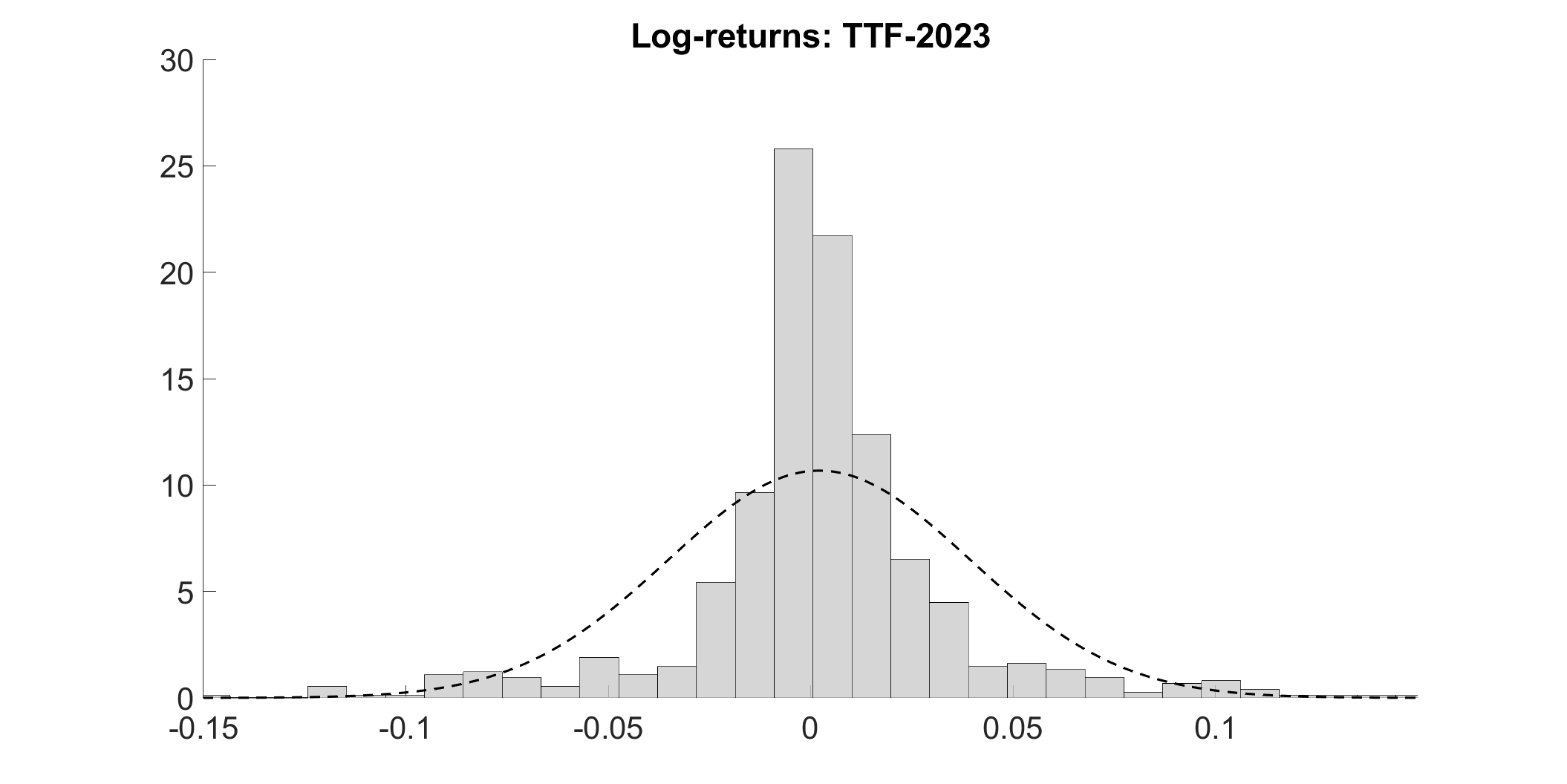}}
	\subfigure{\includegraphics[width=0.48\textwidth]{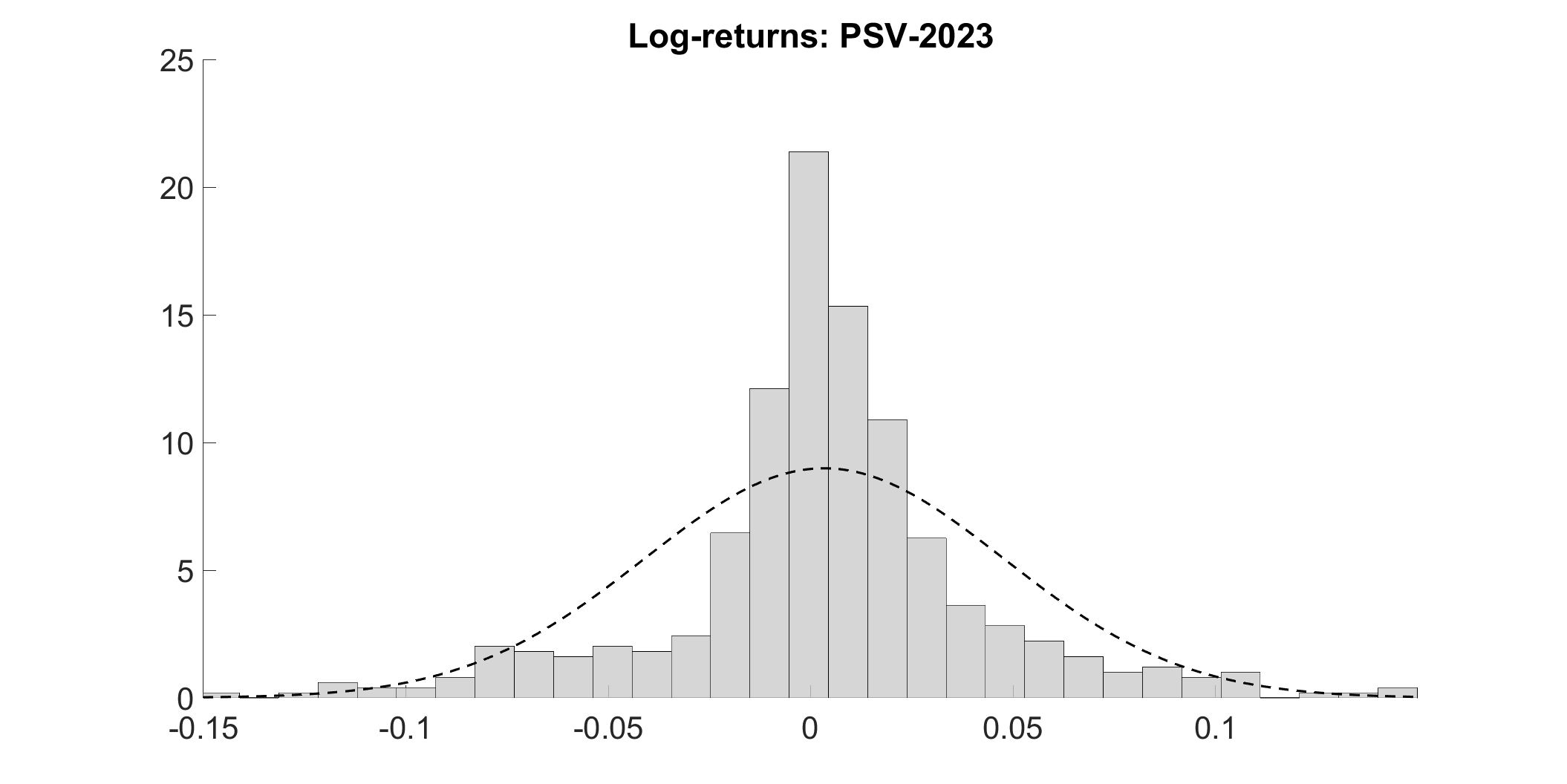}}
	\caption{Daily futures log-returns densities for calendar 23 products.}
	\label{fig:density_log_ret}
\end{figure}

 Within the classical HJM framework log-returns are assumed to be normally distributed. As observed by many authors (\citet{BenthKoekebakker05}, \citet{Frestad2007} and \citet{green2006})  in energy markets log-returns are not normally distributed, but their distribution is often skewed and presents heavy tails or fat tails effect. Furthermore, the log-returns volatility is not constant but often clusters appear. In Figure \ref{fig:density_log_ret} we plot the empirical probability density function of log-returns compared with the normal one fitted on the same set of data as above. In all cases we observe that the real log-returns distribution is peaked and presents tails which are heavier than the normal distribution ones. Furthermore, all distributions result to be skewed. 
\par In light of this results, by choosing a normal distribution we might lose some market peculiarities. On the other hand, in order to get a simple and stable calibration methodology the hypothesis of normality in log-returns is commonly accepted by practitioners. 
\par Gaussianity hypothesis in log-returns might be relaxed by using \Levy\ processes (see \citet{Benth2008b}) or \Levy\ copulas, as proposed by \citet{Panov2019} and \citet{CT2003}, but this approach is very hard to handle from a practical point of view, \textcolor{black}{as pointed out by the author themselves}. Indeed, the calibration step is hard to tackle, even if the multidimensional distribution is versatile as the Normal Inverse Gaussian one, especially if the number of underlying asset is large. Several authors, such as \citet{SL2010}, \citet{Schoutens03}, \citet{BB2013}, \citet{Buchman2020Calib} and \citet{gardini2021}, investigated other techniques in order to consider dependence in log-returns remaining in a \Levy\ framework. Most of them work quite well if the number of the risky assets to remains small, but complications arise when one deals with many risky underlying assets since the number of parameters rapidly grows as the number of underlying assets increases. Consequently calibration becomes hard to perform in practice and its results might be unreliable. For these reasons, \textcolor{black}{despite its limitations}, the Gaussian framework remains a milestone among practitioners in multi-commodity energy markets. On the other hand, if a focus on a single product is required, a more general model among the ones we listed should be considered.\\

\par In the next sections, following the idea in \citet{Benth2008b}, we apply the HJM framework to energy markets, with a particular focus on European power and natural gas markets. Nevertheless, the approach is very general and could be easily adapted to other commodity markets such as oil, precious metals and  agricultural products.

\clearpage

\section{The model}
\label{sec:model}
\textcolor{black}{In this section we discuss in detail the HJM framework applied to energy markets. We briefly discuss the mathematical setting and we refer the interested reader to \citet[Chapters~6,8]{Benth2008b}. Furthermore, in the didactic spirit of the article, we discuss some \enquote{toy-models} which are useful to fix the main modeling concepts. Once that the dynamics and the calibration procedure for these simple models are clear, we introduce the most general framework which turns out to be a simple extension of the previous setting.}

\subsection{\textcolor{black}{Modeling aspects and mathematical setting}}
The key diver for the swap price is the underlying spot $S=\left\{S(t); t \ge 0\right\}$. Assume that we have entered a forward contract delivering the spot at time $\tau$ and that the interest rate $r\ge0$ is constant. Denoting $f(t,\tau)$ the forward price at time $t\le \tau$ of entry the contract, the payoff is given by:
\begin{equation*}
	S(\tau) - f(t,\tau),
\end{equation*}
at delivery time $\tau$. Assuming that the market does not admit arbitrages, from the theory of mathematical finance we know that there exists a measure $\Q$, called \emph{risk-neutral measure}, such that the price of any derivatives is given by the discounted expectation under $\Q$ of its payoff. Since enter a forward contract is free we have that:
\begin{equation*}
	0 = e^{-r(T-\tau)} \E^{\Q}\left[S(\tau) - f(t,\tau)|\mathcal{F}_{t}\right],
\end{equation*}
where $\left(\mathcal{F}(t); t \ge 0\right)$ is the filtration containing all market information up to time $t$. The forward price is set at time $t$ and hence $f(t,\tau)$ is adapted to $\left(\mathcal{F}(t); t \ge 0\right)$ and hence:
\begin{equation*}
	f(t,\tau) = \E^{\Q}\left[S(\tau)|\mathcal{F}_{t}\right].
\end{equation*}
It is easy to check that the process $f=\left\{f(t,\tau),t\le \tau \right\}$ is a $\Q$-martingale.\\

\par Having that is mind, assume that $T^{*}$ is the final time for the market and, in the spirit of the HJM approach, for $0\le t\le \tau \le T^{*}$ we model the forward prices dynamics under $\Q$ as:
\begin{equation}
	f(t,\tau) = f(0,\tau)\exp\left\{\int_{0}^{t}a(u,\tau)du + \sum_{k=1}^{\tilde{N}}\sigma_{k}(u,\tau)dW_{k}(u)\right\}
	\label{eqn:fwd_contracts_dynamics}
\end{equation}
where $a\left(u,\tau\right)$ and $\sigma_{k}\left(u,\tau\right)$ are real-valued continuous functions on $\left[0,\tau\right]\times \left[0,T^{*}\right]$, with the additional hypothesis that $\sigma_{k}$ are positive functions. Since $f(t,u)$ must be $\Q$-martingales some restrictions on the drift $a\left(u,\tau\right)$ must be imposed. By \citet[Proposition~6.1]{Benth2008b} we get the following drift condition:
\begin{equation}
	\int_{0}^{t}a(u,\tau)du + \frac{1}{2} \sum_{k=1}^{\tilde{N}}\int_{0}^{t}\sigma_{k}^{2}(u,\tau)du=0
	\label{eqn:drift_condition}
\end{equation}
If Equation \eqref{eqn:drift_condition} does not hold arbitrages may appear. 
\par In energy markets it is customary to find contracts with overlapping delivery periods. For example, one can observe the calendar contract delivering electricity on the whole year, together with four contract on the same year each of them delivering electricity on a quarter. In order to avoid arbitrages, one needs to satisfy certain conditions between prices of swaps contracts. It could be show that in continuous time we have to satisfy \emph{no-arbitrage conditions} of the form:
\begin{equation}
	F(t,\tau^{s},\tau^{e}) = \int_{\tau^{s}}^{\tau^{e}} \hat{w}\left(u,\tau^{s},\tau^{e}\right)f(t,u)du,
	\label{eqn:noa_conditions}
\end{equation}
where $\hat{w}\left(u,\tau^{s},\tau^{e}\right)$ is a weighting function of the form:
\begin{equation*}
	\hat{w}\left(u,s,t\right) = \frac{w(u)}{\int_{t}^{s}w(v)dv},
\end{equation*}
where $w(u)=1$ if the settlement of the swap takes place at the end of the delivery period (see \citet[Chapter~4]{Benth2008b}).
\par By analogy to the modeling adopted for future contracts $f(t,u)$ we set the dynamics for $F\left(t,\tau^{s},\tau^{e}\right)$ as:
\begin{equation*}
	F(t,\tau^{s},\tau^{e}) = F(0,\tau^{s},\tau^{e})\exp\left\{\int_{0}^{t}A(u,\tau^{s},\tau^{e})du + \sum_{k=1}^{\tilde{N}}\int_{0}^{t}\Sigma_{k}\left(u,\tau^{s},\tau^{e}\right)dW(u)\right\},
\end{equation*}
where $A(u,\tau^{s},\tau^{e})$ and $\Sigma_{k}\left(u,\tau^{s},\tau^{e}\right)$ are continuous real-valued functions and $F(0,\tau^{s},\tau^{e})$ is the initial forward curve. In order to have that swap prices are $\Q$-martingale we must have that:
\begin{equation*}
	\int_{0}^{t}A(u,\tau^{s},\tau^{e})du + \frac{1}{2}\sum_{k=1}^{\tilde{N}}\int_{0}^{t}\Sigma_{k}(u,\tau^{s},\tau^{e})du=0.
\end{equation*}

Unfortunately it turns out that such a model for the swap price $F\left(t,\tau^{s},\tau^{e}\right)$ does not respect the no-arbitrage conditions in continuous time in Equation \eqref{eqn:noa_conditions}.
\par For this reason, other models must be consider. A simple way to obtain such models is to generate them from forward contracts $f(t,\tau)$. Otherwise, one can be inspired by the LIBOR market models for fixed income theory (see \citet{BrigoMercurio2001} and \citet{osterleegrzelak2020})

\subsubsection{\textcolor{black}{Swap models on forward}}
The first attempt to model the dynamics of the swaps contracts $F(t,\tau^{s},\tau^{e})$ is to start from the dynamics of forward prices $f(t,\tau)$ in Equation \eqref{eqn:fwd_contracts_dynamics} together with the drift condition in Equation \eqref{eqn:drift_condition} and derive the swap price as:
\begin{equation*}
	F(t,\tau^{s},\tau^{e}) = \int_{\tau^{s}}^{\tau^{e}} \hat{w}\left(u,\tau^{s},\tau^{e}\right)f(t,u)du,
\end{equation*}
so that non-arbitrage relations of Equation \eqref{eqn:noa_conditions} hold. Unfortunately this approach leads to non-Markovian models (except for trivial cases) which are hard to calibrate and heavy to simulate as was discussed in \citet[Chapter~6]{Benth2008b}. A different approach relies upon the so called \emph{market models}.
\subsubsection{\textcolor{black}{Market models}}
Using the same approach of the LIBOR models, one can construct a dynamics for the \emph{traded} contracts matching the observed volatility term structure. Unlike the HJM approach, we only consider only those products which are traded in the market, avoiding the continuous-time no-arbitrage conditions. First, contract which cannot be decomposed into contracts with smaller delivery period must be singled out. For example, we model the monthly contract and we do not consider year and quarter products. These contract are called basic contracts and they have delivery periods given by:
\begin{equation*}
	\left\{\left[\tau_{1}^{s},\tau_{1}^{e}\right],\dots,\left[\tau_{\tilde{N}}^{s},\tau_{\tilde{N}}^{e}\right]\right\},
\end{equation*}
where $\tilde{N}$ is the total number of the contracts we consider. Hence we state that the dynamics for each of them is given by:
\begin{equation*}
	F(t,\tau_{m}^{s},\tau_{m}^{e}) = F(0,\tau_{m}^{s},\tau_{m}^{e})\exp\left\{\int_{0}^{t}A_{m}(u,\tau^{s},\tau^{e})du + \sum_{k=1}^{\tilde{N}}\int_{0}^{t}\Sigma_{c,k}(u,\tau^{s},\tau^{e})dW_{k}(u)\right\},
\end{equation*}
where $A_{m}$ and $\Sigma_{c,k}$ are continuous real-valued functions on $\left[0,\tau_{m}^{b}\right]$, together with the drift condition:
\begin{equation*}
	\int_{0}^{t}\left(A_{m}(u,\tau_{m}^{s},\tau_{m}^{e}) + \frac{1}{2}\sum_{k=1}^{\tilde{N}}\Sigma_{c,k}^{2}(u,\tau_{m}^{s},\tau_{m}^{e})\right)du=0,
\end{equation*}
which ensures that swap price dynamics is a $\Q$-martingale.

\par In the next session we introduce some \emph{toy-models} which will help the reader to better understand the concept and hence we gradually move through the general modeling approach.

\subsection{A single factor toy-model}
\label{sec:toy_model}
Consider the price of the fixed \textcolor{black}{swap contract $F\left(t,\tau^{s},\tau^{e}\right)$} with maturity $T\le \tau^{s}$\footnote{\textcolor{black}{In order to simplify the discussion, we usually assume that $T=\tau^{s}$, namely that the contract expires as soon as the delivery period starts. This assumption is no so far from what happens in real markets.}}, signed at $t_{0}$, for $t \in \left[t_{0},T\right]$. Assume that we have only a single source of uncertainty and, assuming to work under the risk neutral measure $\Q$, consider a dynamics of the following type:   

\begin{equation}
	\frac{dF(t,\tau^{s},\tau^{e})}{F(t,\tau^{s},\tau^{e})} = \sigma(t,T)dW(t),\quad F(t_{0},\tau^{s},\tau^{e}) = F(0,\tau^{s},\tau^{e})\; a.s.,\quad t\le T,
	\label{eqn:dynamic}
\end{equation}
where the volatility $\sigma(t,T)$ is assumed to be a deterministic time dependent function $\sigma(t,T): \left[t_{0},T\right] \mapsto \R^{+}$ and $F(0,\tau^{s},\tau^{e}) \in \R^{+}$ is the value of the future contract at time $t_{0}$.
\par According to \citet{Samuelson1963} the term structure of commodity forward price volatility typically declines with contract horizon: this is what is commonly known as Samuelson's effect. Therefore, it is customary to assume a volatility function which depends on the time to maturity $T-t$, namely $\sigma\left(t,T\right)= \sigma(T-t)$. In particular, $\sigma(T-t)$ will be decreasing in $T-t$, reflecting the fact that contract with longer maturities are less volatile than contracts with a shorter one. 
\par
\begin{figure}[h]
	\centering
	\includegraphics[width=1\textwidth]{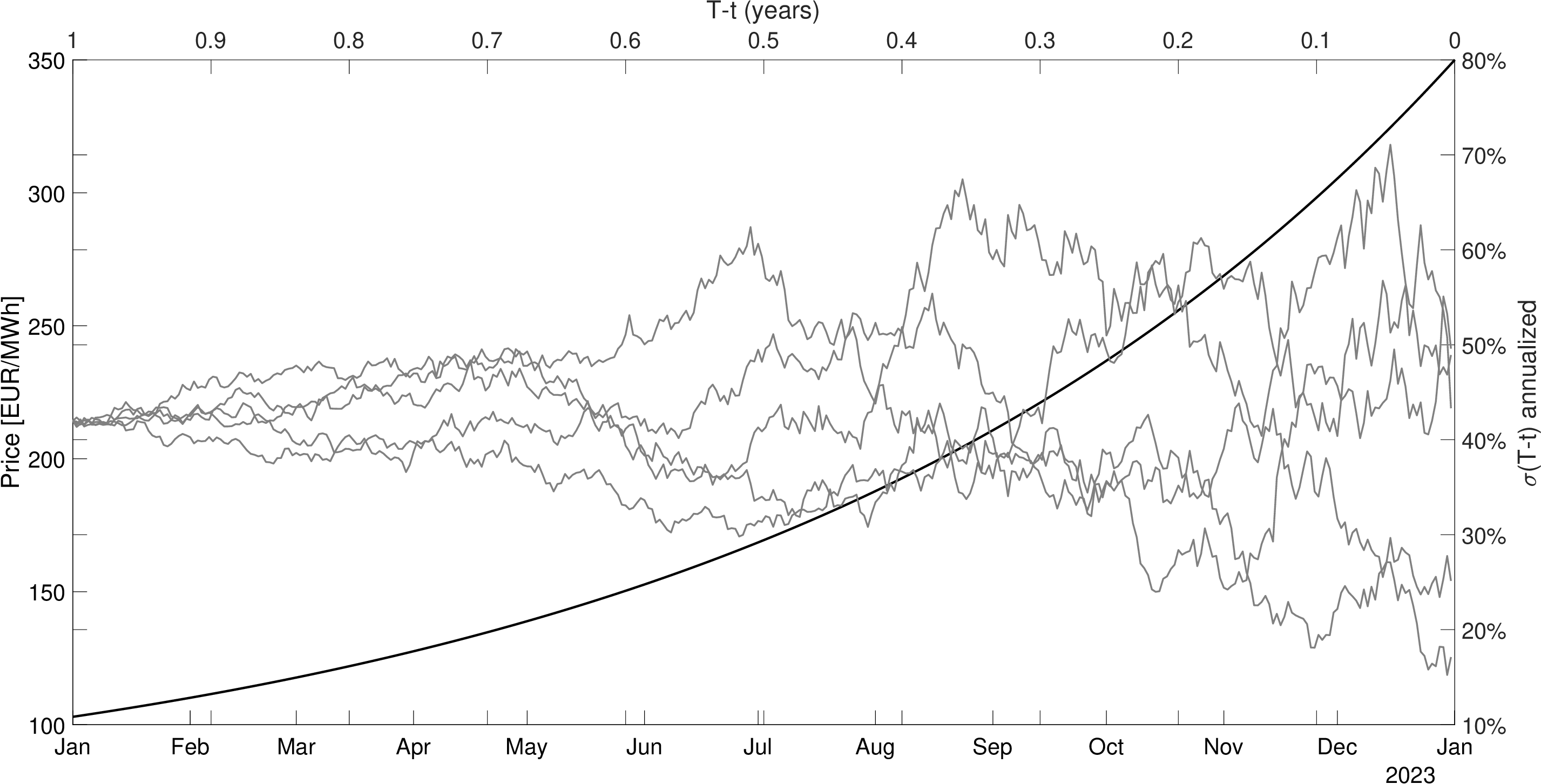}
	\caption{Possible simulations of the forward prices $F(t,\tau^{s},\tau^{e})$.}
	\label{fig:simulation_path_toy_model}
\end{figure}

 Using very basic \ito's calculus we can solve the Equation \eqref{eqn:dynamic}, obtaining
\begin{equation*}
	F(t,\tau^{s},\tau^{e}) = F(t_{0},\tau^{s},\tau^{e})\exp\left\{-\frac{1}{2} \int_{t_{0}}^{t}\sigma\left(T-s\right)^{2}ds + \int_{t_{0}}^{t}\sigma(T-s)dW(s)\right\}.
\end{equation*}
The existence of an explicit solution for the stochastic differential Equation \eqref{eqn:dynamic} is extremely important in order to exactly simulate the trajectories of process $F = \left\{F(t,\tau^{s},\tau^{e}); t_{0}\le t\le T\right\}$. Some possible realizations of the process over one year time horizon for a volatility function of the form:
\begin{equation*}
	\sigma(T-t) = \gamma e^{-2k(T-t)},
\end{equation*}
are shown in Figure \ref{fig:simulation_path_toy_model}. Observe that, at the beginning the volatility is low, since the time to maturity is large, whereas as $T-t \to 0$ the volatility increases, according to the Samuelson's effect.

\begin{figure}[h]
	\centering
	\includegraphics[width=1\textwidth]{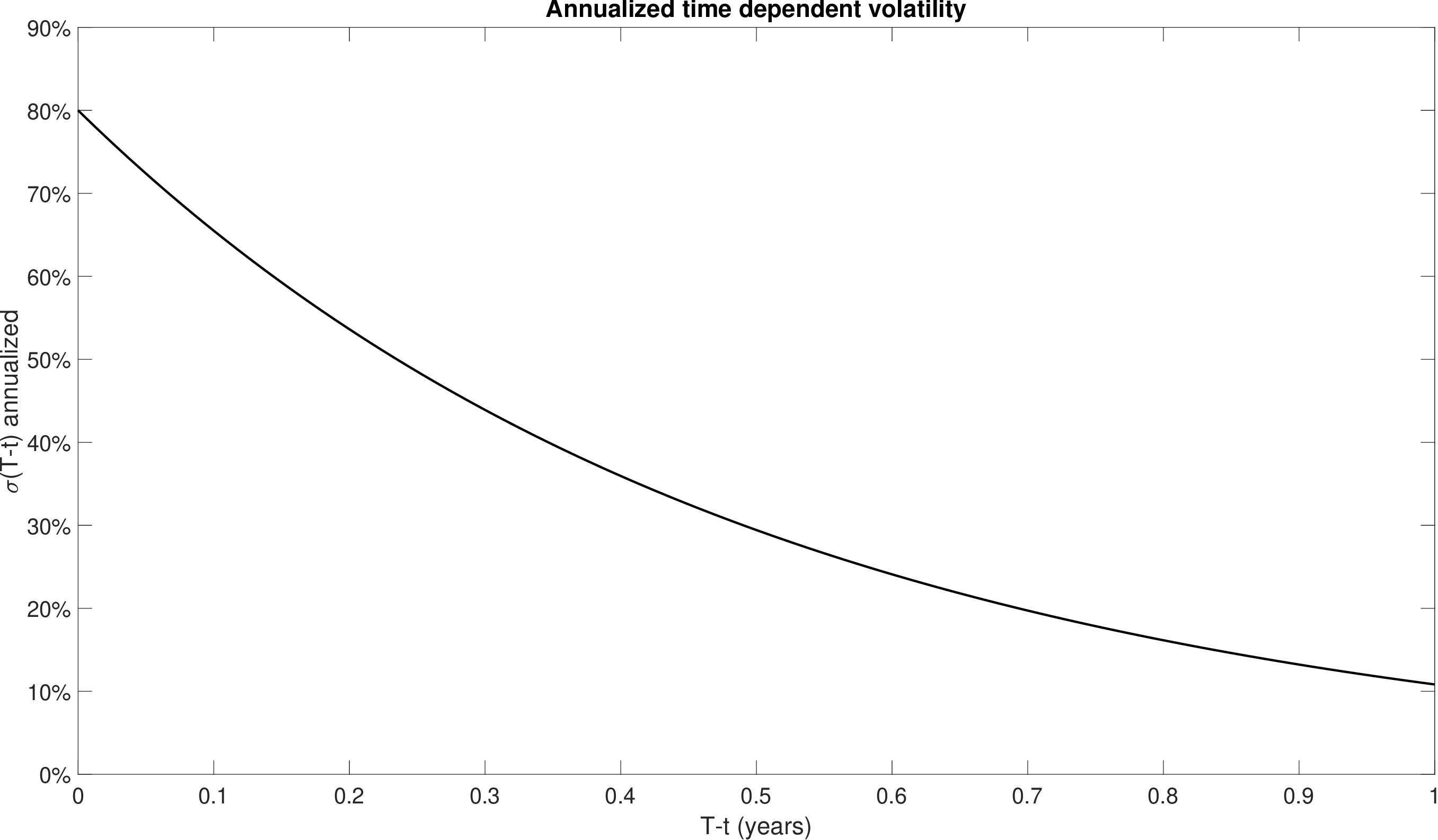}
	\caption{Annualized volatility of the toy model including the Samuelson's effect: $\gamma=0.8$ and $k=1$.}
	\label{fig:volatility_toy_model}
\end{figure}

\begin{figure}[h]
	\centering
	\includegraphics[width=1\textwidth]{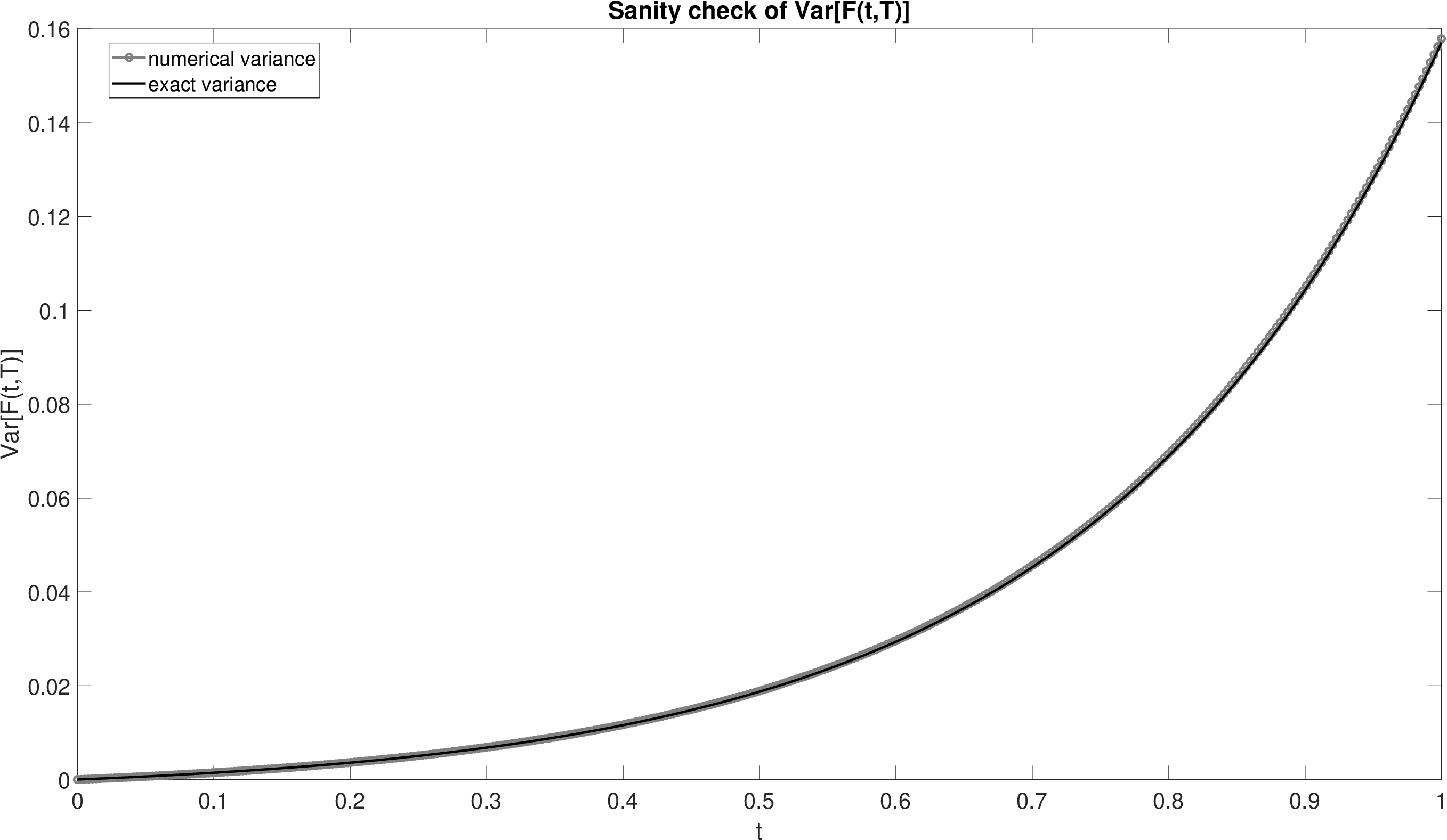}
	\caption{Sanity check of the volatility of log-prices in the toy-model by using $10^{6}$ simulations.}
	\label{fig:sanity_check_volatility_toy_model}
\end{figure}

When one performs simulations, \enquote{sanity checks} are important. By \enquote{sanity check} we mean the comparison of a numerical computation with a theoretical one. For example, by \ito's isometry, it is very easy to check that:
\begin{equation}
	Var\left[\ln F(t,\tau^{s},\tau^{e})\right] = Var\left[ \int_{t_{0}}^{t} \sigma(T-s)dW(s) \right] = \int_{t_{0}}^{t} \sigma\left(T-s\right)^{2}ds,\; t \in \left[t_{0},T\right].
	\label{eqn:exact_expression_variance_log_prices_toy}
\end{equation}
Hence, one can estimate numerically the variance of $F(t,T)$ for each $t \in \left[t_{0},T\right]$ and compare this result with the one given by Equation \eqref{eqn:exact_expression_variance_log_prices_toy}. From Figure \ref{fig:sanity_check_volatility_toy_model} we observe that the numerical quantities and the theoretical ones are very close and hence we are guaranteed that the numerical simulation scheme is correctly implemented.

\begin{figure}[h]
	\centering
	\includegraphics[width=1\textwidth]{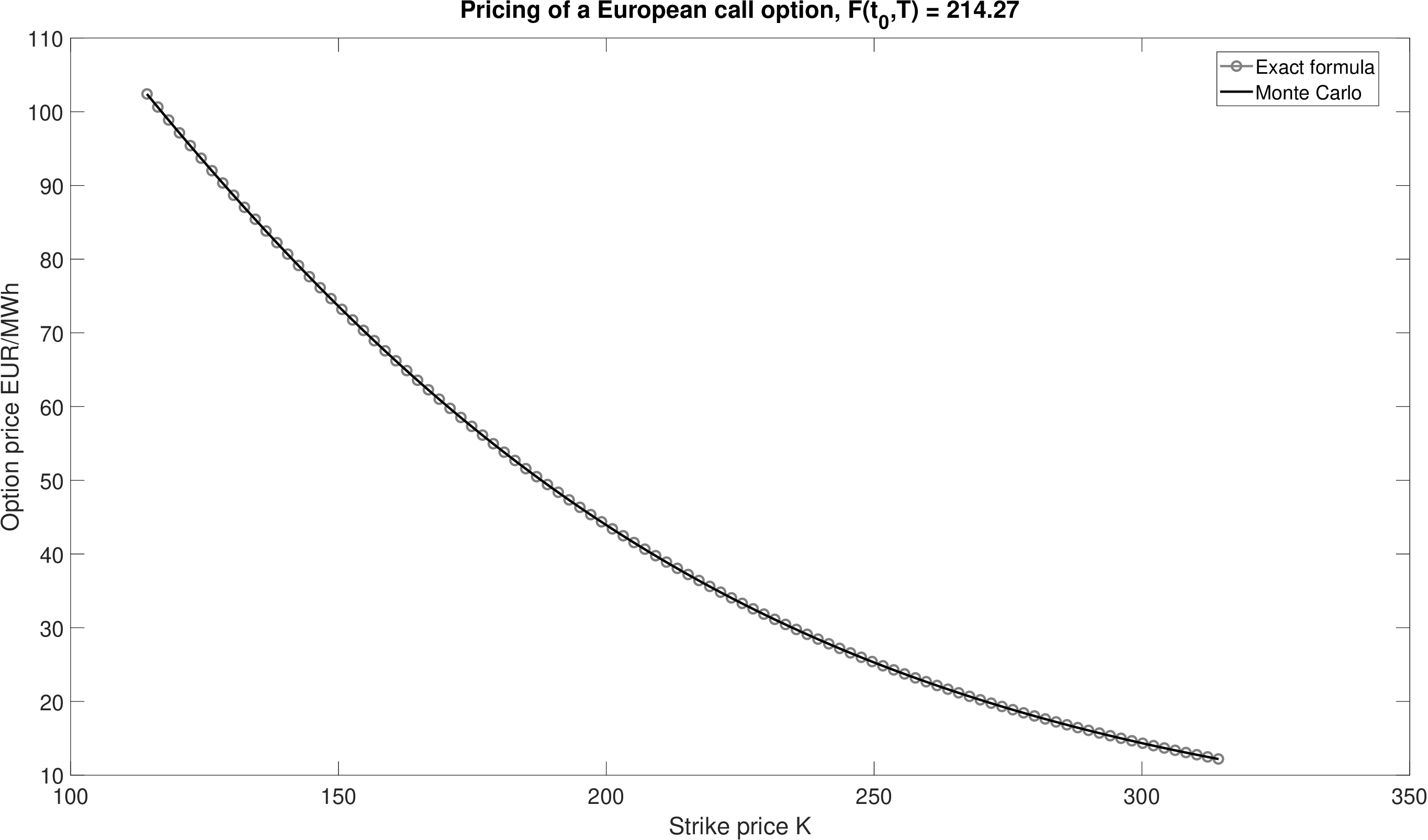}
	\caption{Pricing of European call options with different strike prices. Exact formula is compared to a Monte Carlo pricing with $10^{7}$ simulations.}
	\label{fig:fig_call_option_pricing}
\end{figure}

The knowledge of and explicit expression for variance as the one in Equation \eqref{eqn:exact_expression_variance_log_prices_toy} is important also for option pricing. Following \citet{Boeger2009}, if we assume the following dynamics for the swap price:
\begin{equation*}
	\frac{dF(t,\tau^{s},\tau^{e})}{F(t,\tau^{s},\tau^{e})} = \boldsymbol{\sigma}(t,T) \cdot d\boldsymbol{W}(t), 
\end{equation*}
where $\boldsymbol{\sigma}(t,T) = \left(\sigma_{1}(t,T),\dots,\sigma_{n}(t,T)\right)$ and $\boldsymbol{W} = \left(W_{1},\dots,W_{n}\right)$ is a $n$-dimensional standard Brownian motion with independent components, the price of a call option with maturity $T_{0} \le T$ and strike price $K$ is given by the standard Black formula:
\begin{equation}
	C(t_{0},T_{0},K) = e^{-r(T_{0}-t_{0})} \left(F(t_{0},\tau^{s},\tau^{e})\mathcal{N}\left(d_{1}\right) - K \mathcal{N}\left(d_{2}\right)\right),
	\label{eqn:black_formula}
\end{equation}
where $r\ge 0$ is the risk-free rate, $\mathcal{N}\left(\cdot\right)$ is the cumulative distribution function of a standard normal random variable and:
\begin{align*}
	d_{1} & = \frac{\ln \frac{F(T_{0},\tau^{s},\tau^{e})}{K} + \frac{1}{2} Var\left[\log F(T_{0},\tau^{s},\tau^{e})\right]}{\sqrt{Var\left[\log F(T_{0},\tau^{s},\tau^{e})\right]}} \\
	d_{2} & = d_{1} - \sqrt{Var\left[\log F(T_{0},\tau^{s},\tau^{e})\right]},
\end{align*}
where $Var\left[\log F(T_{0},\tau^{s},\tau^{e})\right]$ can be easily computed by \ito's isometry and depends on the form we chose for $\boldsymbol{\sigma}(T-t)$. \\
For example, if we consider a bi-dimensional standard Brownian motion and we assume that:
\begin{align*}
	\sigma_{1}(t,T) & = 0.8e^{-2(T-t)},\\
	\sigma_{2}(t,T) & = 0.2,\\
\end{align*}
we obtain:
\begin{equation*}
	Var\left[\log F(T_{0},\tau^{s},\tau^{e})\right] = \frac{0.64}{4}\left[e^{-4(T-T_{0})} - e^{-4(T-t_{0})} \right] + 0.04 \left(T_{0} - t_{0}\right).
\end{equation*}
The price of European call option with different strike prices $K$ are shown in Figure \ref{fig:fig_call_option_pricing}

\par \textcolor{black}{Despite any form can be assumed for $\sigma(T-t)$, from a practical point of view a step-wise volatility structure is usually considered. Indeed, if we take a monthly partition of the time interval $\left[t_{0},T\right]$ with dates $t_{0}=T_{0}<\dots,T_{j-1}<T_{j}<\dots<T_{M}=T$, the volatility function assumes the following form:}
\textcolor{black}{
\begin{equation*}
	\sigma(T-t) = \sum_{j=1}^{M}\sigma_{j}\mathbbm{1}_{I_{j}}(T-t),
\end{equation*}
where $\sigma_{j} \in \R^{+}$ and $I_j = \left(T_{j-1},T_{j}\right]$. The key idea to calibrate the parameters $\sigma_{j}$ is the following:
\begin{itemize}
	\item Given a total number $N$ of  swap prices $F(t,\tau_{n}^{s},\tau_{n}^{e}),\; n=1,\dots,N$ construct the products $F_{M_{j}}(t,T_{j}),\; j=1,\dots,M$ where $M_{j}$ means that the product delivery is $j$ months later than the current time $t$. 
	For example, if $F(t,\tau_{n}^{s},\tau_{n}^{e})$ delivers on a quarter you can construct three different products $F_{M_{p}}\left(t,T_{p}\right)$, $F_{M_{q}}\left(t,T_{q}\right)$, $F_{M_{r}}\left(t,T_{r}\right)$ with monthly delivery, such that their weighted mean is equal to the value $F(t,\tau_{n}^{s},\tau_{n}^{e})$\footnote{In this way arbitrages opportunities are not introduced.}.
	\item Each of the products $F_{M_{j}}(t,T_{j})$ has a dynamics of the form:
	\begin{equation*}
		\frac{dF_{M_{j}}(t,T_{j})}{F_{M_{j}}(t,T_{j})} = \sigma_{j}dW(t).
	\end{equation*}
	\item Compute $\sigma_{j},\; j=1,\dots,M$ as the standard deviation of the log-returns of $F_{M_{j}}(t,T_{j})$.
\end{itemize}
This procedure, can be applied to a more general model setting as the one we present in the following section. The data preparation methodology and the calibration algorithm for the general case will be discussed in Section \ref{sec:calibration}.}
 
\subsection{A model for a multi-commodity market}
\textcolor{black}{Now we extend the \enquote{toy-model} we presented in the previous section to a multi-commodity framework. Assume we have $k=1,\dots,K$ markets (DE, F7, IT, TTF and so on) and a final time for the market $T^{*}$. Introduce a partition $t_{0}=T_{0}<\dots <T_{d-1}<T_{d}<\dots<T_{M}=T^{*}$ on $\left[t_{0},T^{*}\right]$. Starting from the quoted products of the form $F^{k}(t,\tau_{n}^{s},\tau_{n}^{e})$, $n=1,\dots,N_{k}$, where $N_{k}$ is the total number of contracts in the market $k$, for each market $k$ we can construct the same number $M_{k}=M$ of contracts of the form $F^{k}_{M_{d}}(t,T_{d}),\; d=1,\dots,M$ with monthly delivery periods. This procedure is described in details in Section \ref{sec:data_preparation}. }  As a first step of our modeling framework we start by defining the dynamics of the fixed-delivery futures contract $F_{M_{d}}^{k}(t,T_{d})$. Assume that each of the the contracts is a \enquote{random factor} which might have an impact on the whole market dynamic. Hence we have a total number of $\tilde{N}=M\cdot K$ random factors. The dynamics of a future product $k=1$ with fixed-delivery $T_{d},\; d=1,\dots,M$ is given by:

\begin{equation}
	\frac{dF_{M_{d}}^{k}(t,T_{d})}{F_{M_{d}}^{k}(t,T_{d})} = \sum_{j=1}^{\tilde{N}} \sigma^{k}_{dj} dW_{j}(t),\quad F_{M_{d}}^{k}(t_{0},T_{d}) = F_{M_{d}}^{k}(0,T_{d})\; a.s.,\quad t \in \left[t_{0},T_{d}\right].
	\label{eqn:dynamic_multi_commodity}
\end{equation}
where $W_{j} = \left\{W_{j}(t); t \ge 0\right\}\; j=1,\dots,\tilde{N}$ are independent Brownian motions. From Equation \eqref{eqn:dynamic_multi_commodity} observe that the dynamics of the single monthly product $F_{M_{d}}^{k}(t,T_{d})$ potentially depends on the ones of all other monthly futures products. From empirical evidence, energy markets are strongly co-integrated and hence it is clear that considering $\tilde{N}$ independent Brownian motions appears to be unreasonable. A possible methodology, based on the dimensional reduction inherited from the Principal Component Analysis (PCA), to select a lower number of stochastic factors, will be presented in Section \ref{sec:calibration}. \\
Observe that Equation \eqref{eqn:dynamic_multi_commodity} can be written in a matrix form as:
\begin{equation*}
	\frac{d\boldsymbol{F}(t,T)}{\boldsymbol{F}(t,T)} = \boldsymbol{\sigma}\cdot d\boldsymbol{W}(t),
\end{equation*} 
where $\boldsymbol{W} = \left(W_{1},\dots,W_{\tilde{N}}\right)$ is a standard Brownian motion with independent components,  $\boldsymbol{\sigma}$ is a $\tilde{N} \times \tilde{N}$ matrix of the form:
\begin{equation*}
	\boldsymbol{\sigma} = 
		\begin{matrix}
			\coolleftbrace{\boldsymbol{\sigma}^{1}}{\sigma_{1,1} \\ \sigma_{2,1}\\ \dots \\ \sigma{M,1} \\ \hline} \\
			\coolleftbrace{\boldsymbol{\sigma}^{2}}{\sigma_{M+1,1} \\ \sigma_{M+2,1}\\ \dots \\ \sigma{2M,1} \\ \hline} \\
			\coolleftbrace{\boldsymbol{\sigma}^{k}}{\vdots \\ \vdots\\ \vdots \\ \hline } \\
			\coolleftbrace{\boldsymbol{\sigma}^{K}}{\sigma_{M(K-1)+1,1} \\ \sigma_{M(K-1)+2,1}\\ \dots \\ \sigma{MK,1} \\} \\
		\end{matrix}%
	\begin{bmatrix}
		\sigma^{1}_{1,1}       & \sigma^{1}_{1,2} & \sigma^{1}_{1,3} & \dots & \sigma^{1}_{1\tilde{N}} \\
		\sigma^{1}_{2,1}       & \sigma^{1}_{2,2} & \sigma^{1}_{M+2,3} & \dots & \sigma^{1}_{2,\tilde{N}} \\
		\dots & \dots & \dots & \dots & \dots \\
		\sigma^{1}_{M,1}       & \sigma^{1}_{M,2} & \sigma^{1}_{M,3} & \dots & \sigma^{1}_{2M,\tilde{N}} \\
		\hline
		\sigma^{2}_{1,1}       & \sigma^{2}_{1,2} & \sigma^{2}_{1,3} & \dots & \sigma^{2}_{1,\tilde{N}} \\
		\sigma^{2}_{2,1}       & \sigma^{2}_{2,2} & \sigma^{2}_{2,3} & \dots & \sigma^{2}_{2,\tilde{N}} \\
		\dots & \dots & \dots & \dots & \dots \\
		\sigma^{2}_{M,1}       & \sigma^{2}_{M,2} & \sigma^{2}_{M,3} & \dots & \sigma^{2}_{M,\tilde{N}} \\
		\hline
		\vdots & \vdots & \vdots & \vdots & \vdots \\
		\vdots & \vdots & \vdots & \vdots & \vdots \\
		\vdots & \vdots & \vdots & \vdots & \vdots \\
		\hline	
		\sigma^{K}_{1,1}       & \sigma^{K}_{1,2} & \sigma^{K}_{1,3} & \dots & \sigma^{K}_{1,\tilde{N}} \\
		\sigma^{K}_{2,1}       & \sigma^{K}_{2,2} & \sigma^{K}_{2,3} & \dots & \sigma^{K}_{2,\tilde{N}} \\
		\dots & \dots & \dots & \dots & \dots \\
		\sigma^{K}_{M,1}       & \sigma^{K}_{M,2} & \sigma^{K}_{M,3} & \dots & \sigma^{K}_{M,\tilde{N}} 
	\end{bmatrix}
\end{equation*}
where $\boldsymbol{\sigma}^{k} \in \R^{M \times \tilde{N}}$ denotes the matrix associated to random factors of $k$-th market product.

\par \textcolor{black}{Once we have defined the dynamics for $F^{k}_{M_{d}}\left(t,T_{d}\right)$} that of the $F^{k}_{n}(t,\tau_{n}^{s},\tau_{n}^{e})$ is a direct consequence. The dynamics of $F^{k}_{n}(t,\tau_{n}^{s},\tau_{n}^{e})$ for $t \in \left[t_{0},\tau_{n}^{s}\right]$ is given by: 
\begin{equation}
	\frac{dF^{k}(t,\tau_{n}^{s},\tau_{n}^{e})}{F^{k}(t,\tau_{n}^{s},\tau_{n}^{e})} =  \sum_{j=1}^{\tilde{N}} \sum_{i=1}^{M}\sigma_{ij}^{k}\mathbbm{1}_{I_{i}}(T^{*}-t)dW_{j}(t).
	\label{eqn:dynamic_absolute_product_multicommodity}
\end{equation}

\par The explicit solution of the Equation \eqref{eqn:dynamic_absolute_product_multicommodity} for $t \in \left[t_{0},\tau_{n}^{s}\right]$ is given by:
\begin{equation}
	\begin{split}
	F^{k}(t,\tau_{n}^{s},\tau_{n}^{e}) = F^{k}(t_{0},\tau_{n}^{s},\tau_{n}^{e}) \exp\left\{ -\frac{1}{2}\sum_{j=1}^{\tilde{N}}\sum_{i=1}^{M}\left(\sigma_{ij}^{k}\right)^{2}\int_{t_{0}}^{t}\mathbbm{1}_{I_{i}}\left(T^{*}-s\right)ds \right. \\
	\left. 
	+\sum_{j=1}^{\tilde{N}}\sum_{i=1}^{M}\sigma_{ij}^{k}\int_{t_{0}}^{t}\mathbbm{1}_{I_{i}}\left(T^{*}-s\right)dW_{j}(s)\right\}.
	\end{split}
	\label{eqn:fwd_exact_solution}
\end{equation}
The existence of a close form solution allows us to simulate the paths of the process in an exact way, without using any discretization method, such as the Euler's of Millstein's one (see \citet{seydel2004}). Once again, observe that if $t \approx t_{0}$ then $T^{*}-t$ is \enquote{large} and hence we are considering the volatility of the products with a large maturity. As soon as time $t$ goes on increases $T^{*}-t$ becomes smaller and smaller hence we are taking into account the volatility of the fixed delivery products with a short time to maturity.
\par Also in this case a sanity check can be performed. Assume we simulate the process following Equation \eqref{eqn:fwd_exact_solution} for a very short period, namely $T^{*}- t \approx T^{*}-t_{0}$. Hence Equation \eqref{eqn:fwd_exact_solution} simplifies to:
\begin{equation}
	F^{k}(t,\tau_{n}^{s},\tau_{n}^{e}) =  F^{k}(t_{0},\tau_{n}^{s},,\tau_{n}^{e}) \exp\left\{-\frac{1}{2} \Delta t \sum_{j=1}^{\tilde{N}} \left(\sigma_{m^{*}j}^{k}\right)^{2} + \sum_{j=1}^{\tilde{N}}\sigma_{m^{*}j}^{k}W_{j}(t) \right\},
	\label{eqn:simplified_solution_relative_prod_simulation}
\end{equation}
where $\Delta t = t-t_{0}$ and $m^{*}$  \textcolor{black}{is such that $\sigma_{m^{*}j},\; j =1,\dots,\tilde{N}$ are the volatilities associated to $F_{M_{m^{*}}}^{k}(t,T_{m^{*}})$ with $T_{m^{*}} = \tau_{n}^{s}$. This simply means that, if we want to simulate the product $F^{k}(t,\tau_{n}^{s},\tau_{n}^{e})$ with monthly delivery period $\left[\tau_{n}^{s},\tau_{n}^{e}\right]$ after three months from now, for say $\Delta t=1$ day, you have to take into account only the volatility associated to the monthly product $F^{k}_{M_{3}}(t,T_{3})$, which is the fixed-delivery product expiry after three months from now.} 
By computing the log-return $x^{k}(t) = \ln F(t,T_{m^{*}}) - \ln F(t_{0},T_{m^{*}})$ we get:
\begin{equation*}
	x^{k}(t) = -\frac{1}{2}\Delta t \sum_{j=1}^{\tilde{N}}\left(\sigma_{m^{*}j}^{k}\right)^{2} + \sum_{j=1}^{\tilde{N}}\sigma_{m^{*}j}^{k}dW_{j}(t),
\end{equation*}
and if we compute its variance we get:
\begin{equation*}
	Var\left[x^{k}(t)\right] = \Delta t \sum_{j=1}^{\tilde{N}}\left(\sigma_{m^{*}j}^{k}\right)^{2}.
\end{equation*}
Hence, we have a simple way to check that the variance of the simulated product is correct, by summing up the squares of the entries of the matrix $\boldsymbol{\sigma}^{k}$ and multiplying by $\Delta t$. \\
Clearly if the approximation $T^{*}-t \approx T^{*} - t_{0}$ does not hold, namely for larger $t$ such expression is no longer valid. On the other hand, simple computations show that:
\begin{equation}
	Var\left[\log F^{k}(t,\tau_{n}^{s},\tau_{n}^{e})\right] = \sum_{j=1}^{\tilde{N}} \sum_{i=1}^{M} \left(\sigma_{ij}^{k}\right)^{2} \int_{t_{0}}^{t} \mathbbm{1}_{I_{i}}\left(T^{*}-s\right) ds,
	\label{eqn:var_log_price}
\end{equation}
which can be easily used to check the correctness of the simulations.
\par As stated in Section \ref{sec:toy_model} once that the expression of $Var \left[\log F^{k}(t,\tau_{n}^{s},\tau_{n}^{e})\right]$ is known, a Black's style pricing formula of the form shown for European call options as the one shown in Equation \ref{eqn:black_formula} can be easily derived.

\subsection{The spot dynamic}

\par Starting from Equation \eqref{eqn:fwd_exact_solution}  we can retrieve the dynamics for the spot prices by defining the spot prices $S(t)$ as:
\begin{equation*}
	S(t) = \lim_{\substack{\tau^{s}\to t \\ \tau^{e} \to t}} F(t,\tau^{s},\tau^{e}).
\end{equation*}
Passing to the limit in Equation \eqref{eqn:fwd_exact_solution} we have that:
\begin{equation}
	\begin{split}
		S^{k}(t) = F^{k}(t_{0},t) \exp\left\{ -\frac{1}{2}\sum_{j=1}^{\tilde{N}}\sum_{i=1}^{M}\left(\sigma_{ij}^{k}\right)^{2}\int_{t_{0}}^{t}\mathbbm{1}_{I_{i}}\left(t-s\right)ds \right. \\
		\left. 
		+\sum_{j=1}^{\tilde{N}}\sum_{i=1}^{M}\sigma_{ij}^{k}\int_{t_{0}}^{t}\mathbbm{1}_{I_{i}}\left(t-s\right)dW_{j}(s)\right\}.
	\end{split}
	\label{eqn:spot_exact_solution}
\end{equation}
Observe that it is important to understand how volatility behaves as time $t$ increases. If $t \approx t_{0}$ then the volatility of the spot price $S^{k}(t)$ is the one of the fixed delivery futures products with the shorter time to maturity. As time $t$ increases, the spot price $S^{k}(t)$ somehow includes all the volatility effects from the products with shorter time to maturity to the ones with a longer one. This is reasonable from an economical point of view. Indeed, if we want to simulate the process $S^{k}(t)$ in a year it would have been the $M_{12}$, the $M_{11}$ and so on up to $M_{0}$ and hence we must sum-up all their volatility contributions. Also in this case, an expression for $Var\left[S^{k}(t)\right]$ similar to the one in Equation \eqref{eqn:var_log_price} can be easily obtained.
\par From a practical point of view, especially for power prices, it is customary to consider a hourly granularity. $F^{k}(t_{0},t)$ represents the power forward price today for hour $t$ and hence the function $F^{k}(t_{0},t)$ for $t \in \left[t_{0},T^{*}\right]$, represents the hourly forward curve for the spot market $k$. In simulations routines, in order to simplify the simulation, we assume the the \enquote{shocks} with respect to the hourly forward curve are daily. For commodities with daily granularity, such as the natural gas, everything is performed on a daily basis. The hourly or daily forward curve should be inferred from futures market products of the form $F(t,\tau^{s},\tau^{e})$ and can be obtained by using different methodologies, such as the one proposed by \citet[Chapter~7]{Benth2008b}.

\subsection{A multidimensional toy example: modeling and calibration}
\label{sec:multi_commodity_to_example}
In order to make things clearer, in this section we present a toy version of the model we presented above. Let be $T^{*}$ the upper-bound of the maturities time in the market and assume we have only four products with non-overlapping delivery periods $m$ (for example consecutive years $Y_{m}$) and maturities $T_{m},\; m=1,\dots,M$ with $M=4$. Moreover, assume $K=1$ hence we are focusing on a single market (for example the power futures market in Germany) and we omit the superscript $k$. The fixed delivery products
have the following dynamics\footnote{Once again, $F_{d}(t,T_{m})$ is the forward products with maturity $T_{m}$ and delivery period $d$, which can be year, quarter or month. For example $F_{Y_{1}}(t,T_{1})$ the forward product with maturity $T_{1}$ and with delivery period the following year. }:
\begin{equation*}
	\frac{dF_{Y_{m}}\left(t,T_{m}\right)}{F_{Y_{m}}(t,T_{m})} = \sigma_{m1}dW_{1}(t) + \sigma_{m2}dW_{2}(t) + \sigma_{m3}dW_{3}(t) + \sigma_{m4}dW_{4}(t) = \sum_{j=1}^{\tilde{N}} \sigma_{mj}dW_{j}(t).
	\label{eqn:dynamic_multi_commodity_toy}
\end{equation*}
with $m=1,\dots,M$ and $t\le T_{m}$. Observe that the components of the multidimensional standard Brownian motion $\boldsymbol{W} = \left\{\left(W_{1}(t), W_{2}(t), W_{3}(t), W_{4}(t) \right); t \in \left[t_{0},T^{*}\right]\right\}$,  can be both correlated and independent. The dynamics in Equation \eqref{eqn:dynamic_multi_commodity_toy} can be written in a matrix form as:

\begin{equation*}
	\frac{d\boldsymbol{F}(t,T)}{\boldsymbol{F}\left(t,T\right)} = \boldsymbol{\sigma} \cdot d\boldsymbol{W}(t),\quad t \in \left[t_{0},T\right] 
\end{equation*}
where :
\begin{equation*}
	\begin{split}
	\boldsymbol{F}(t,T) = \begin{bmatrix}
		F_{Y_{1}}\left(t,T_{1}\right) \\
		F_{Y_{2}}\left(t,T_{2}\right) \\
		F_{Y_{3}}\left(t,T_{3}\right) \\
		F_{Y_{4}}\left(t,T_{4}\right)
	\end{bmatrix},  \qquad T= \min_{m \in \left[1,M\right]}T_{m}, \qquad
	\boldsymbol{\sigma} = \begin{bmatrix} 
		\sigma_{11} & \sigma_{12} & \sigma_{13} & \sigma_{14} \\
		\sigma_{21} & \sigma_{22} & \sigma_{23} & \sigma_{24} \\
		\sigma_{31} & \sigma_{32} & \sigma_{33} & \sigma_{34} \\
		\sigma_{41} & \sigma_{42} & \sigma_{43} & \sigma_{44} \\
	\end{bmatrix}
	\end{split}
\end{equation*}

In the market you do not directly observe the matrix $\boldsymbol{\sigma}$ but it can be efficiently estimated from real market data. In order to do so, we assume that $N$ equally spaced daily observation at times $t_{0} \le t_{1} \le \dots \le t_{N}$ of the process $\boldsymbol{F}$ are given and let $\Delta t = t_{i+1} - t_{i}$ defined in fraction of years, for example $\Delta t = \frac{1}{260}$. Define the log-return for the product with delivery $T_{m}$ as:
\begin{equation*}
	X_{i}^{m} = \ln \frac{F_{m}(t_{i+1},T_{m})}{F_{m}\left(t_{i},T_{m}\right)}
\end{equation*} 
and assume that the vector $\boldsymbol{X} = \left[X^{1},X^{2},X^{3},X^{4}\right]$ is normally distributed with mean $\boldsymbol{\mu}$ and covariance $\boldsymbol{\Sigma}$.  
Since $\boldsymbol{\Sigma}$ is a covariance matrix it is a symmetric and positive-definite matrix and hence it factorizes as:

\begin{equation*}
	\boldsymbol{\Sigma} = \boldsymbol{C} \boldsymbol{\Gamma} \boldsymbol{C}^{T} = \boldsymbol{C} \boldsymbol{\Gamma}^{\frac{1}{2}} \left(\boldsymbol{C} \boldsymbol{\Gamma}^{\frac{1}{2}}\right)^{T}
\end{equation*}
for $\boldsymbol{\Gamma} \in \R^{M \times M}$ diagonal matrix and $\boldsymbol{C} \in \R^{M \times M}$ is an orthogonal matrix such that $\boldsymbol{C}^{T} \boldsymbol{C} = \boldsymbol{I}$, where $\boldsymbol{I}$ is the identity matrix. On the other hand we have that $\boldsymbol{\Sigma} = \boldsymbol{X}^{T}\boldsymbol{X}$ and it can be proved that:
\begin{equation*}
	\boldsymbol{\Sigma} = \Delta t \boldsymbol{\sigma} \boldsymbol{\sigma}^{T}.
\end{equation*}
Therefore, $\boldsymbol{\sigma}$ can be estimated as:
\begin{equation*}
	\boldsymbol{\sigma} = \frac{\boldsymbol{C}\boldsymbol{\Gamma}^{\frac{1}{2}}}{\Delta t^{\frac{1}{2}}}.
\end{equation*}
This procedure is summarized in Algorithm \ref{alg:esitmate_sigma}.

\begin{algorithm}
	\caption{Estimation of $\boldsymbol{\sigma}$}
	\label{alg:esitmate_sigma}
	\begin{algorithmic}[1]
		\State Assume to observe $N$ realization of the random vector $\boldsymbol{X}$ and list them into the matrix $\hat{\boldsymbol{X}}$.
		\State Compute the sample covariance matrix $\hat{\boldsymbol{\Sigma}}$ as $\hat{\boldsymbol{\Sigma}}=\hat{\boldsymbol{X}}^{T} \hat{\boldsymbol{X}}$.
		\State Diagonalize the matrix $\hat{\boldsymbol{\Sigma}} = \boldsymbol{C} \boldsymbol{\Gamma} \boldsymbol{C}^{T}$.
		\State $\hat{\boldsymbol{\sigma}}$ is given by: $\hat{\boldsymbol{\sigma}} = \boldsymbol{C}\boldsymbol{\Gamma}^{1/2}\Delta t^{-1/2}$.
	\end{algorithmic}
\end{algorithm}

\begin{rem}
Observe that the matrix $\hat{\boldsymbol{\sigma}}$ defined in Algorithm \ref{alg:esitmate_sigma} is not unique, but it is unique up to unitary transformation. Indeed assume that $\boldsymbol{U} \in \R^{M\times M}$ is such that $\boldsymbol{U}\boldsymbol{U}^{T} = \boldsymbol{I}$. Then:
\begin{equation*}
	\boldsymbol{\Sigma} = \Delta t \boldsymbol{\sigma}\boldsymbol{\sigma} = \Delta t \boldsymbol{\sigma}\boldsymbol{U}\boldsymbol{U}^{T}\boldsymbol{\sigma}^{T} = \Delta t \boldsymbol{\sigma}\boldsymbol{U}\left(\boldsymbol{\sigma}\boldsymbol{U}\right)^{T}
\end{equation*}
and hence we can define $\bar{\boldsymbol{\sigma}} = \boldsymbol{\sigma}\boldsymbol{U}$ and this is a $M \times M$ matrix with the property that:
\begin{equation*}
	\boldsymbol{\Sigma} = \Delta t \bar{\boldsymbol{\sigma}} \bar{\boldsymbol{\sigma}}^{T}.
\end{equation*}
\end{rem}

\begin{figure}[h]
	\centering
	\includegraphics[width=1\textwidth]{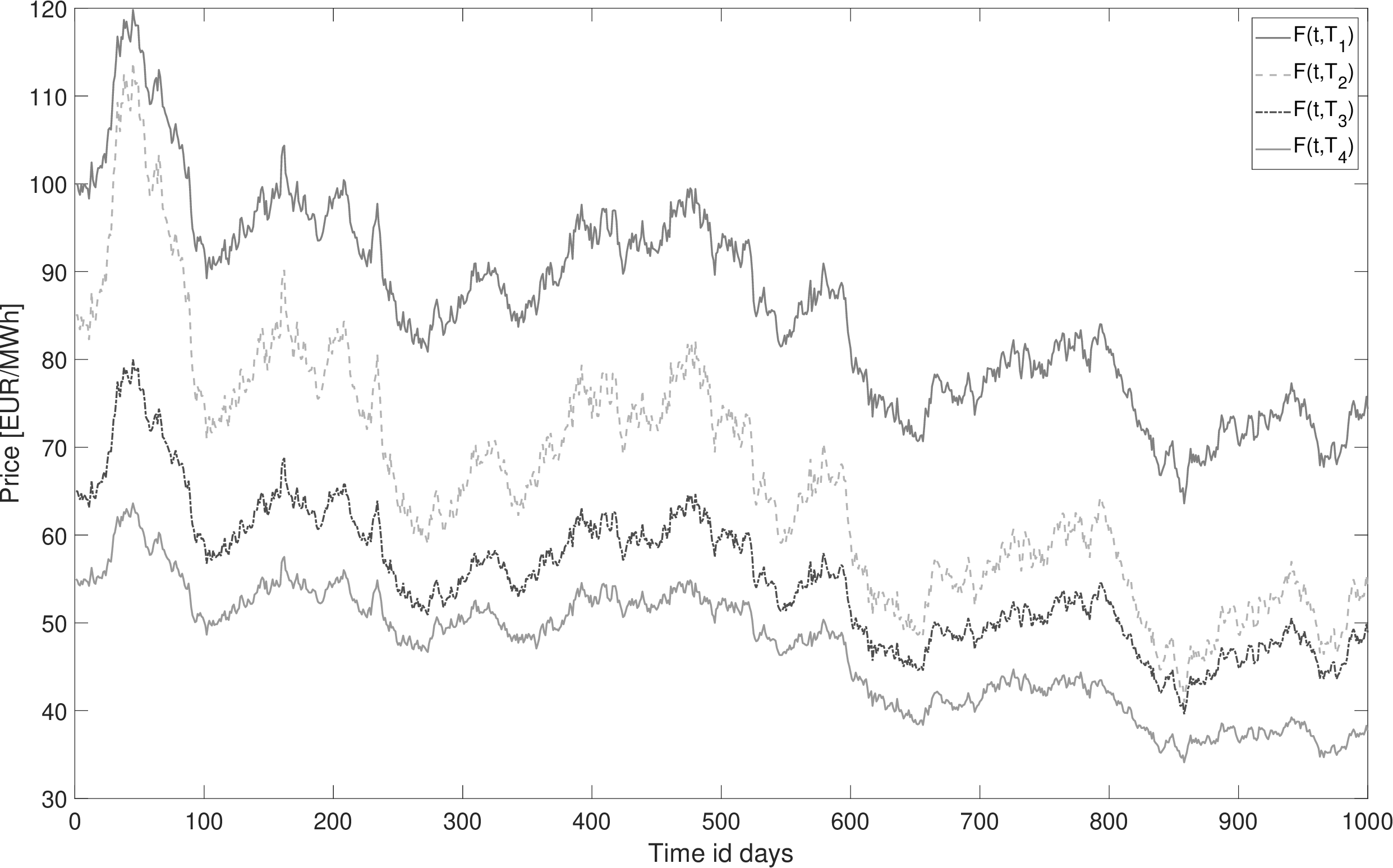}
	\caption{Possible simulation of the forward prices process $\boldsymbol{F}$ .}
	\label{fig:simulation_path_toy_model_multicommodity}
\end{figure}

Within our toy example assume that:
\begin{equation*}
	\boldsymbol{\sigma} = \begin{bmatrix} 
		0.15 & 0.019 & -0.13 & 0.018 \\
		0.25 & 0.014 & -0.19 & 0.015 \\
		0.185 & 0.012 & -0.13 & 0.018 \\
		0.125 & 0.044 & -0.131 & 0.043 \\
	\end{bmatrix},
\end{equation*}
and hence simulate the trajectories of the process $\boldsymbol{F}$ for a given time horizon $\left[t_{0},T\right]$, with $T\le T^{*}$. If we plot the trajectories obtained by simulating from in Equation \eqref{eqn:dynamic} we obtain the results in Figure \ref{fig:simulation_path_toy_model_multicommodity}. Hence compute the log-returns and get $\hat{\boldsymbol{\sigma}}$ following Algorithm \ref{alg:esitmate_sigma}, which gives:
\begin{equation*}
	\hat{\boldsymbol{\sigma}} = \begin{bmatrix} 
		-0.0004 &  -0.0056  & -0.0054 &    0.1964 \\
		0.0003  & -0.0037 &   0.0203  &  0.3090 \\
		-0.0002 &   0.0087 &   0.0141 &   0.2228 \\
		0.0001 &   0.0017 &  -0.0462 &   0.1809 \\
	\end{bmatrix}.
\end{equation*}
Depending on the number of time step you choose and on the seed of random number generator you could obtain a different matrix $\hat{\boldsymbol{\sigma}}$. %

From Figure \ref{fig:simulation_path_toy_model_multicommodity} it is clear that the whole market can be properly described by using less than four risk factors. We get the same conclusion by looking at the matrix correlation $\rho$ of log-returns and we obtain:
\begin{equation}
	\rho = \begin{bmatrix} 
		1.00  &  0.99  &  0.99 &   0.98 \\
		0.99  &  1.00 &   0.99  &  0.95 \\
		0.99  &  0.99  &  1.00 &   0.95 \\
		0.98  &  0.95 &   0.95 &   1.00 \\
	\end{bmatrix},
\label{eqn:correlation_matrix}
\end{equation}
If we perform the PCA analysis we get that the eigenvalues of the covariance matrix $\hat{\boldsymbol{\Sigma}}$ are:
\begin{equation*}
	\lambda = \left[0.8325,	0.0107, 0.0005, 0.0000\right] \cdot 10^{-3}.
\end{equation*}
It turns out that the first two principal components (eigenvectors) associated to the first two eigenvalues explains more than the $99\%$ of the variance. For this reason we can use only two Brownian motions to properly model the whole market, instead of the original four. If we define: 
\begin{equation*}
	\boldsymbol{\sigma}^{*} = \boldsymbol{C}^{*} \left(\boldsymbol{\Gamma}^{*}\right)^{1/2} \Delta t^{-1/2},
\end{equation*} 
and we simulate the processes by using only two stochastic factors and $\boldsymbol{\sigma}^{*}$ and compute the empirical matrix of log-returns correlations $\rho^{*}$, we get:

\begin{equation}
	\rho^{*} = \begin{bmatrix} 
	1.00  &  0.99 &   0.99 &   0.97 \\
	0.99 &   1.00  &  0.99 &   0.95 \\
	0.99  &  0.99  &  1.00  &  0.95 \\
	0.97 &   0.95 &   0.95  &  1.00 \\
	\end{bmatrix},
\end{equation}
which is extremely close to the one shown in Equation \eqref{eqn:correlation_matrix}.

\begin{figure}[h]
	\centering
	\includegraphics[width=1\textwidth]{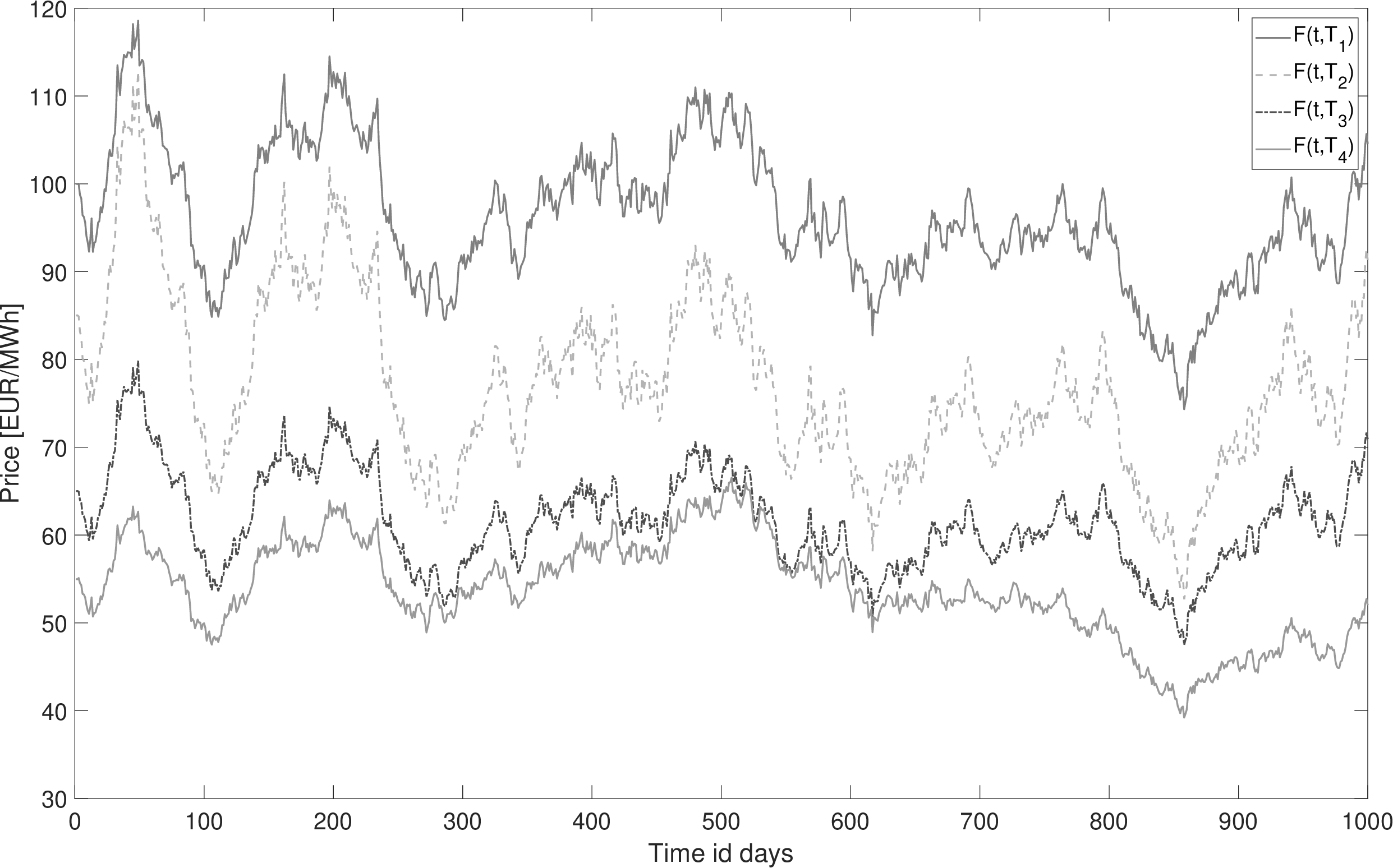}
	\caption{Possible simulation of the forward prices process $\boldsymbol{F}$ .}
	\label{fig:simulation_path_toy_model_multicommodity_sigma_hat}
\end{figure}

Furthermore, if we compute the original covariance matrix $\hat{\boldsymbol{\Sigma}}$ and $\boldsymbol{\Sigma}^{*}$ given by:
\begin{equation*}
	\boldsymbol{\Sigma}^{*} = \Delta t \boldsymbol{\sigma}^{*} \left(\boldsymbol{\sigma}^{*}\right)^{T}
\end{equation*}
 we get:

\begin{align*}
	\hat{\boldsymbol{\Sigma}} & = \begin{bmatrix} 
		1.4859 &	2.3309 &	1.6781 &	1.3756 \\
		2.3309 &	3.6897 &	2.6580 &	2.1138 \\
		1.6781 &	2.6580 &	1.9197 &	1.5256 \\
		1.3756 &	2.1138 &	1.5256 &	1.3405 \\
	\end{bmatrix} \cdot 10^{-4},\\
	\boldsymbol{\Sigma}^{*} & = \begin{bmatrix} 
		1.4846 &  2.3302 &  1.6800 &  1.3760 \\
		2.3302 &   3.6892 &  2.6592 &   2.1140 \\
		1.6800 &  2.6592 &  1.9169 &   1.5250 \\
		1.3760 &   2.1140 &  1.5250 &   1.3404 \\
	\end{bmatrix} \cdot 10^{-4},
\end{align*}
which confirms that only two Brownian motions are enough to explain the most part of the variance of log-returns and hence to model the market. In the following section, we formalize the above concepts.

\subsection{PCA for dimension reduction}
\label{sec:PCA_dimension_reduction}
In this section we formalize the example in Section \ref{sec:multi_commodity_to_example} and we show how PCA can be used in order to identify a relatively small number of stochastic factors which drive the whole market. Assume we have $K$ markets and for each of them we have the same number of contracts with given maturities $F_{m}^{k}(t,T_{m})$, with $k=1,\dots,K$ and $m=1,\dots,M$.
Consider a matrix consisting of $n_{obs}$ rows and $\tilde{N}=M \cdot K$ columns. We compute the matrix of log-returns $\boldsymbol{X} \in \R^{(n_{obs}-1) \times \tilde{N}}$. Now we apply the PCA and we compute the eigenvalues and the associated eigenvectors. We have the following proposition.

\begin{thm}{[\citet[Result~8.2]{JWichern2002}]}
	Let $\boldsymbol{X}=\left(X_{1},\dots,X_{d}\right)$ be a random vector with covariance matrix $\Sigma$ with eigenvalue-eigenvector pair $\left(\lambda_{1},v_{1}\right),\dots,\left(\lambda_{d},v_{d}\right)$ where $\lambda_{1}\ge \lambda_{2}\ge \dots,\ge \lambda_{d}\ge 0$. Let $Y_{i} = v_{1}\boldsymbol{X}$ for $i=1,\dots,d$ be the principal components. Then:
	\begin{equation*}
		\sigma_{11} + \dots + \sigma_{dd} = \sum_{i=1}^{d}var\left(X_{i}\right) = \lambda_{1} + \dots + \lambda_{d} = \sum_{i=1}^{d}var(Y_{i})
	\end{equation*}
\end{thm}
This result states that the total population variance is given by the sum of the eigenvalue $\lambda_{i}$, for $i=1,\dots,d$. Hence the percentage of the variance explained by the principal component $Y_{k}$ is given by:
\begin{equation*}
	\frac{\lambda_{k}}{\sum_{i=1}^{d}\lambda_{i}},\quad k=1,\dots,d.
\end{equation*}
The number of principal components $k$ is chosen such that they are enough to explain a sufficiently large percentage of the variance. Usually this choice is made heuristically, namely there is not a rigorous way to chose the number $k$. See \citet{JWichern2002} for details.
\par Once that we have chosen how many principal components to consider, say $N$, we can define the matrix $\boldsymbol{\sigma}^{*}$ as:
\begin{equation*}
	\boldsymbol{\sigma}^{*} = \boldsymbol{C}^{*} \left(\boldsymbol{\Gamma}^{*}\right)^{1/2}\Delta t^{-1/2},
\end{equation*}
where $\boldsymbol{C}^{*}$ is the $\R^{\tilde{N} \times N}$ matrix consisting on the first $N$ eigenvector associated to the eigenvalues $\lambda_{1},\dots,\lambda_{N}$, $\boldsymbol{\Gamma}^{*}$ is the diagonal $\R^{N \times N}$ matrix containing $N$ eigenvalues and $\Delta t$ is the time interval between quotations (usually a day: $\Delta t=1/252$).
Once we have fitted the matrix $\boldsymbol{\sigma}^{*}$ we can use $N$ independent stochastic factors to simulate both forward and spot pricing by using Equations \eqref{eqn:fwd_exact_solution} and \eqref{eqn:spot_exact_solution}.

\section{Real-world example: calibration}
\label{sec:calibration}
In the previous section, we have shown how to calibrate the model on a toy example. In this section, we move to a real market application, showing how to calibrate the model on futures power and gas markets quotations. In most power markets both peak-load and base-load products are present. The difference between them is that peak-load contracts delivers electricity only is some specific hours (typically from 8 to 20) during the working days, whereas a base-load contract delivers power for all the hours of the delivery periods. An off-peak contract, which is not typically traded, delivers energy in non peak-load periods and can be obtained from peak-load and base-load quotations. Usually traders and risk managers are interested in considering both type of products. So far, we have not specified the difference between base-load and peak-load contracts, but we observe we can easily include them. In particular, in spot simulations both peak-load and base-load products' volatility might be included. A possible strategy is to simulate peak-load and base-load spot prices separately and hence merging together. On the other hand, in many power futures market peak-load contracts are less liquid that the corresponding base-load ones and hence \enquote{expertise adjustment} has to be done to properly retrieve peak-load quotations when they are missing. This leads to the introduction of arbitrary choices and hence to questionable spot simulations behavior. For these reason and in order to simplify the exposition we focus only on base-load quotation.
\par Before calibrating the model we need some data preparation which is crucial to properly fit the parameters. This part is the most delicate one and great attention should be payed to it: if the data are not properly prepared, calibration might leads to wrong parameters estimation and output results might be misleading. This section is split in two parts. First  we show how to prepare the data, then we use the PCA technique to calibrate the model.

\subsection{Data preparation}
\label{sec:data_preparation}
As discussed in Section \ref{sec:model} we have assumed that the volatility $\boldsymbol{\sigma}(t,T)$ is a \textcolor{black}{step-wise} function of the time to maturity $T-t$. As we have shown in Section \ref{sec:multi_commodity_to_example}, such volatility function must be fitted on products of the form $F_{D}^{k}(t,T)$ which can be reconstructed from what we observe in the market, namely  $F\left(t,\tau^{s},\tau^{e}\right)$ (see Table \ref{tbl:absolute_product_structure}).

\begin{table}
	\centering
	\begin{tabularx}{1.0\textwidth}{l*{10}{c}}
		Trading date & Jan-20 &	Feb-20 &	Mar-20 &	Q2-20 &	Q3-20 &	Q4-20 &	2021 &	2021  \\
		\hline
		2020-01-02 &	36.05 &	\textcolor{darkred}{39.76} &	37.15 &	35.50 &	39.05 &	45.30 &	43.85 &	46.55
	\end{tabularx}
	\caption{Some of the products $F(t,\tau^{s},\tau^{e})$ available on German futures market for the trading date $4/1/2020$.}
	\label{tbl:absolute_product_structure}
\end{table}

\par Consider $K$ markets, each of them having the same number $N_{k},\; k=1,\dots,K$ of products $F^{k}\left(t,\tau_{n}^{s},\tau_{n}^{e}\right)$, for $n=1,\dots,N_{k}$, with monthly, quarterly and yearly delivery periods. Hence we can introduce the following set of \emph{syntetic-futures} $F_{D}^{k}(t,T)$, such that:
\begin{itemize}
	\item $F_{M_{h}}^{k}(t,T)$ is a contract which delivers in the following $h^{th}$ month, with delivery $T$.
	\item $F_{Q_{h}}^{k}(t,T)$ is a contract which delivers in the following $h^{th}$ quarter, with delivery $T$.
	\item $F_{Y_{h}}^{k}(t,T)$ is a contract which delivers in the following $h^{th}$ year, with delivery $T$.
\end{itemize} 
\textcolor{black}{We can assume that $T=\tau_{n}^{s}$. By observing in Table \ref{tbl:absolute_product_structure} the product $F(t,\tau^{s},\tau^{e})$ with delivery on February 2020, we observe in Table \ref{tbl:relative_product_structure} that it is translated in the product $F_{M_{1}}(t,T)$ (see red values). From Table \ref{tbl:relative_product_structure} we can also observe what in Section \ref{sec:market_structure_and_analysis} we called \emph{rolling-mechanism}: on the last day of march, the $F_{M_{1}}\left(t,T\right)$ product is the one with delivery period on April, whereas, on the first day of April it is the one with delivery period on May. Hence, for all markets $k=1,\dots,K$ we construct the synthetic products $F^{k}_{M_{h}}(t,T)$ and we obtain a data-set as shown in Table \ref{tbl:relative_product_structure}.}

\begin{table}
	\centering
\begin{tabularx}{1.0\textwidth}{l*{10}{c}}
	Trading date & $M0$ &	$M1$ &	$M2$ &	$Q1$ &	$Q2$ &	$Q3$ &	$Y1$ &	$Y2$  \\
	\hline
	2020-01-02 &	36.05 &	\textcolor{darkred}{39.76} &	37.15 &	35.5 &	39.05 &	45.3 &	43.85 &	46.55 \\
	2020-01-03 &	38.06 &	40.4 &	37.8 &	36.55 &	39.85 &	45.97 &	44.85 &	47.08  \\
	2020-01-06 &	36.93 &	39.46 &	37.29 &	36 &	39.36 &	45.5 &	44.55 &	46.89  \\
	\dots &	\dots &	\dots &	\dots &	\dots &	\dots & \dots &	\dots &	\dots  \\
	2020-03-30 &	22.49 &	17.6 &	19.08 &	26.74 &	26.84 &	33.74 &	34.72 &	38.3  \\
	2020-03-31 &	15.74 &	17.06 &	\textcolor{darkgreen}{19.79} &	26.74 &	27.63 &	\textcolor{darkgreen}{34.45} &	35.65 &	39.05  \\
	2020-04-01 &	15.74 &	\textcolor{darkgreen}{19.04} &	23.24 &	26.74 &	\textcolor{darkgreen}{33.76} &	36.24 &	34.95 &	38.56  \\
	2020-04-02 &	16.39 &	19.09 &	23.26 &	26.86 &	33.98 &	36.57 &	35.35 &	39.04  \\
	\dots &	\dots &	\dots &	\dots &	\dots &	\dots & \dots &	\dots &	\dots &	 \\
	2023-02-13 &	129.69 &	131.22 &	126.49 &	131.05 &	149.58 &	177.10 &	158.18 &	128.11  \\
	2023-02-14 &	129.95 &	131.01 &	126.46 &	131.33 &	150.25 &	179.23 &	159.34 &	128.50  \\
\end{tabularx}
\caption{Relative products for the German forward markets.}
\label{tbl:relative_product_structure}
\end{table}
\par At this point one might be tempted to compute the log-returns (opportunely filtered) of the prices in Table \ref{tbl:relative_product_structure} and apply the PCA to identify the number of risky factors.  

\par 
On the other hand, in order to simplify the code implementation, we suggest the following approach. Given $T^{*}$ we introduce a partition on $\left[t_{0},T^{*}\right]$,  $t_{0}=T_{0}<\dots<T_{i-1}<T_{i}<\dots,T_{M}=T^{*}$ such that each interval covers exactly one months. Hence, starting from the products with a coarser granularity, namely quarter and year ($F_{Q_{h}}\left(t,T\right)$ and $F_{Y_{h}}\left(t,T\right)$)  we create monthly products $F_{M_{h}}\left(t,T\right)$. In order to do so, first we have to compute a \emph{flat monthly forward curve} by obtaining the value of each monthly futures contract in a market coherent way, guarantying that no arbitrage opportunities arise\footnote{This simply means that, for example, if three monthly futures are reconstructed starting from a quarter, the weighted mean of monthly products must coincides with the one of the original quarter futures.}. A well known approach for the construction of a smooth forward curve with a seasonal effect which is coherent with the market observed futures quotations has been proposed by \citet[Chapter~7]{Benth2008b}. A similar procedure can be adopted and simplified by substituting the smooth curve with a step-wise one. The algorithm, which reduce to the solution of a linear system, is detailed in the following session.

\subsubsection{Construction of the monthly forward curve}
\label{sec:monthly_flat_fwd_curve}
In this section we show how to construct the monthly forward curve for a given market starting from the futures products available on the market. Details can be found in \citet[chapter~7]{Benth2008b}.
\par
\begin{figure}
	\begin{tikzpicture}[scale=0.7]
		
		
		\draw[very thick,latex-latex] (0,7.75) node[left]{$F(t,\tau_{n}^{s},\tau_{n}^{e})$}
		|- (18,0) node[below]{$T$};
		
		\draw[line width=0.5mm, scale=0.5, domain=0:2, smooth, variable=\x, black] plot ({\x}, {12})  node[below,xshift=-20] {\tiny $\tau_{1}^{s}$} node[below,xshift=0] {\tiny $\tau_{1}^{e}$};
		\draw[line width=0.5mm, scale=0.5, domain=2:4, smooth, variable=\x, black] plot ({\x}, {10}) node[below,xshift=-20] {\tiny $\tau_{2}^{s}$} node[below,xshift=0] {\tiny $\tau_{2}^{e}$};
		\draw[line width=0.5mm, scale=0.5, domain=4:6, smooth, variable=\x, black] plot ({\x}, {14}) node[below,xshift=-20] {\tiny $\tau_{3}^{s}$} node[below,xshift=0] {\tiny $\tau_{3}^{e}$};
		\draw[line width=0.5mm, scale=0.5, domain=4:10, smooth, variable=\x, black] plot ({\x}, {8}) node[below,xshift=-60] {\tiny $\tau_{4}^{s}$} node[below,xshift=0] {\tiny $\tau_{4}^{e}$};
		\draw[line width=0.5mm, scale=0.5, domain=10:16, smooth, variable=\x, black] plot ({\x}, {6.5}) node[below,xshift=-60] {\tiny $\tau_{5}^{s}$} node[below,xshift=0] {\tiny $\tau_{5}^{e}$};
		\draw[line width=0.5mm, scale=0.5, domain=16:22, smooth, variable=\x, black] plot ({\x}, {10}) node[below,xshift=-60] {\tiny $\tau_{6}^{s}$} node[below,xshift=0] {\tiny $\tau_{6}^{e}$};
		\draw[line width=0.5mm, scale=0.5, domain=10:34, smooth, variable=\x, black] plot ({\x}, {5}) node[below,xshift=-240] {\tiny $\tau_{7}^{s}$} node[below,xshift=0] {\tiny $\tau_{7}^{e}$};

		\draw [dashed] (0,0) -- (0,0) node[below]{\tiny $\tau_{0}$};
		\draw [dashed] (1,6) -- (1,0) node[below]{\tiny $\tau_{1}$};
		\draw [dashed] (2,7) -- (2,0) node[below]{\tiny $\tau_{2}$};
		\draw [dashed] (3,7) -- (3,0) node[below]{\tiny $\tau_{3}$};
		\draw [dashed] (5,4) -- (5,0) node[below]{\tiny $\tau_{4}$};
		\draw [dashed] (8,5) -- (8,0) node[below]{\tiny $\tau_{5}$};
		\draw [dashed] (11,5) -- (11,0) node[below]{\tiny $\tau_{6}$};
		\draw [dashed] (17,2.5) -- (17,0) node[below]{\tiny $\tau_{7}$};
	\end{tikzpicture}
\caption{Quoted futures product $F\left(t,\tau_{n}^{s},\tau_{n}^{e}\right)$.}
\label{fig:fwd_curve_toy}
\end{figure}
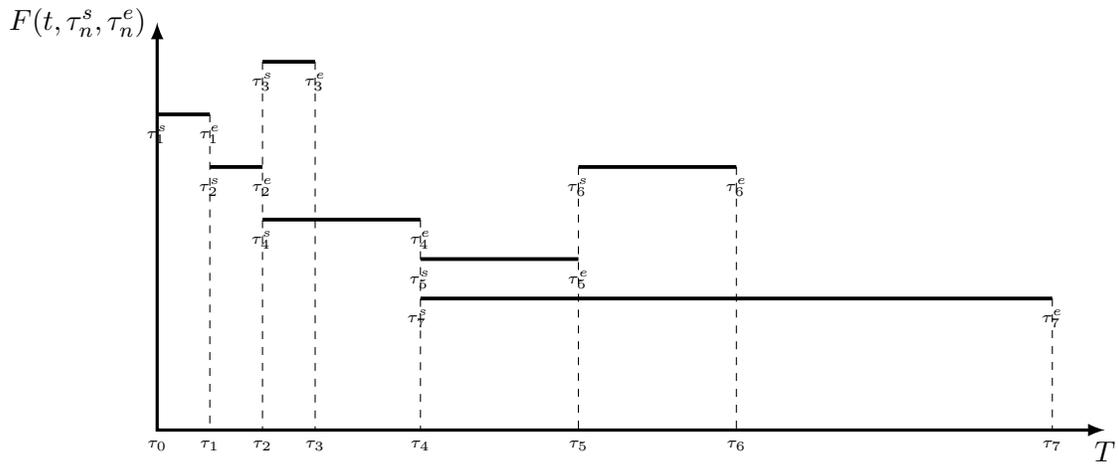

\begin{figure}[!h]
	\centering
	\includegraphics[width=1\textwidth]{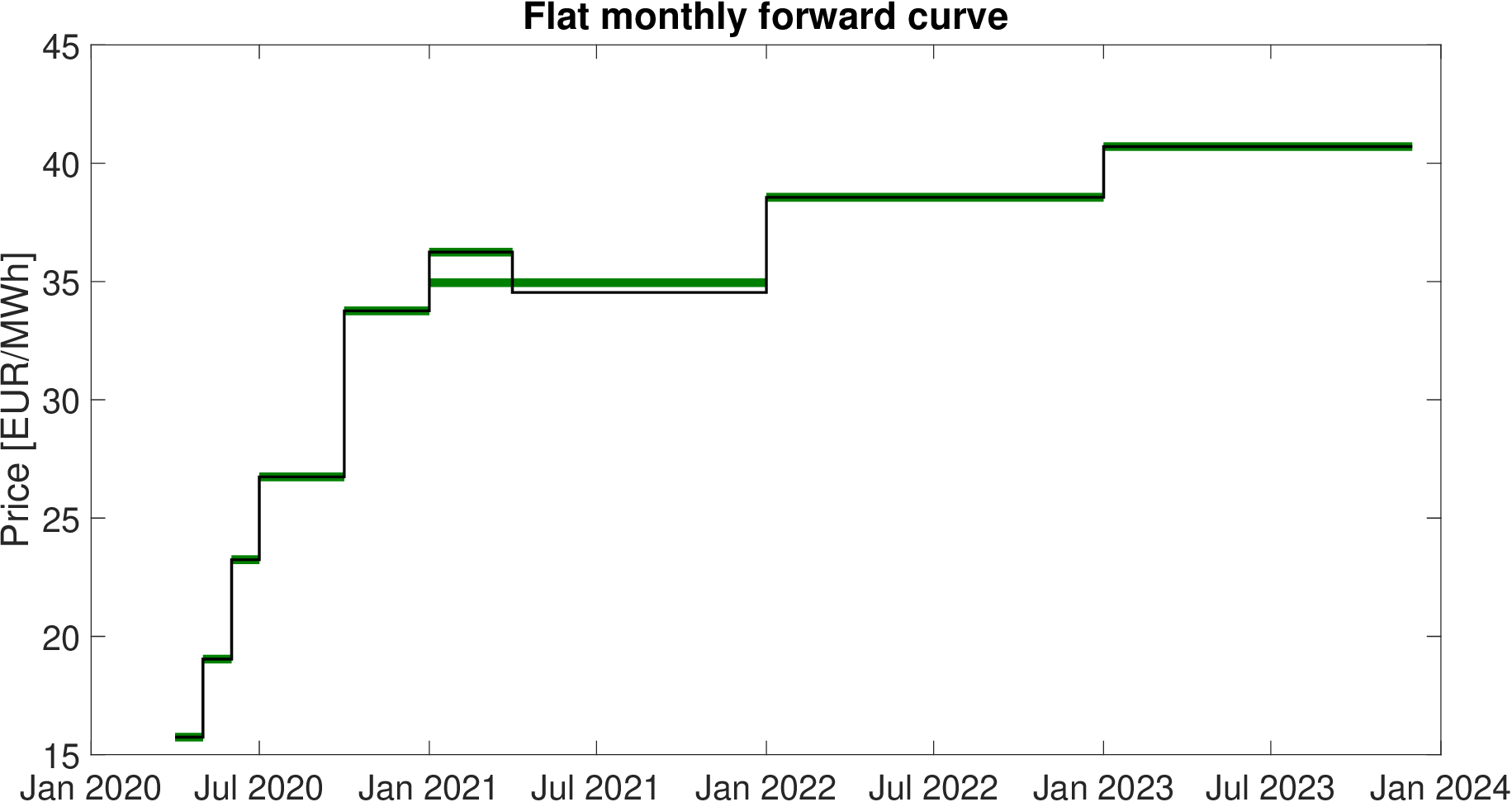}
	\caption{Reconstructed monthly forward curve (black) compared with the quoted forward price for the trading date 2020-04-01 (green). Observe that non arbitrage constraints are satisfied on the year 2021.}
	\label{fig:fwd_curve_real}
\end{figure}

 Consider the example in Figure \ref{fig:fwd_curve_toy} where overlapped products are allowed and suppose we want to define a step-wise forward curve on the intervals $\left[\tau_{i-1},\tau_{i}\right],\; i=1,\dots,M$ with $M=7$ of the form:

\begin{equation*}
	\epsilon\left(u\right) = \sum_{i=1}^{M} a_{i}\mathbbm{1}_{\left[\tau_{i-1},\tau_{i}\right]}(u),
\end{equation*}
where $\left\{a_{i}\right\}_{i=1}^{M}$ are the values we have to fit. Consider $F\left(t_{0},\tau_{i}^{s},\tau_{i}^{e}\right)$ we have to guarantee that non arbitrage constraints are satisfied by satisfying, for each quoted product $i=1,\dots,n$, the following relation:
\begin{equation*}
	F\left(t_{0},\tau_{i}^{s}, \tau_{i}^{e}\right) = \frac{1}{\tau_{i}^{e} - \tau_{i}^{s}} \int_{\tau_{i}^{s}}^{\tau_{i}^{e}}\epsilon(u)du.
\end{equation*}
It is easy to show that finding the values $\left\{a_{i}\right\}_{i=1}^{n}$ reduces to the solution of a linear system which can be done numerically in a very efficient way. Once that the values of $\left\{a_{i}\right\}_{i=1}^{n}$ are available $\epsilon(u)$ is determined. For each trading date and for each market such a monthly forward curve must be computed. If we consider as trading date $4/1/2020$ the resulting monthly forward curve is shown in Figure \ref{fig:fwd_curve_real}. Of course, if during the construction of the forward curve some products are completely overlapped they must be removed preserving those with the finer granularity. Once that the \emph{flat monthly forward curve} has been constructed, products $F_{M_{h}}(t,T_{m}),\; m=1,\dots,M$ can be easily obtained.

\par
\begin{figure}[!h]
	\centering
	\includegraphics[width=1\textwidth]{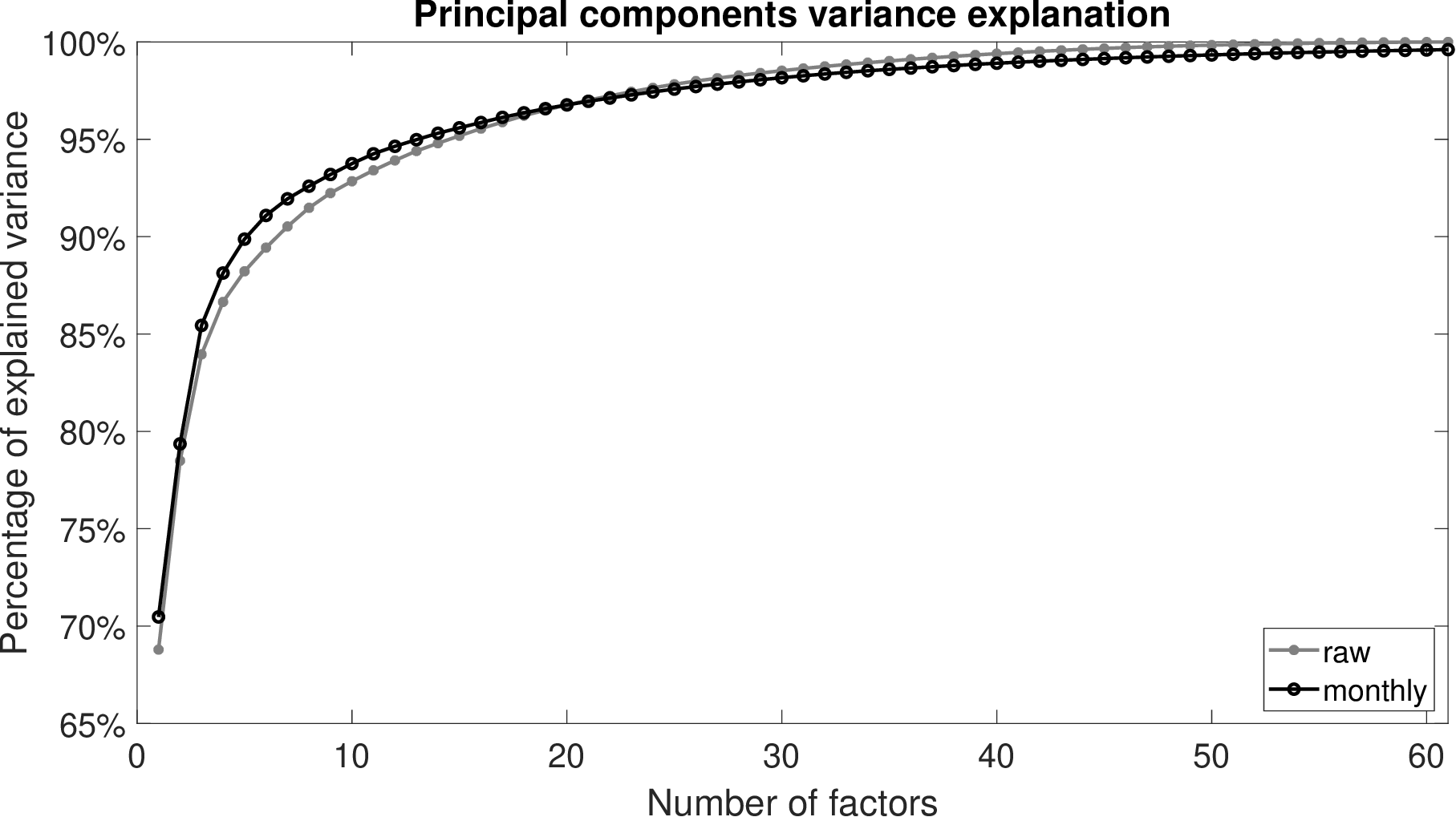}
	\caption{Percentage of the total variance explained by principal components on the raw and on the monthly data.}
	\label{fig:number_principal_components}
\end{figure}

\begin{figure}[!h]
	\centering
	\includegraphics[width=1\textwidth]{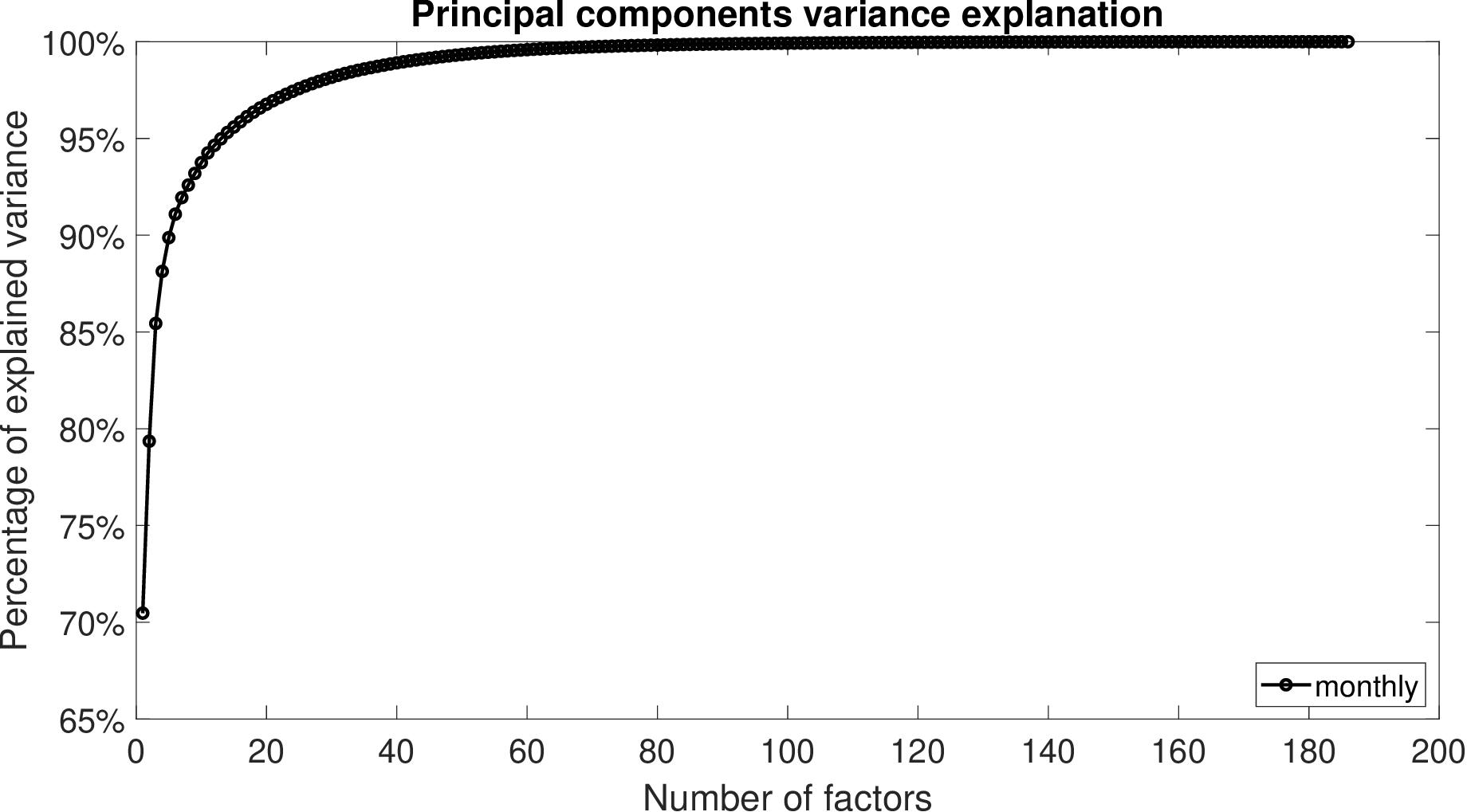}
	\caption{Variance explained from the principal components of the monthly data. One can easily observe that only the first principal components are required to explain a sufficiently large amount of the total variance.}
	\label{fig:principal_components_monthly_data}
\end{figure}

 Finally, we check if the latter methodology has an impact on the PCA we use to detect the number of random factors which drive the market. Since we are only obtaining monthly products form those with a coarser granularity, we are not expected to introduce more information and variability with respect the case in which we work with row date of Table \ref{tbl:relative_product_structure}. In order to verify this claim, we compare the number of factor we need to explain the $95\%$ of the variance both using raw and monthly data from German power futures market over the period from $1/1/2020$ to $31/12/2021$. From Figure \ref{fig:number_principal_components} we observe that in both cases fourteen principal components are needed to explain around the  $95\%$ of the total variance. In case of monthly data the original total number of random factors is much larger than that of the raw data, since many quotations have been replicated on a finer granularity scale. Nevertheless, since many of them are just \enquote{replications} of the exiting ones, they are not needed to explain a significant amount of the whole market variance as confirmed by Figure \ref{fig:principal_components_monthly_data}.  

In light of the preceding results we can claim that the calibration of volatility parameters using the PCA procedure we sketched in Section \ref{sec:PCA_dimension_reduction}, can be applied both on raw data ($F_{Y_{h}}^{k}(t,T)$, $F_{Q_{h}}^{k}(t,T)$, $F_{M_{h}}^{k}(t,T)$) or on reconstructed monthly forward contracts ($F_{M_{h}}^{k}(t,T)$). The latter approach is preferable when commodity with different granularity are present: in this case, working with products having the same granularity leads to cleaner analysis and easier coding.

\section{Numerical results and \textcolor{black}{derivatives pricing}}


\label{sec:numerical_results}
In this section we consider \textcolor{black}{some real applications}. 
\textcolor{black}{We calibrate the model on futures market prices $F(t,\tau^{s},\tau^{e})$ over the period from $1/1/2020$ to $31/12/2020$. Starting from contracts of the form $F(t,\tau^{s},\tau^{e})$, we obtain $F_{M_{h}}(t,T_{m})$ as we have shown in Section \ref{sec:data_preparation}. We consider six different markets: four power futures markets, Germany (DE), France (F7), Italy (IT) and Switzerland (CH) and two natural gas markets, Italy (PSV) and The Netherlands (TTF), with delivery dates up to $M=24$ months ahead. Larger maturities can be also considered, if data are available.}
\par
\begin{figure}
	\centering
	\subfigure{\includegraphics[width=0.48\textwidth]{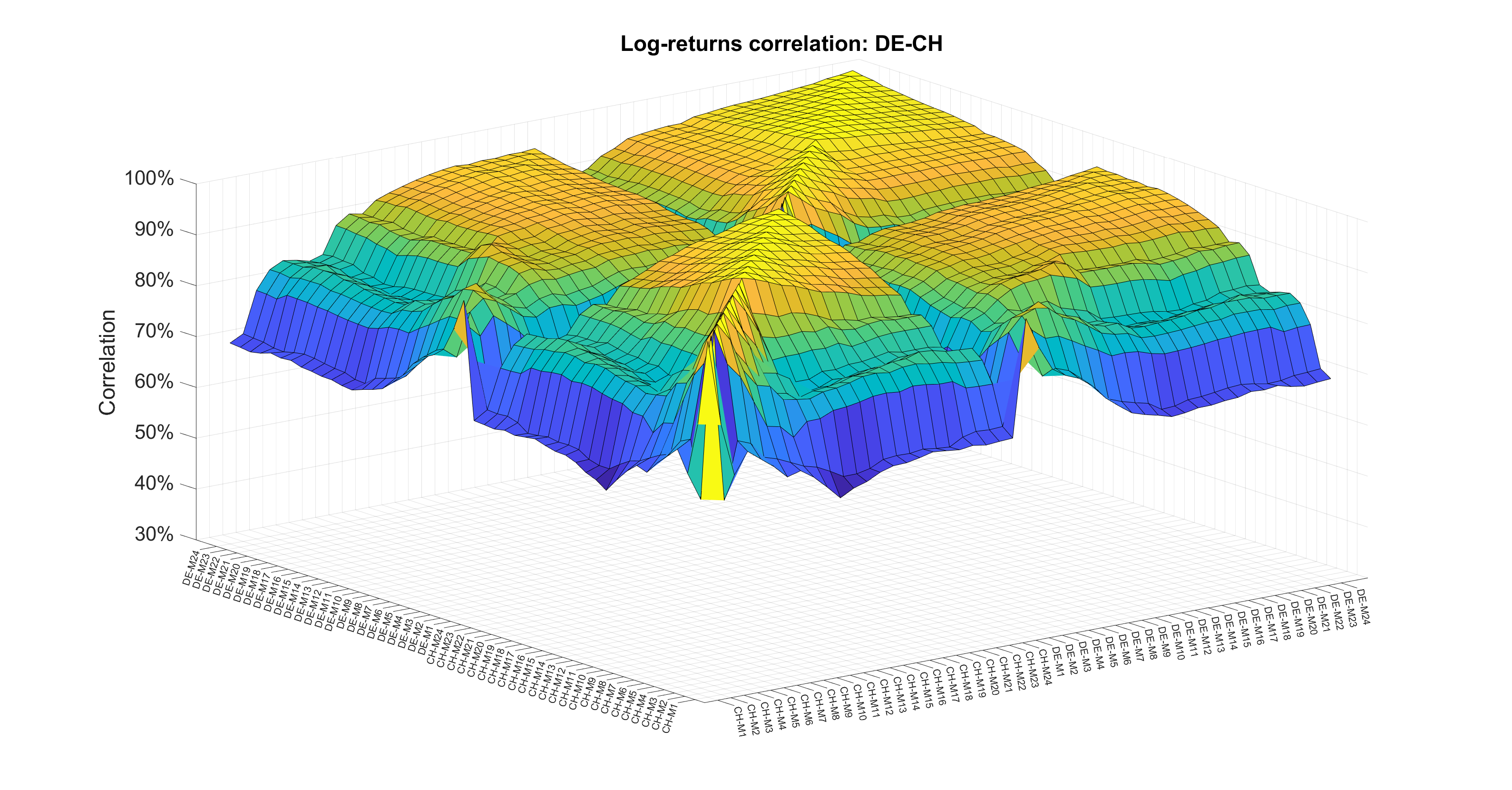}} 
	\subfigure{\includegraphics[width=0.48\textwidth]{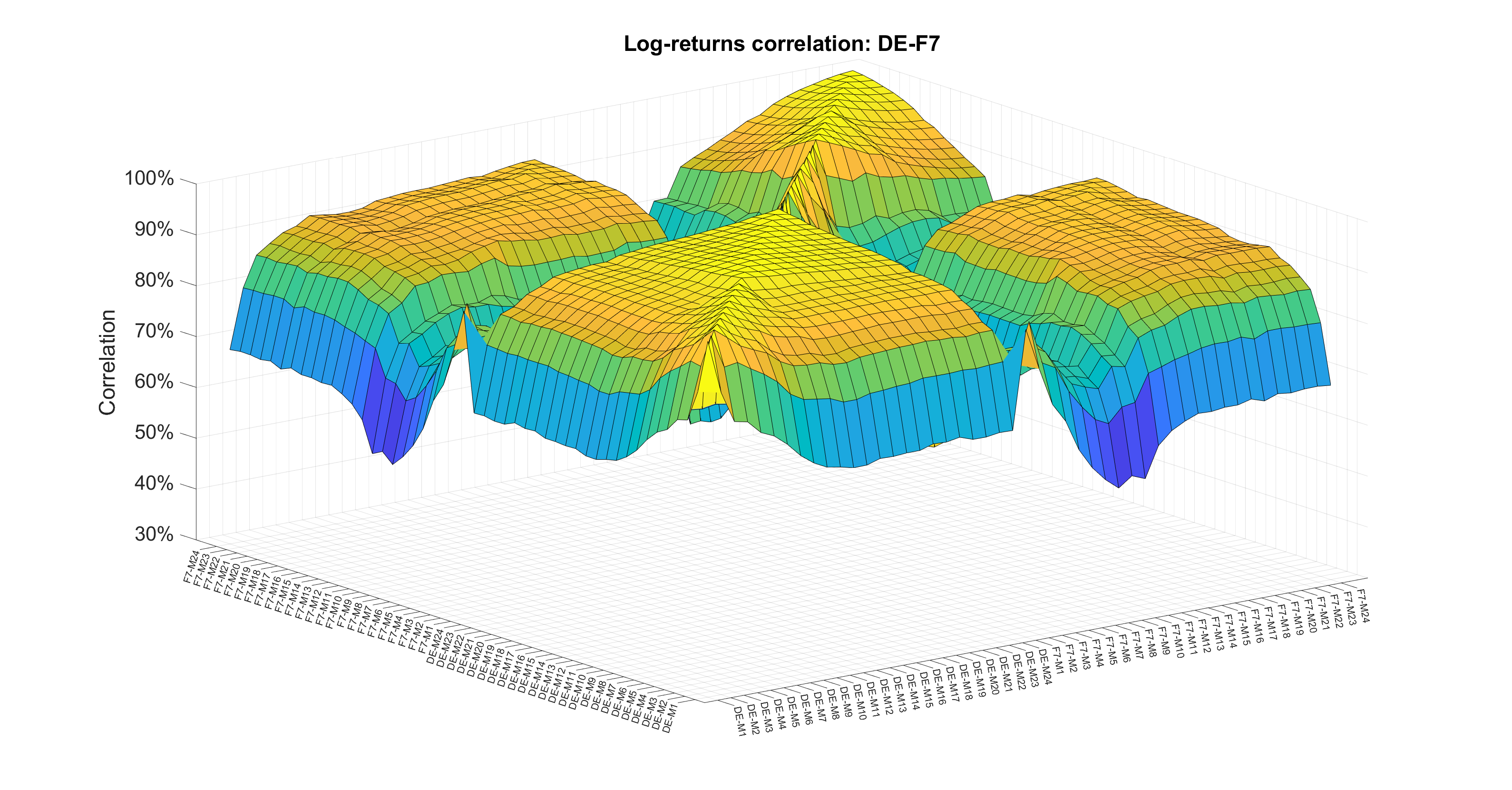}} 
	\subfigure{\includegraphics[width=0.48\textwidth]{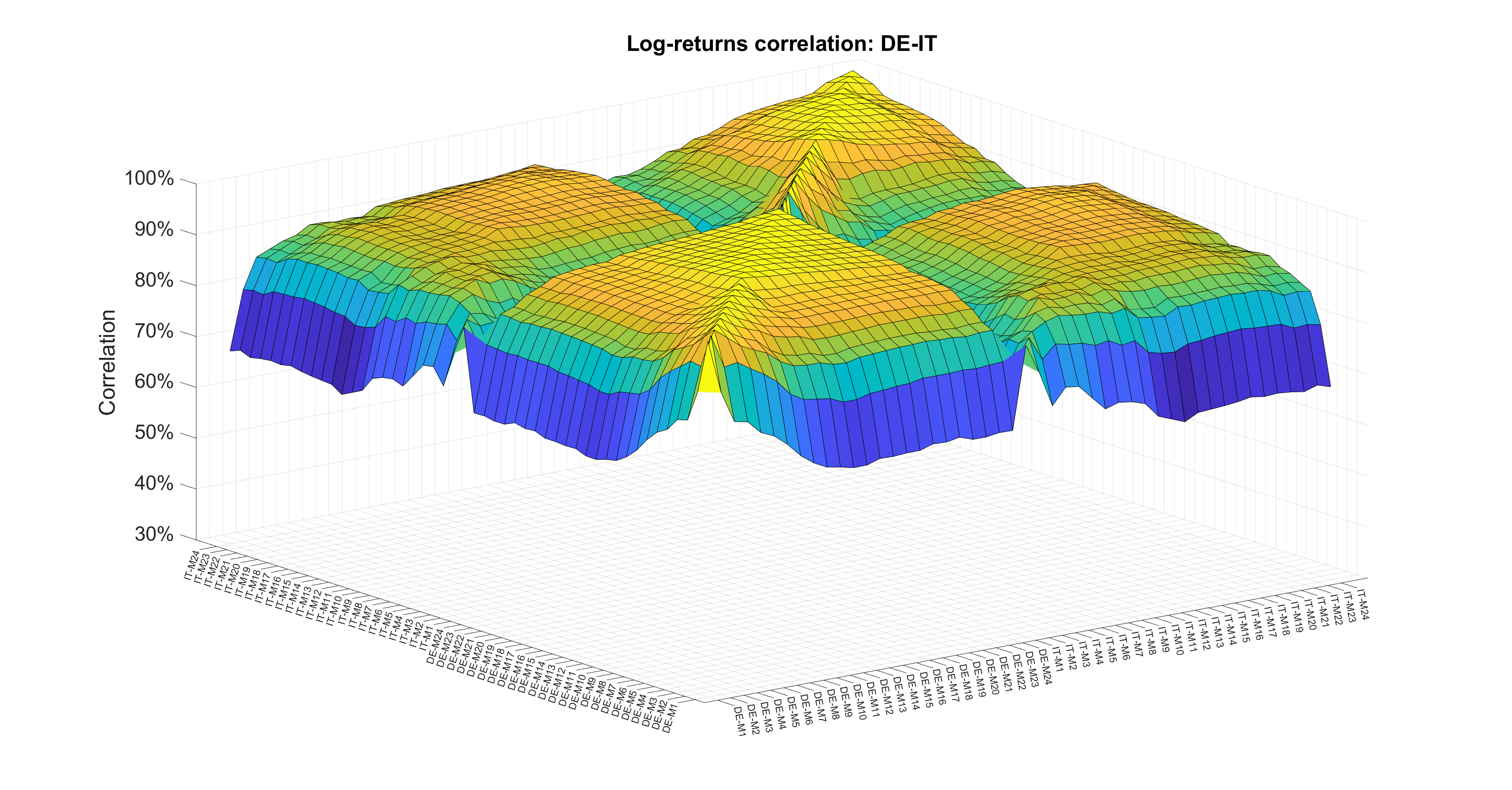}}
	\subfigure{\includegraphics[width=0.48\textwidth]{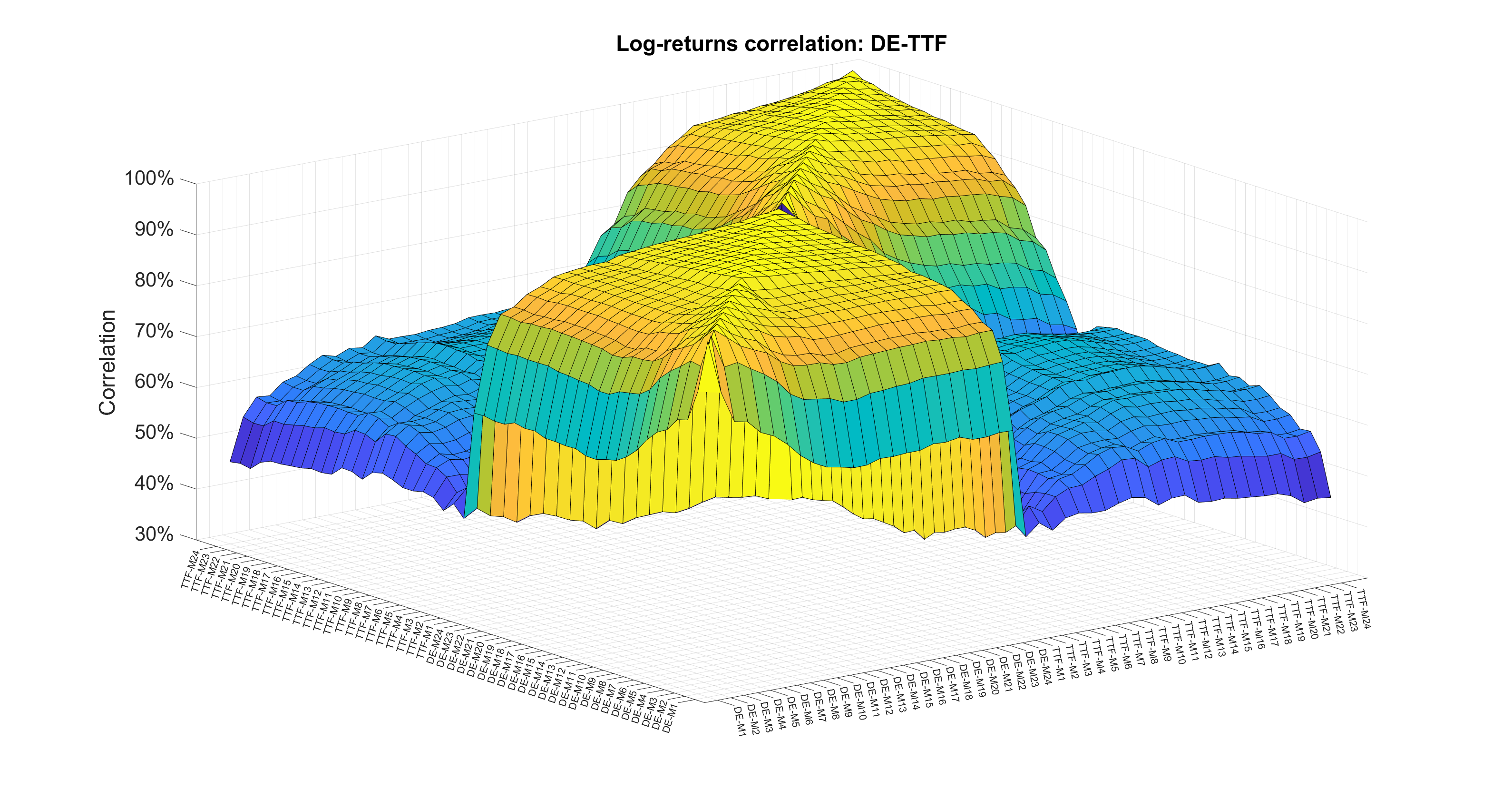}}
	\subfigure{\includegraphics[width=0.48\textwidth]{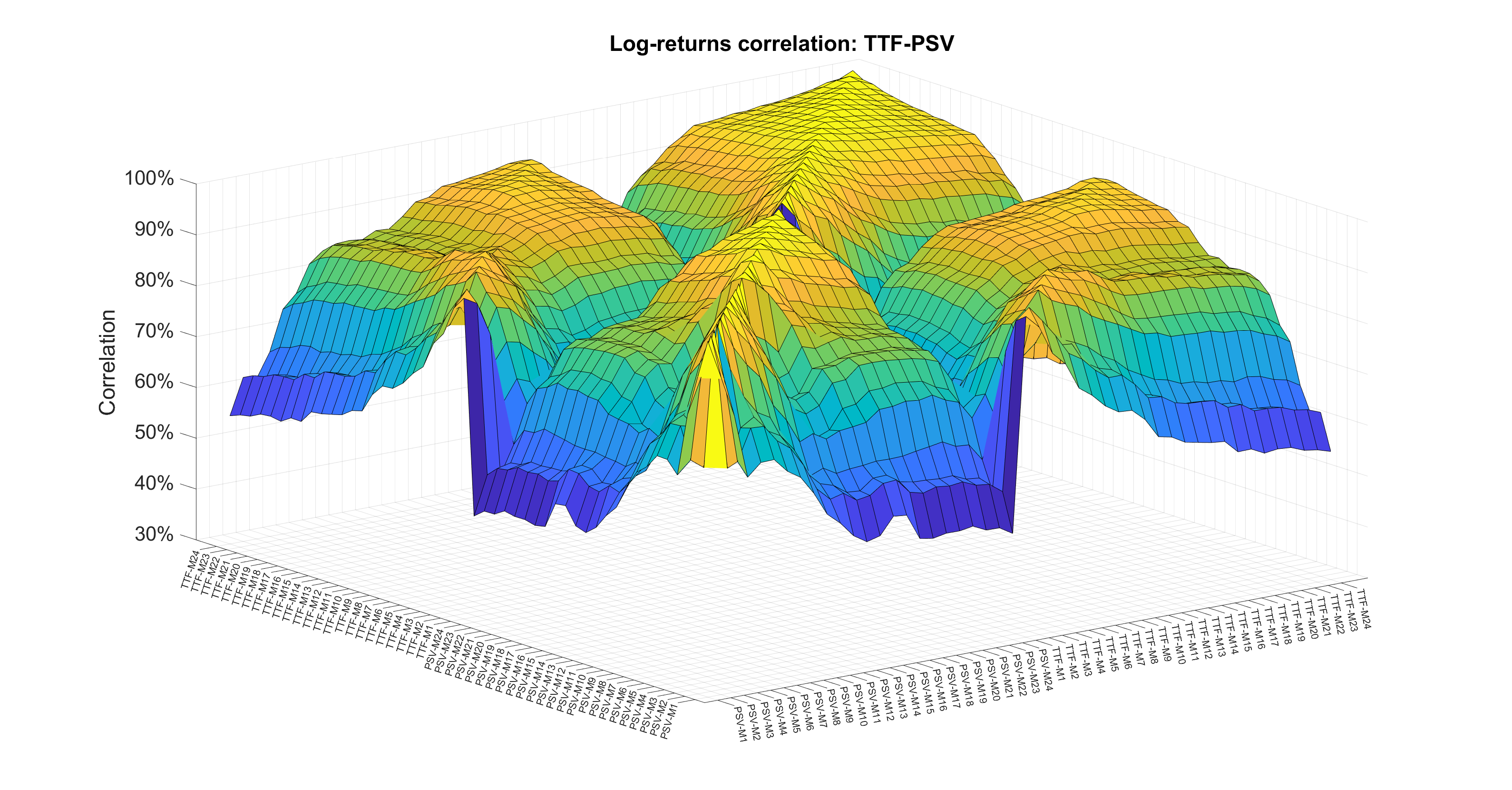}}
	\subfigure{\includegraphics[width=0.48\textwidth]{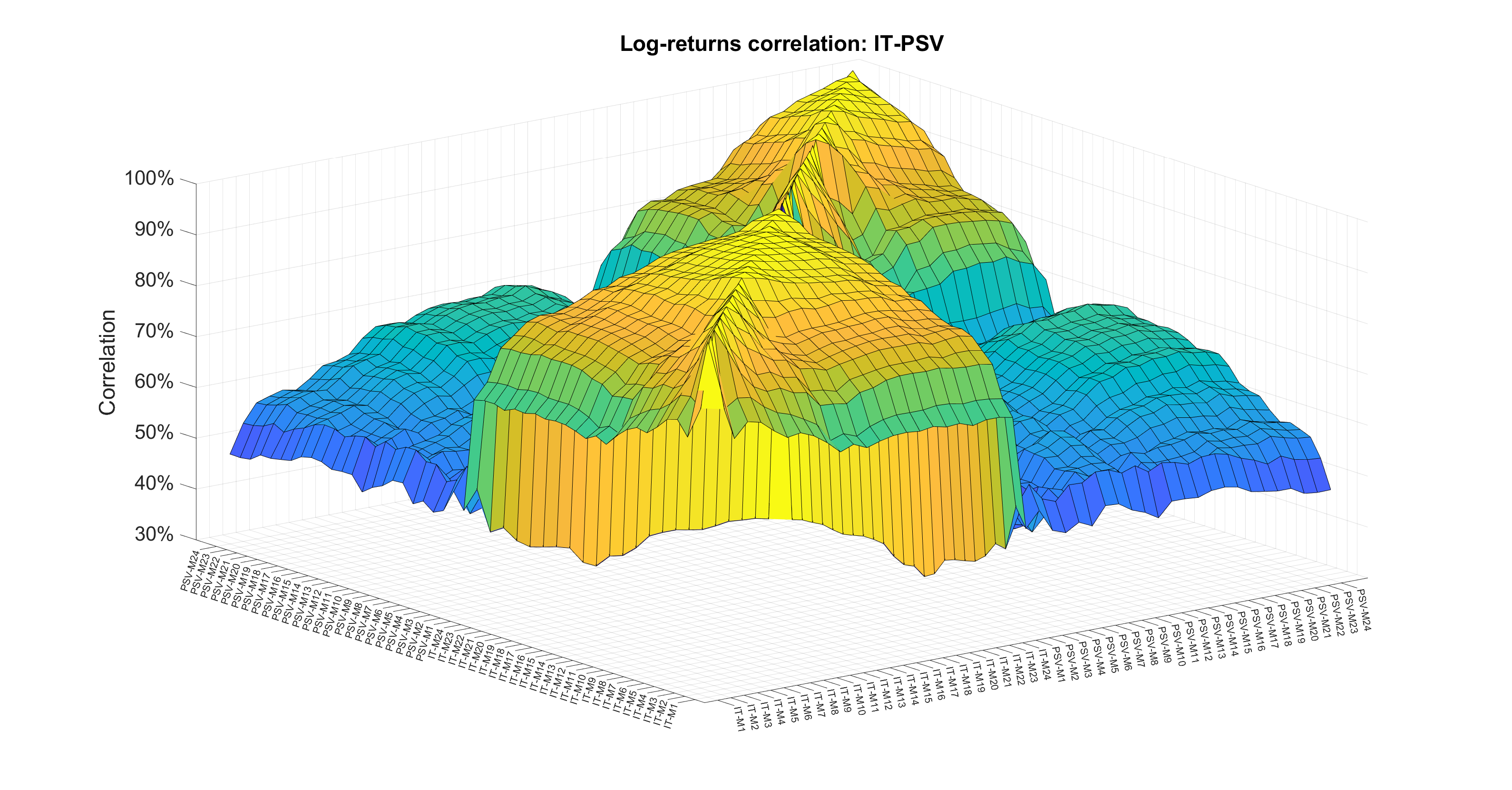}}
	\caption{Historical daily log-returns correlation surfaces for different commodities. }
	\label{fig:correlation_surfaces_historical}
\end{figure}

 As observed in Section \ref{sec:market_structure_and_analysis} we expect a significant level of co-integration between markets: indeed, as we can observe in Figure \ref{fig:correlation_surfaces_historical}, the level of historical correlation is high for especially for the long-term ones. Furthermore, the correlation between power and natural gas is significant too, since the natural gas is commonly used as fuel for electricity production in many European countries.\\ 

\par Starting from the reconstructed quotations of the form $F_{M_{h}}(t,T_{m})$ we compute the log-returns matrix $\boldsymbol{X}$. As discussed in the previous section, \enquote{fake} spikes in log-returns might be created and they must be removed. Moreover, since we are assuming that log-returns are Gaussian, we filter out the outliers the approach proposed by \citet{CarteaFigueroa}. 
Once matrix $\boldsymbol{X}$ has been prepared, we perform a PCA analysis and we select a sufficient bulk of stochastic factors such that they are enough to explain a sufficiently large part of the total market variance. As observed by \citet{Feron2020}, \citet{Geman2005} and \citet{Koekebakker2005} 5-10 factors in power markets are enough to explain a large part of the variance, whereas for other markets, such as metals, 4 are enough. In this example we start from 144 stochastic factors and, after a PCA, analysis we get that 10 factors are enough to explain the 90\% of the total market variance. 

\begin{figure}
	\centering
	\subfigure{\includegraphics[width=0.48\textwidth]{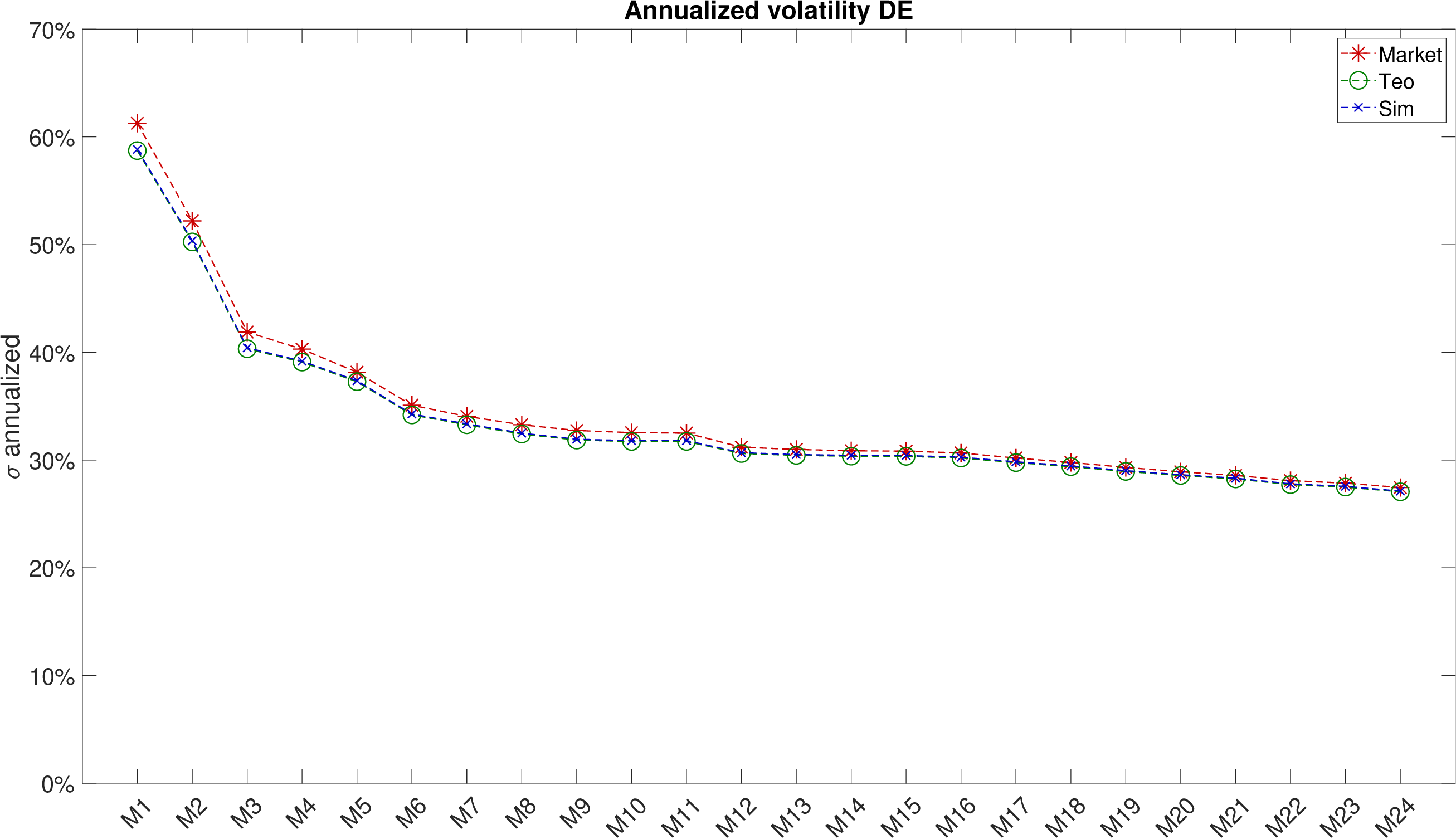}} 
	\subfigure{\includegraphics[width=0.48\textwidth]{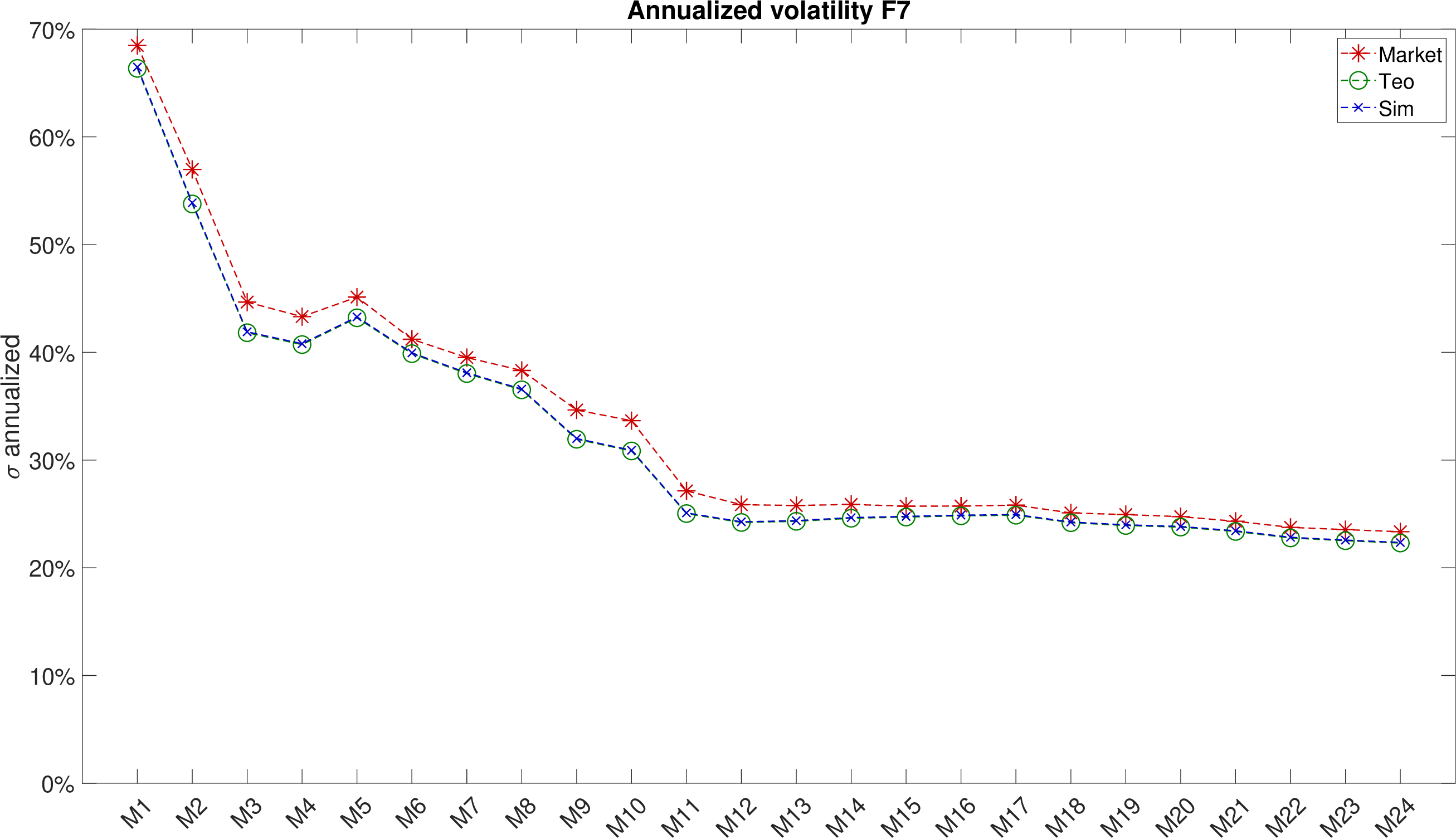}} 
	\subfigure{\includegraphics[width=0.48\textwidth]{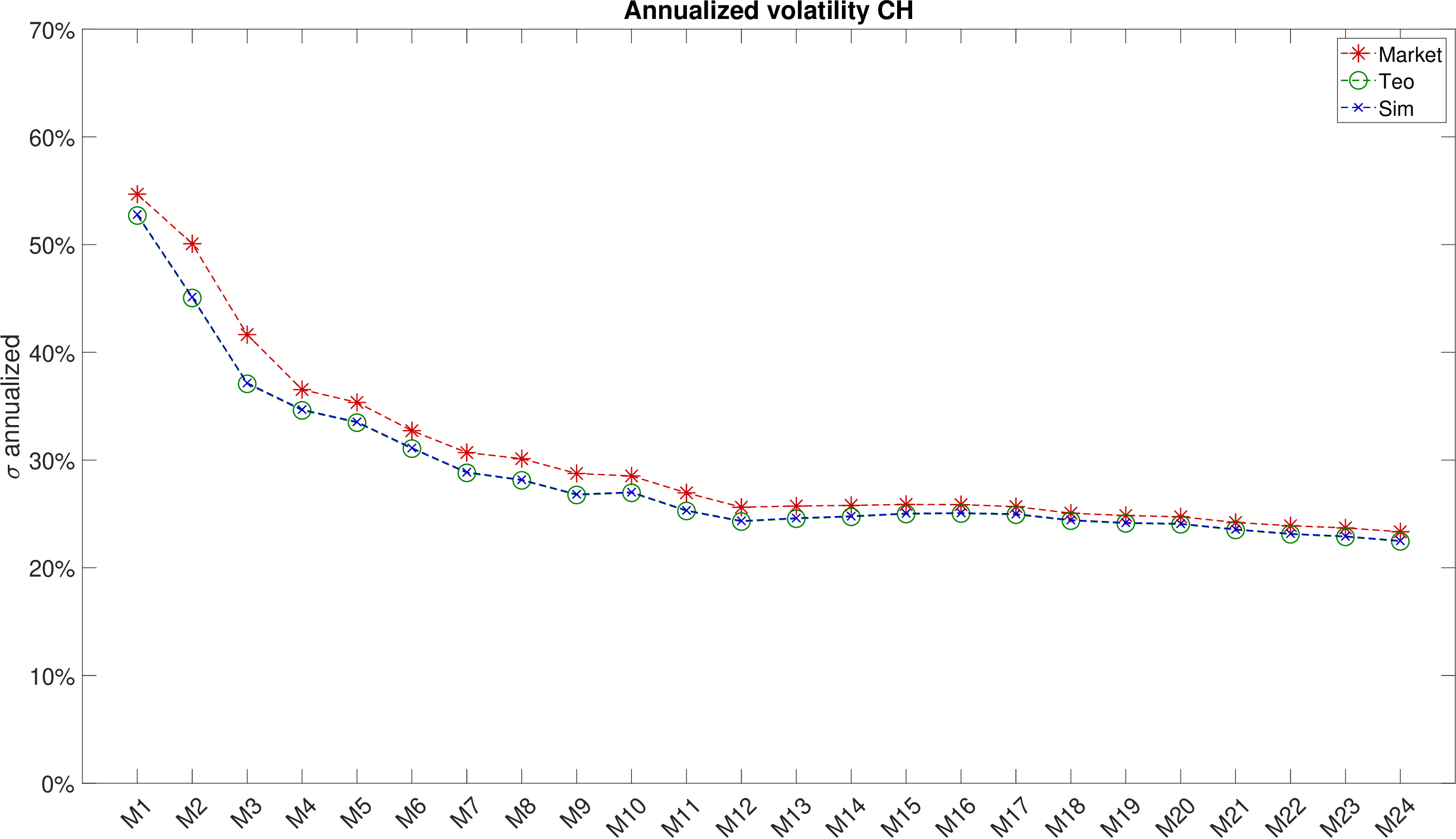}}
	\subfigure{\includegraphics[width=0.48\textwidth]{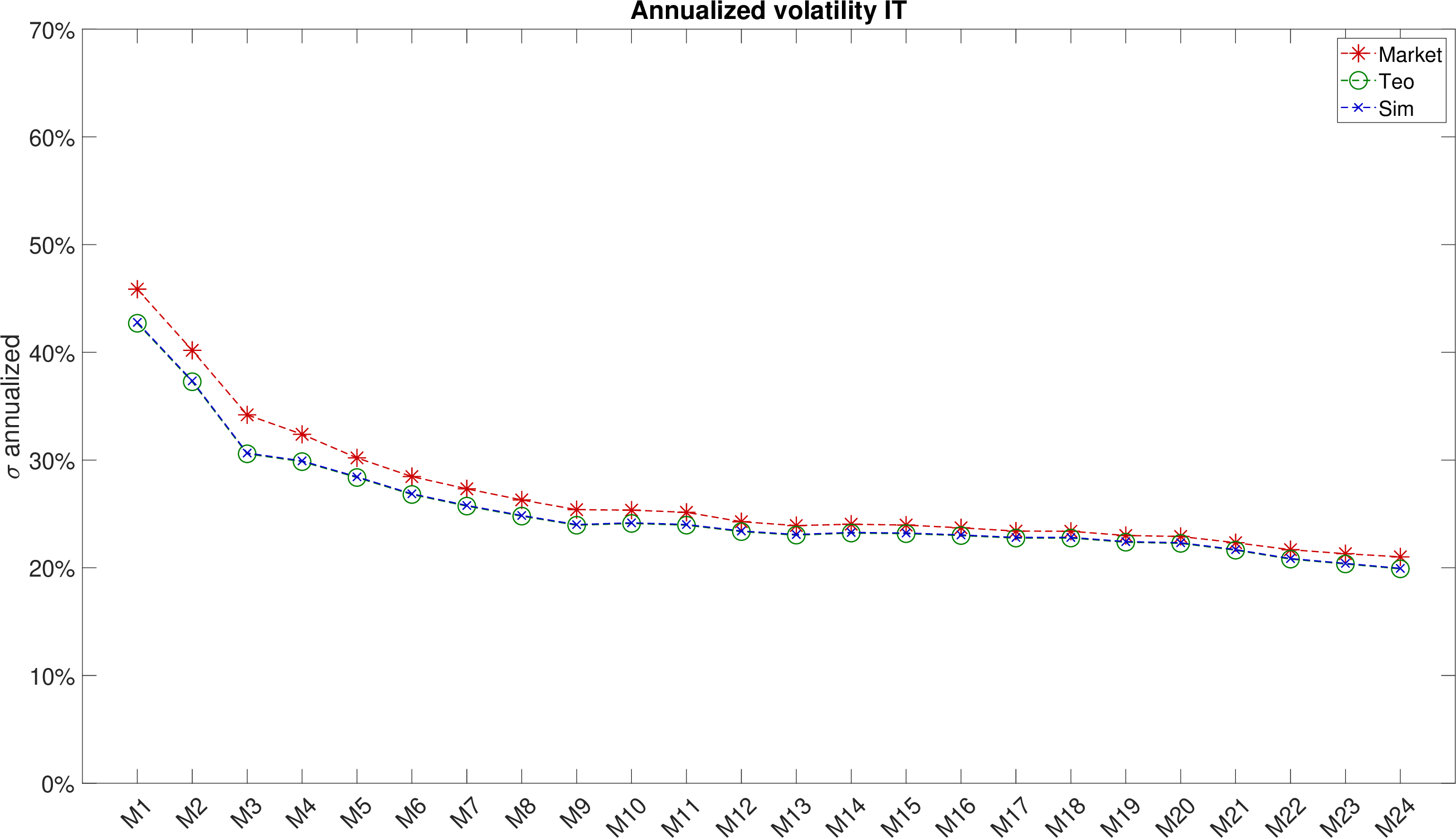}}
	\subfigure{\includegraphics[width=0.48\textwidth]{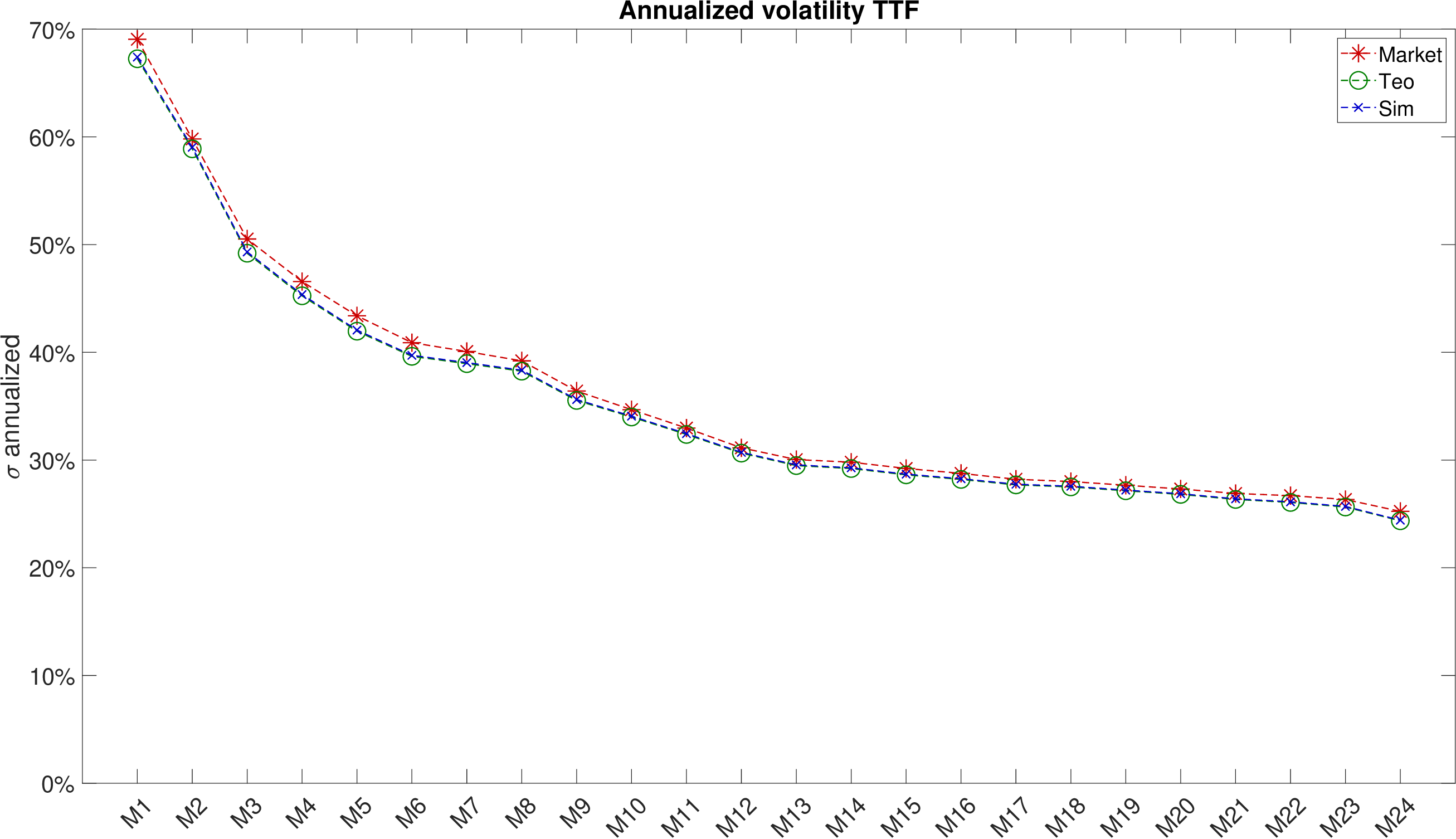}}
	\subfigure{\includegraphics[width=0.48\textwidth]{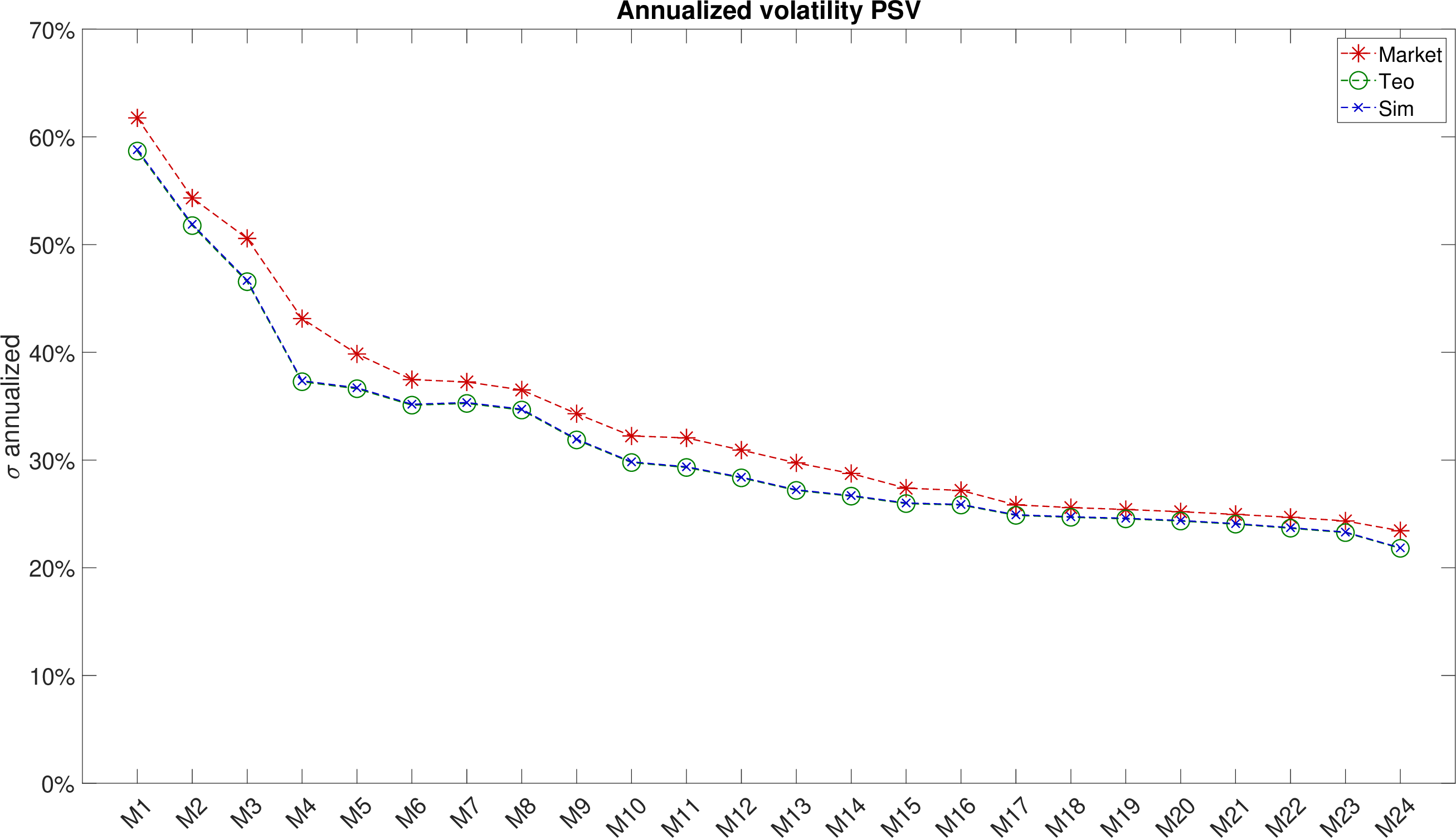}}
	\caption{Simulated, market and theoretical annualized volatility. }
	\label{fig:variance_check}
\end{figure}

\par
\begin{figure}[!h]
	\centering
	\includegraphics[width=1\textwidth]{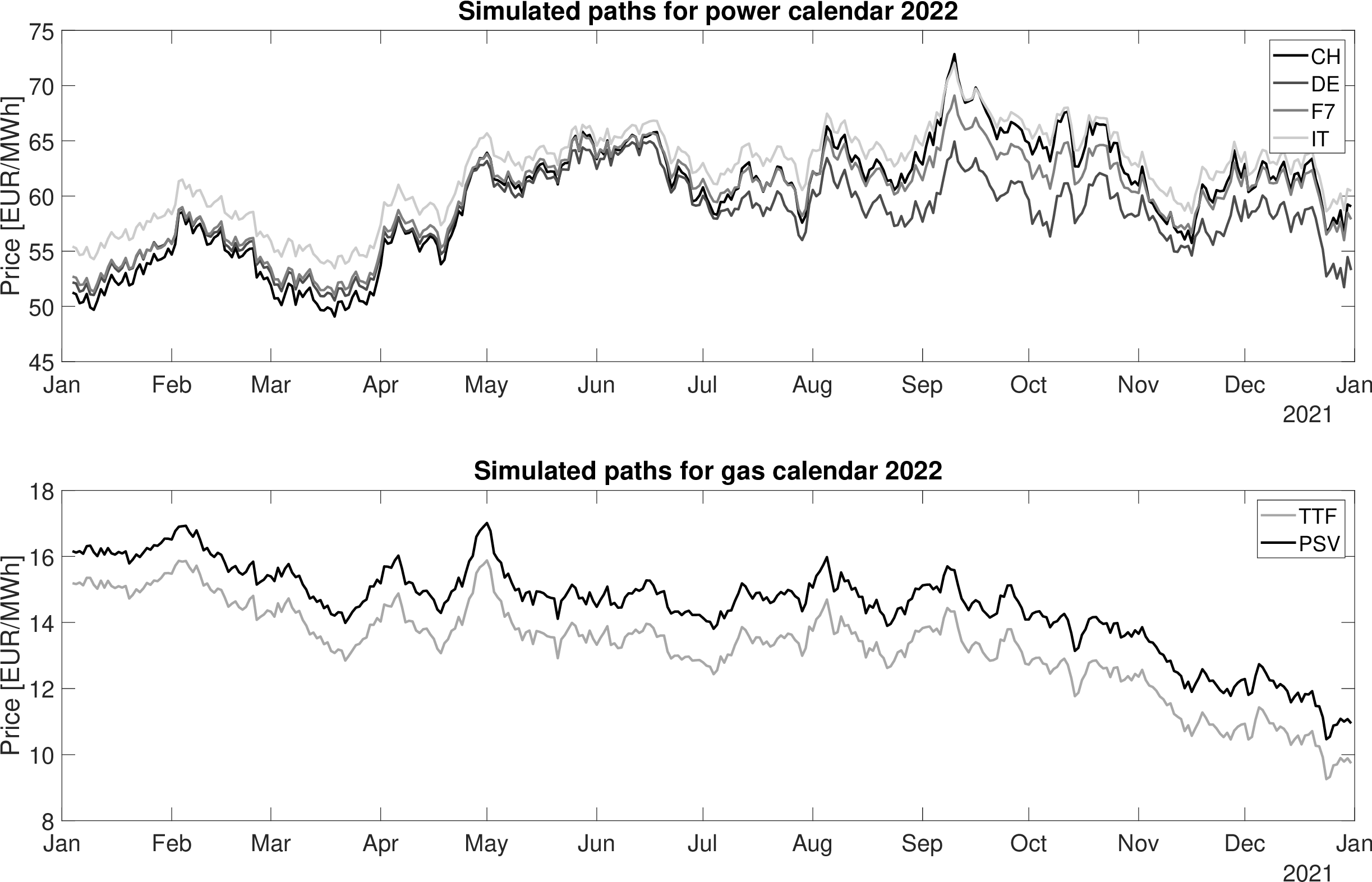}
	\caption{Sample paths for the product calendar 2022 with delivery date from 2022-01-01 to 2022-12-31.}
	\label{fig:all_markets_calendar_2022}
\end{figure}

 Following the procedure we illustrated in Section \ref{sec:multi_commodity_to_example}, by PCA  we calibrate the model by fitting the matrix $\boldsymbol{\sigma}^{*}$. Once this has been done, we are ready to simulate the structure of the forward curve. Considering as $t_{0} = 4/1/2021$, by using Equation \eqref{eqn:fwd_exact_solution} we simulate futures price for the all the markets we consider $F^{k}(t,\tau^{s},\tau^{e}),\; k=1,\dots,K$ with delivery period $\tau^{s} = 1/1/2022$, $\tau^{e} = 31/12/2022$ form $t_{0}$ up to the maturity $T=31/12/2022$. The result is shown in Figure \ref{fig:all_markets_calendar_2022}. The dynamics of the prices exhibits an extremely high level of dependence: in particular, both power and natural gas prices tend to rise or fall together over the whole simulation period.

%

\begin{figure}
	\centering
	\subfigure{\includegraphics[width=0.48\textwidth]{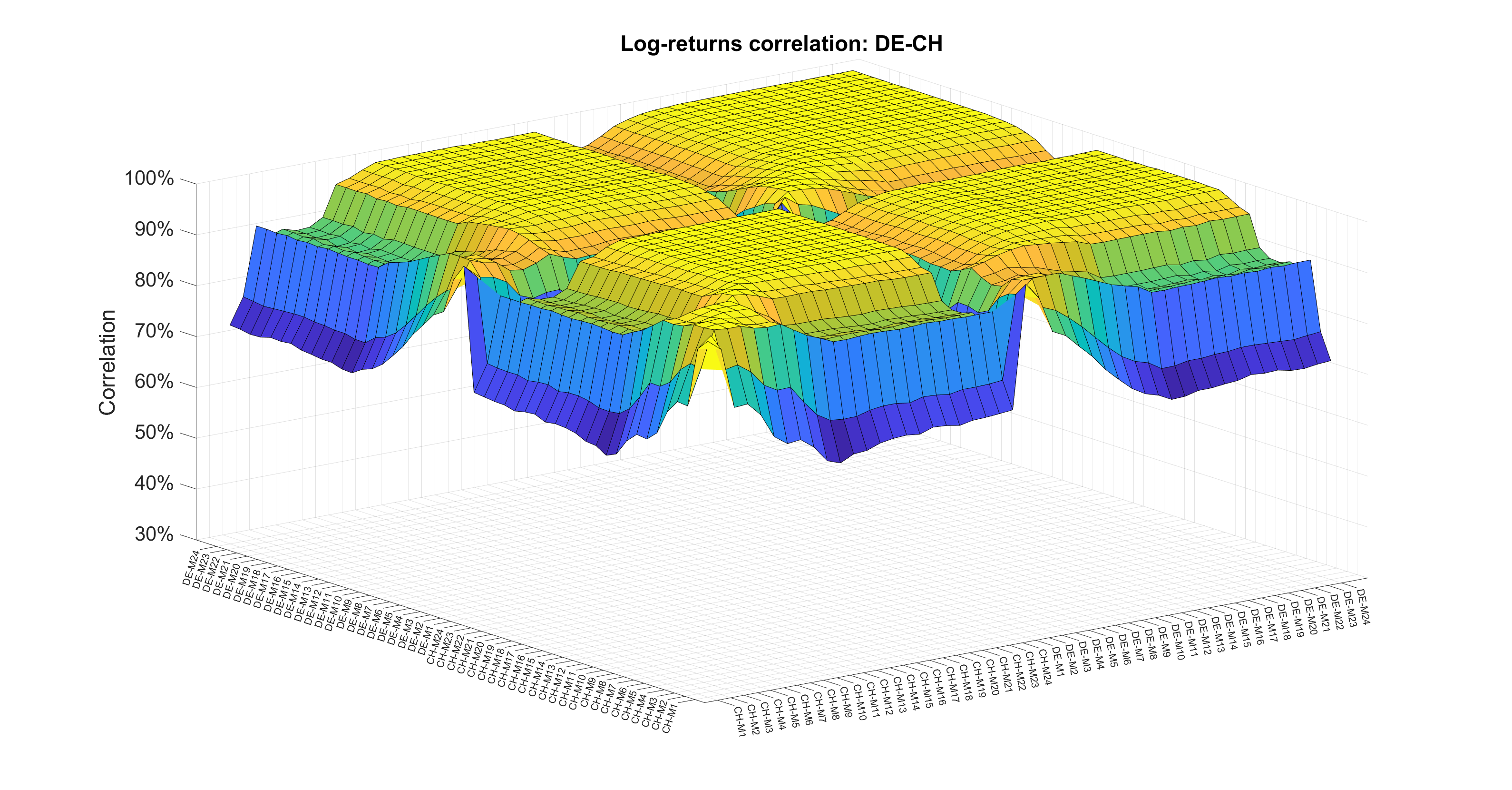}} 
	\subfigure{\includegraphics[width=0.48\textwidth]{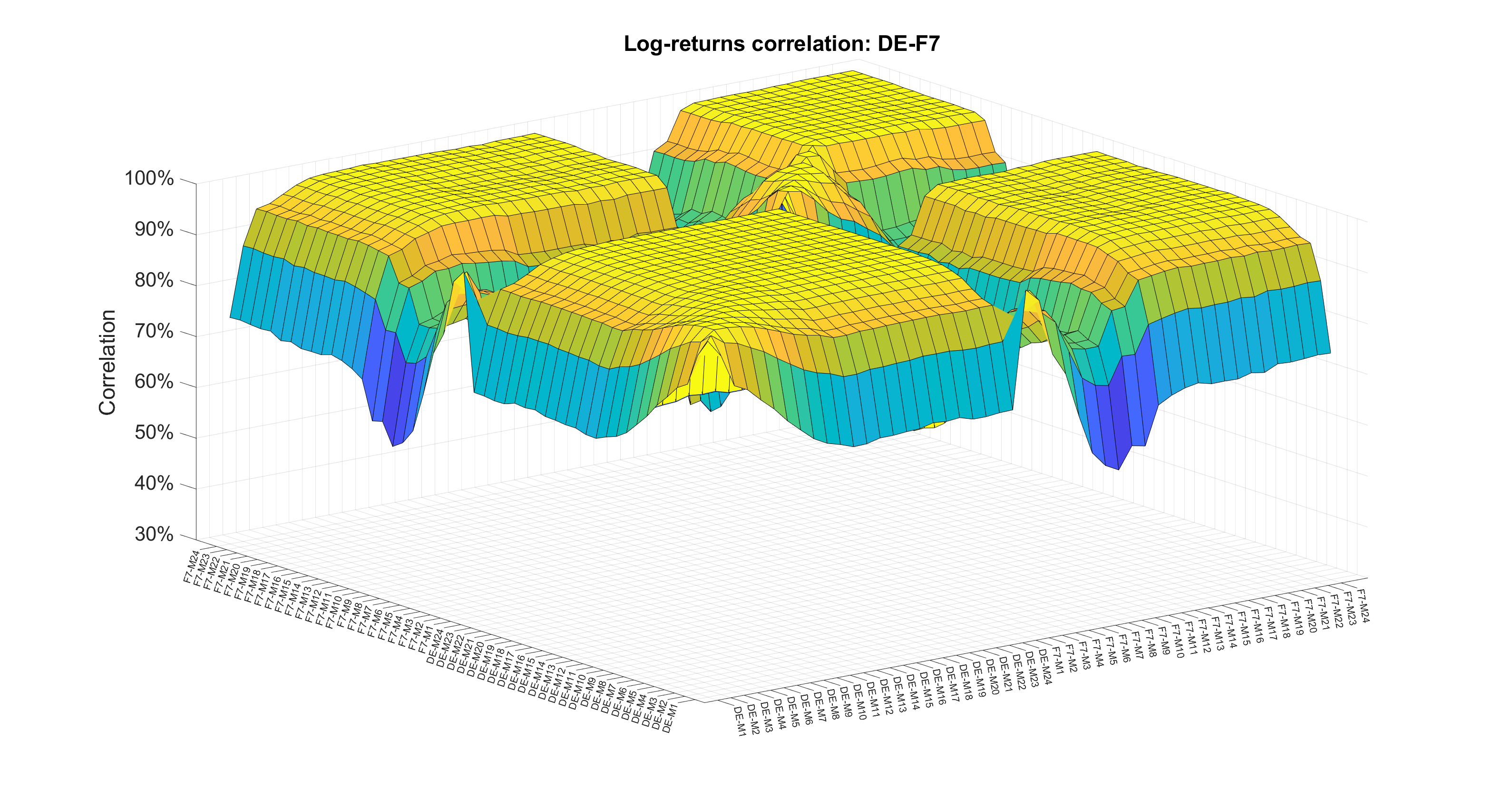}} 
	\subfigure{\includegraphics[width=0.48\textwidth]{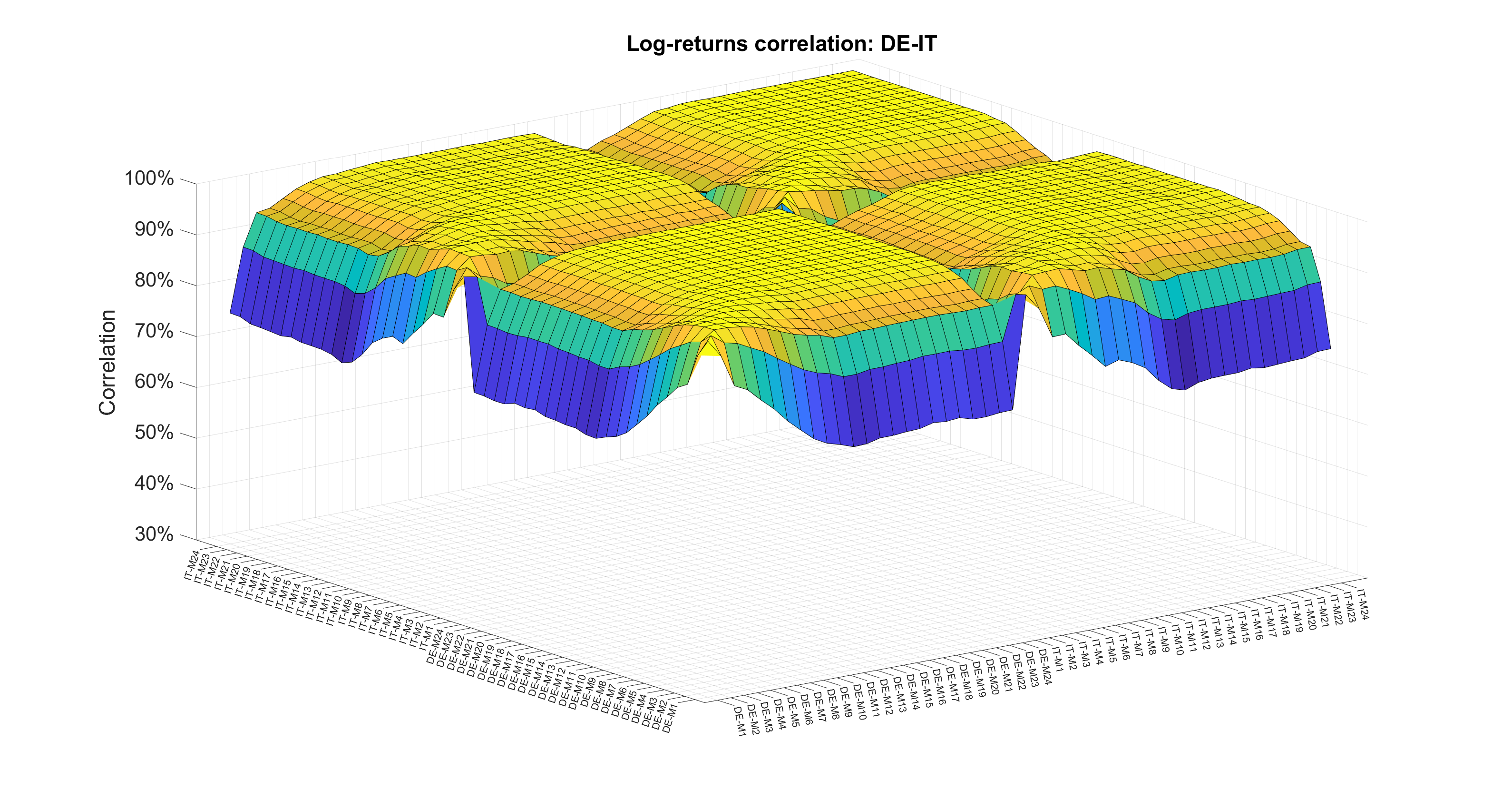}}
	\subfigure{\includegraphics[width=0.48\textwidth]{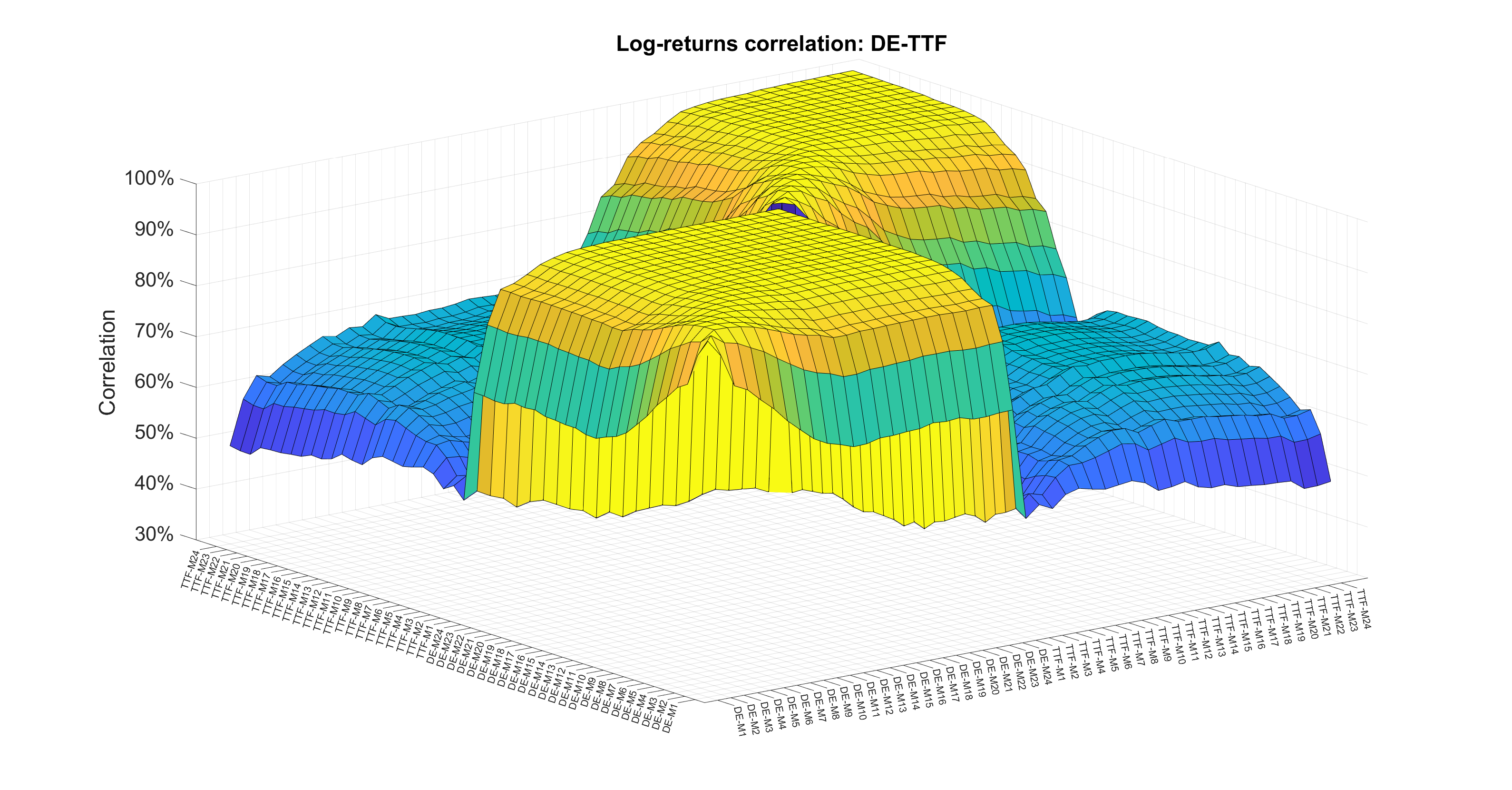}}
	\subfigure{\includegraphics[width=0.48\textwidth]{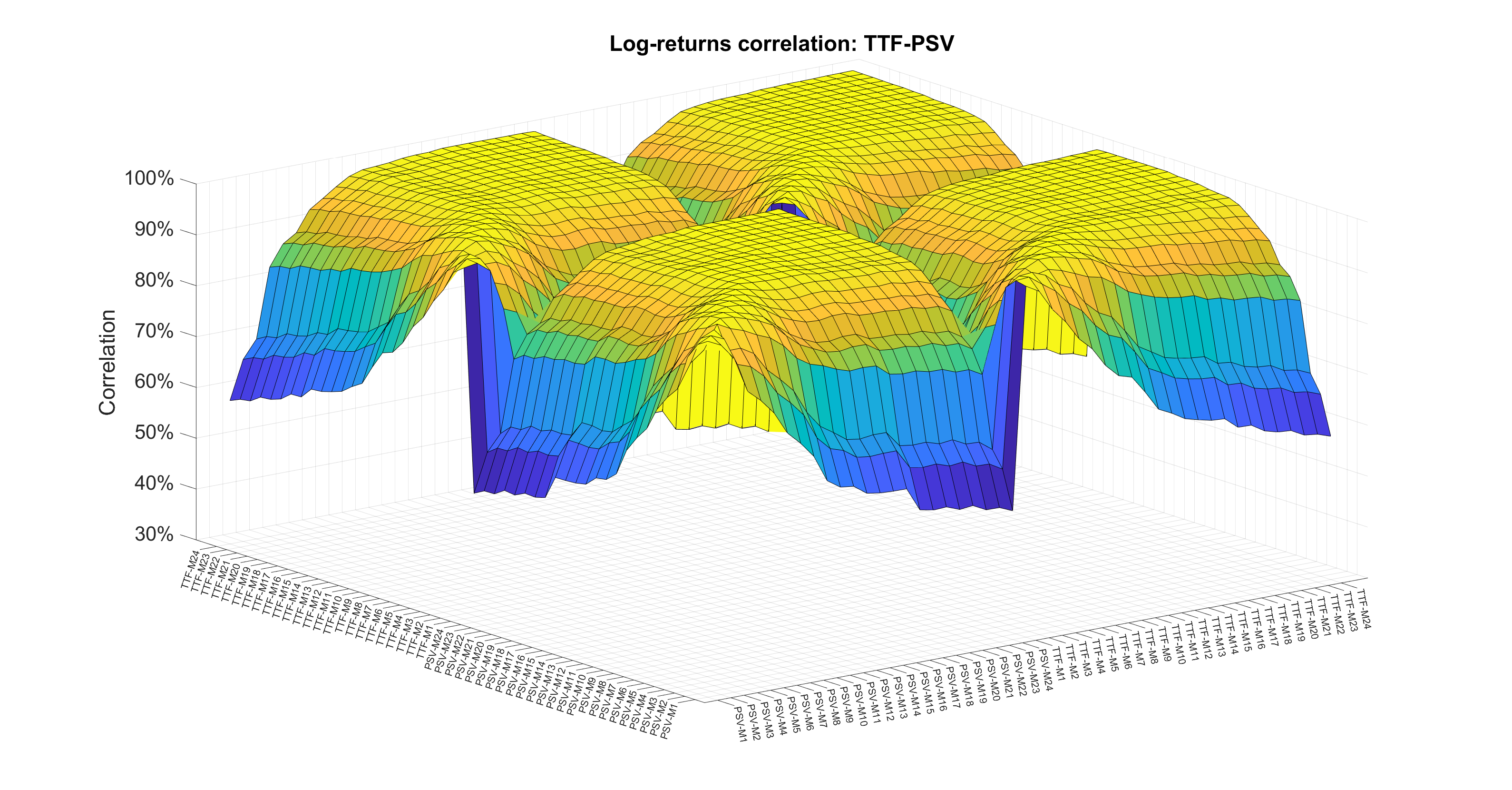}}
	\subfigure{\includegraphics[width=0.48\textwidth]{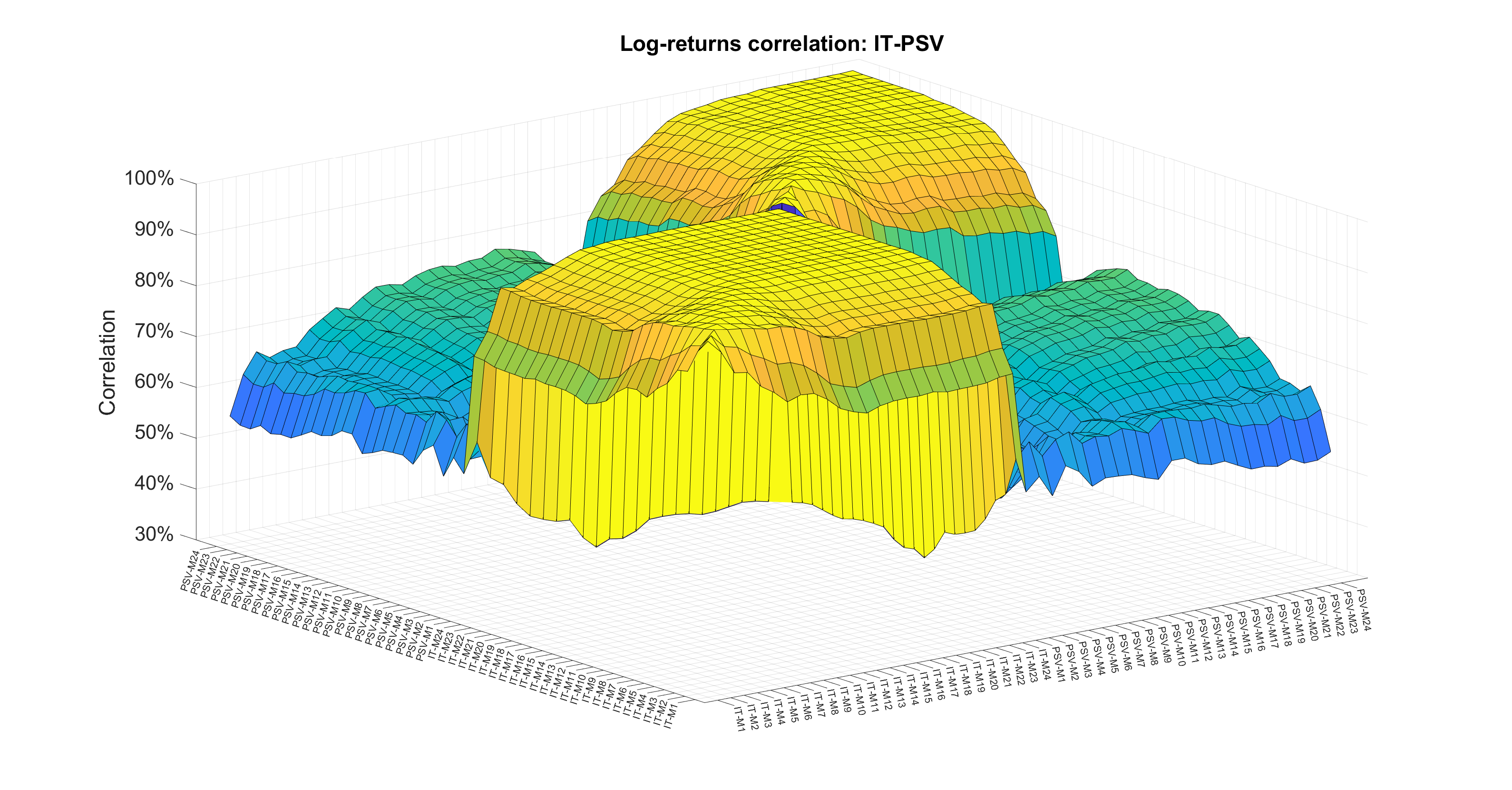}}
	\caption{Simulated daily log-returns correlation surfaces for different commodities.}
	\label{fig:correlation_surfaces_simulations}
\end{figure}

Furthermore, in Figure \ref{fig:correlation_surfaces_simulations} we show the log-returns matrices correlation computed on simulated fixed delivery products $F_{M_{h}}^{k}(t,T_{m}),\; k=1,\dots,K\; m \in \left[1,M\right]$ on a given time interval. A comparison between Figure \ref{fig:correlation_surfaces_historical} and Figure \ref{fig:correlation_surfaces_simulations} shows that the correlation structure is properly replicated from the proposed model. Of course, since we used only 10 stochastic instead of the 144 original ones, the correlation surface computed from the simulation appears to be less varied than the original one. In particular, the correlation surface computed from the simulations is flat for long maturities. This happens because PCA selects a single stochastic factor to move all the long term structure. This is in accordance with empirical evidence since products with long time to maturity are strongly correlated and tend to oscillate in the same way.  

\par
\begin{figure}
	\centering
	\subfigure{\includegraphics[width=0.48\textwidth]{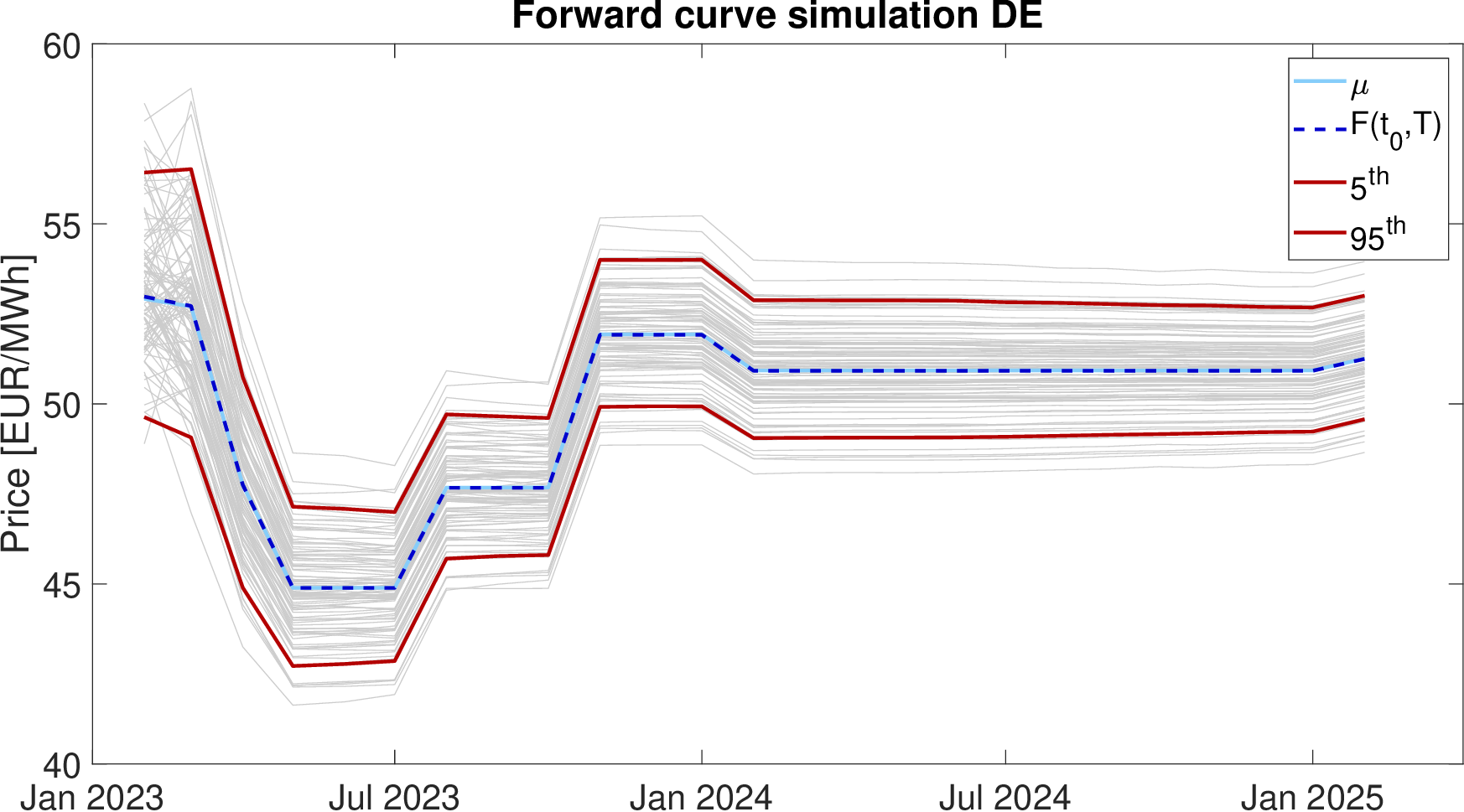}} 
	\subfigure{\includegraphics[width=0.48\textwidth]{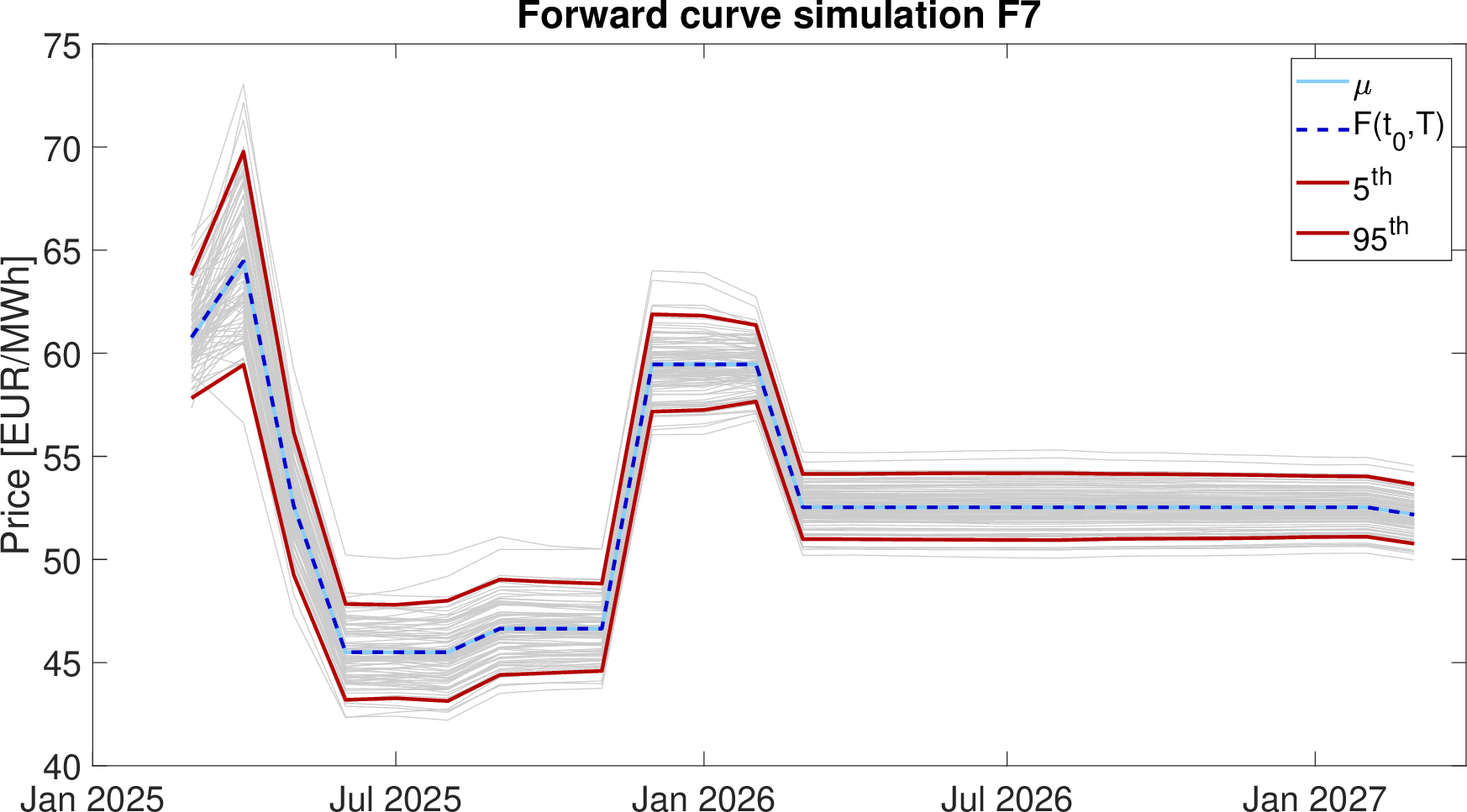}} 
	\subfigure{\includegraphics[width=0.48\textwidth]{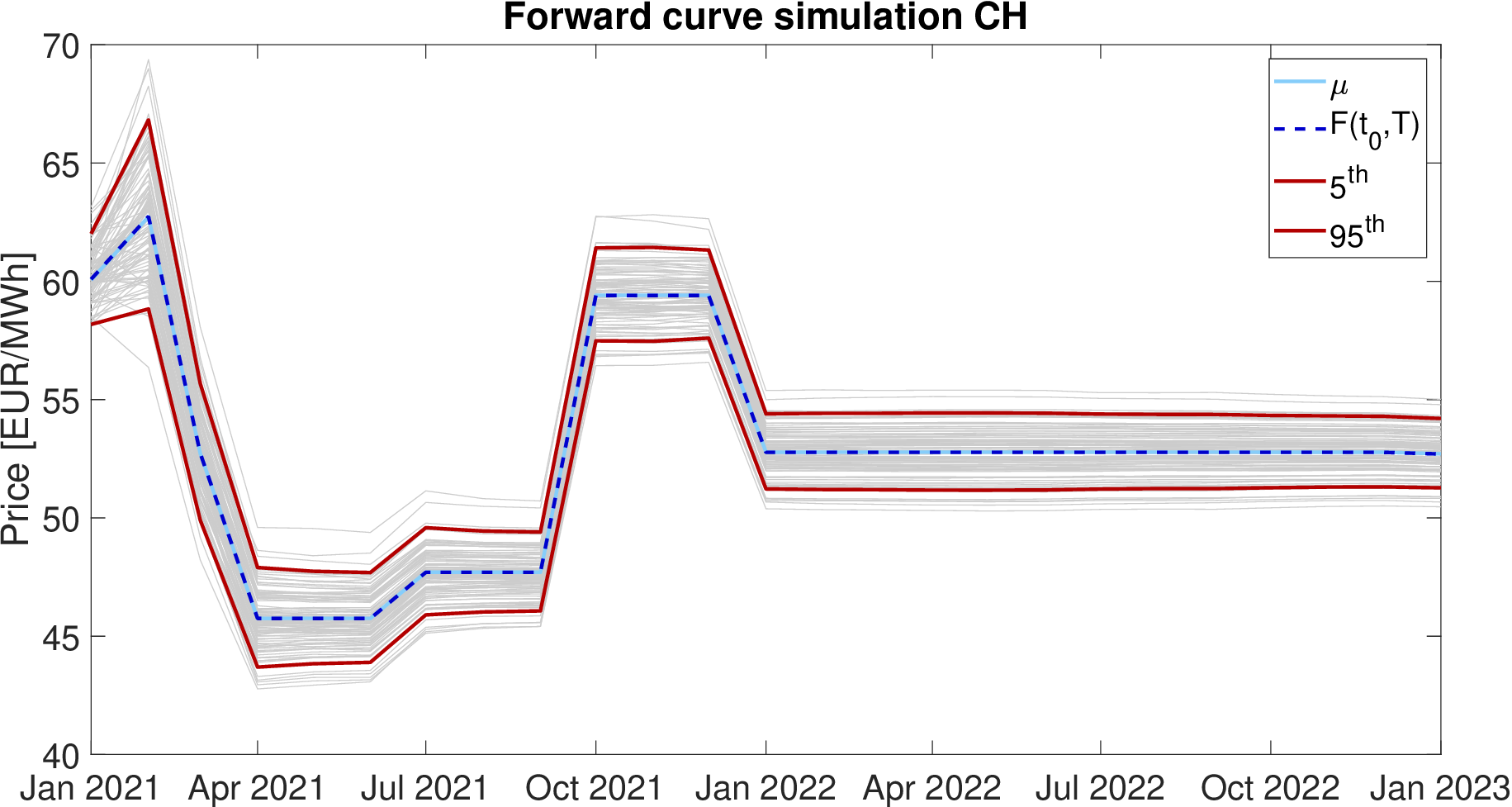}}
	\subfigure{\includegraphics[width=0.48\textwidth]{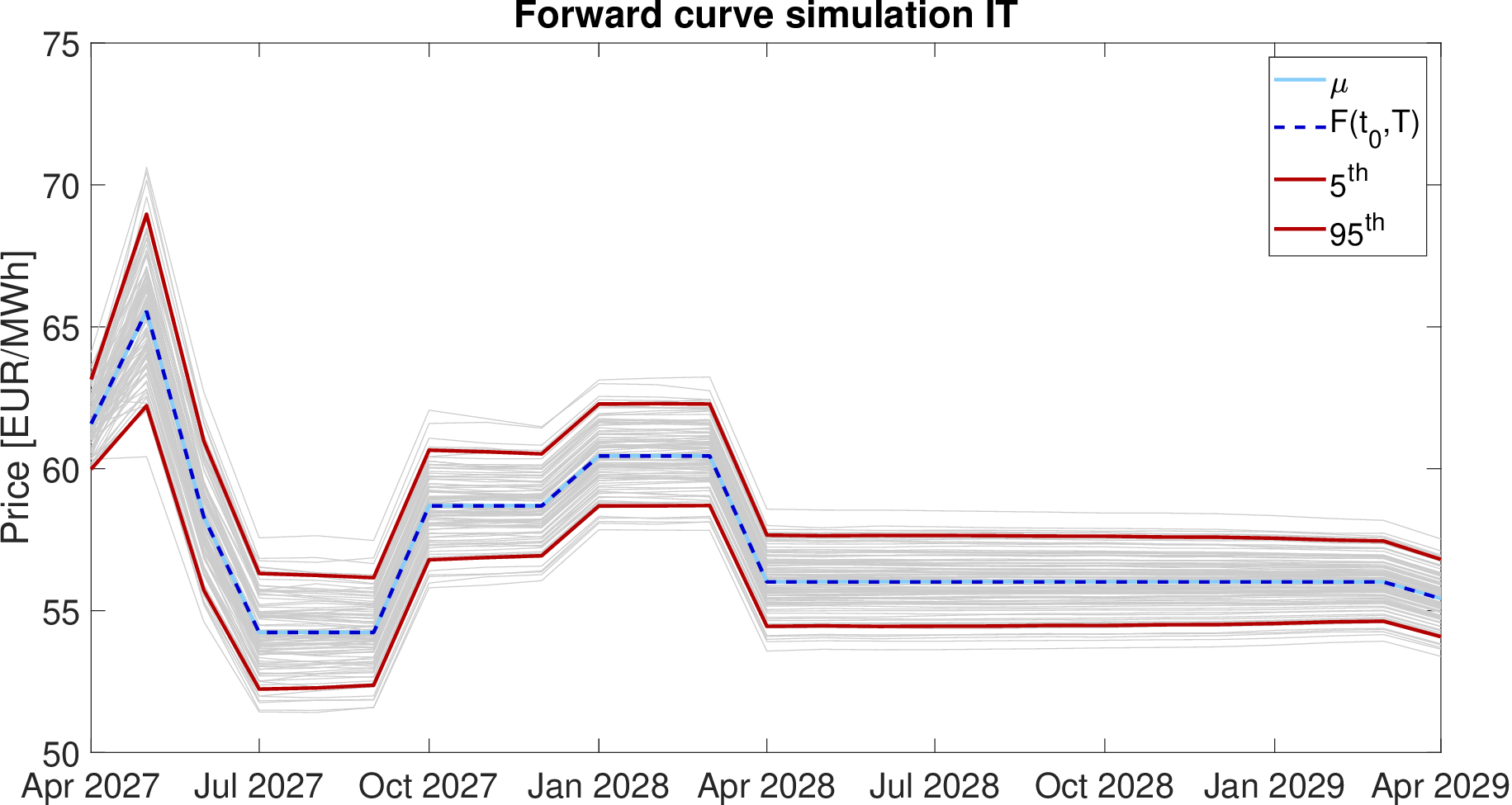}}
	\subfigure{\includegraphics[width=0.48\textwidth]{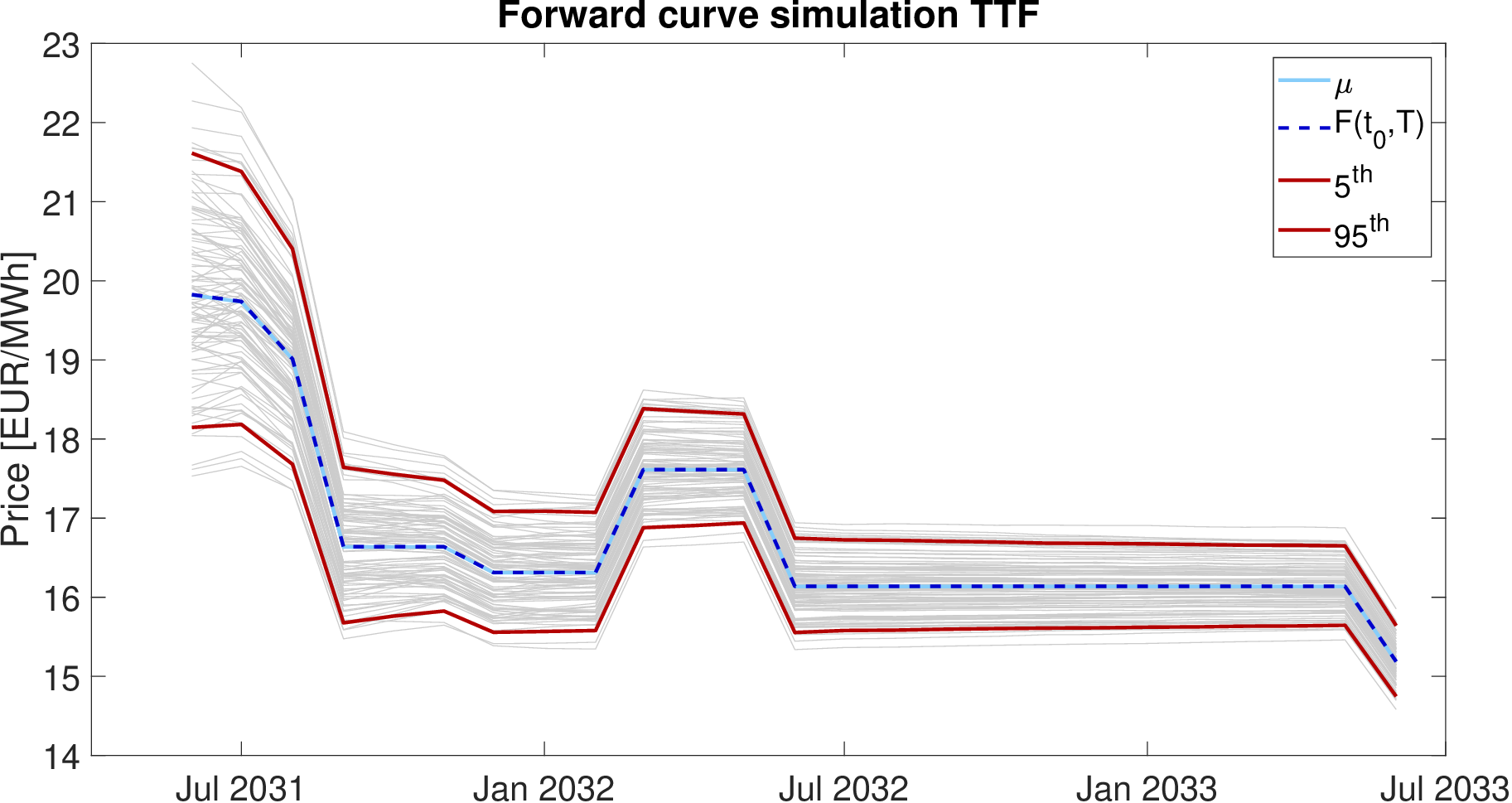}}
	\subfigure{\includegraphics[width=0.48\textwidth]{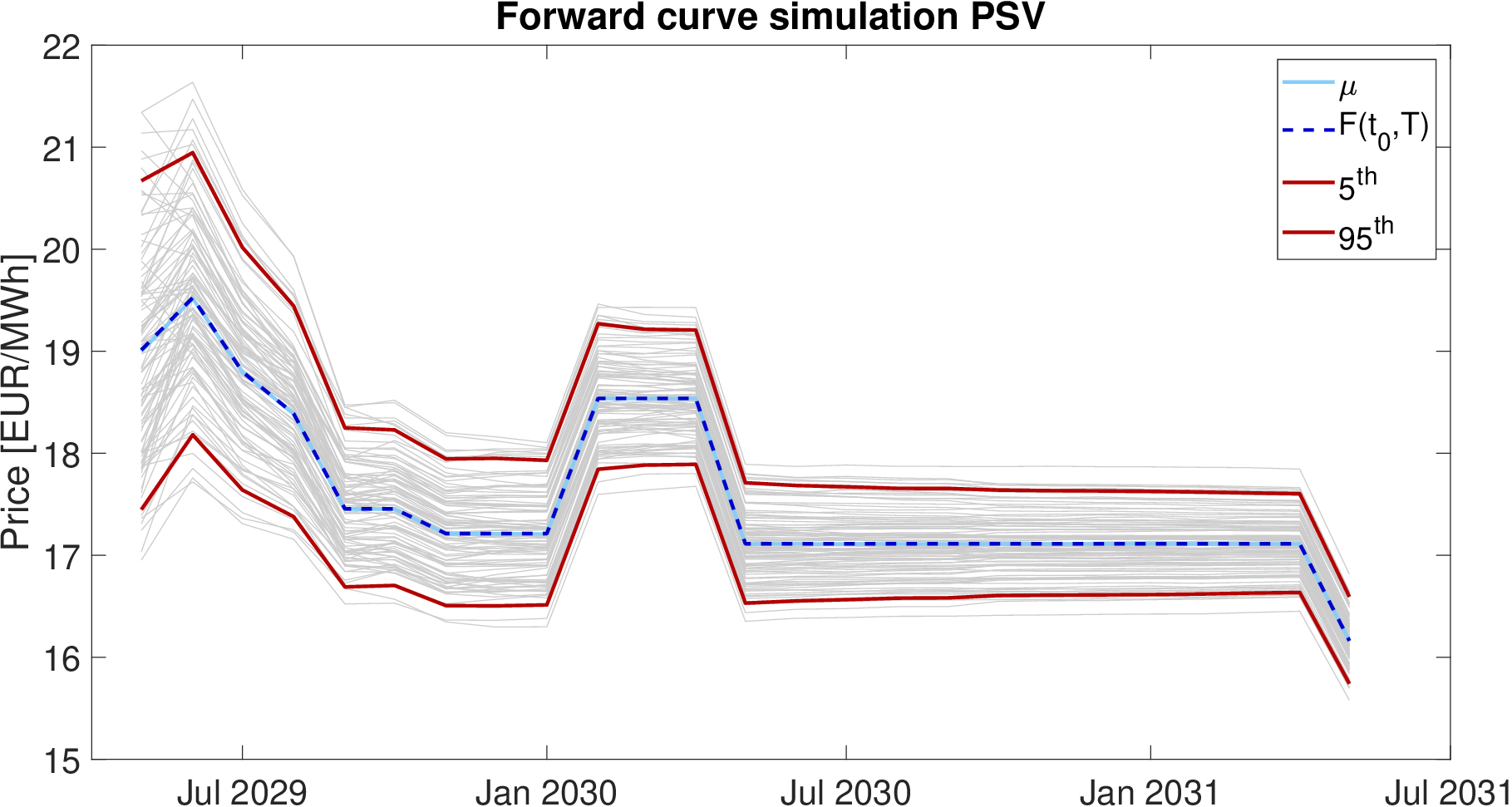}}
	\caption{Simulated two days forward curves: $n_{sim}=400$.}
	\label{fig:forward_curves_simulations}
\end{figure}

 In many cases, practitioners are interested in computing risk metrics (VaR, TVaR, CVaR, ...) of a given portfolio. In order to achieve this task, many approaches, historical, parametric and Monte Carlo among the others, are available. If a Monte Carlo approach is chosen simulations of the forward curve must be performed. In order to do so, we use Equation \eqref{eqn:simplified_solution_relative_prod_simulation} to simulate the forward curve after a few days, typically one or two. Results are shown in Figure \ref{fig:forward_curves_simulations}. In particular, we observe that the expected value of the simulated forward curves $\mu$, converges at today's forward curve $F(t_{0},T)$ as one should expect by construction. The fifth and the ninety-fifth percentiles in red, give an idea of the amplitude of the simulations. \textcolor{black}{Once that simulations are available the desired risk metric can be computed: if non-linear derivatives are present, Delta-Gamma approximation can be used as proposed by \citet{Glass2004}.}

\begin{figure}
	\centering
	\subfigure[Spot prices simulations.]{\includegraphics[width=1\textwidth]{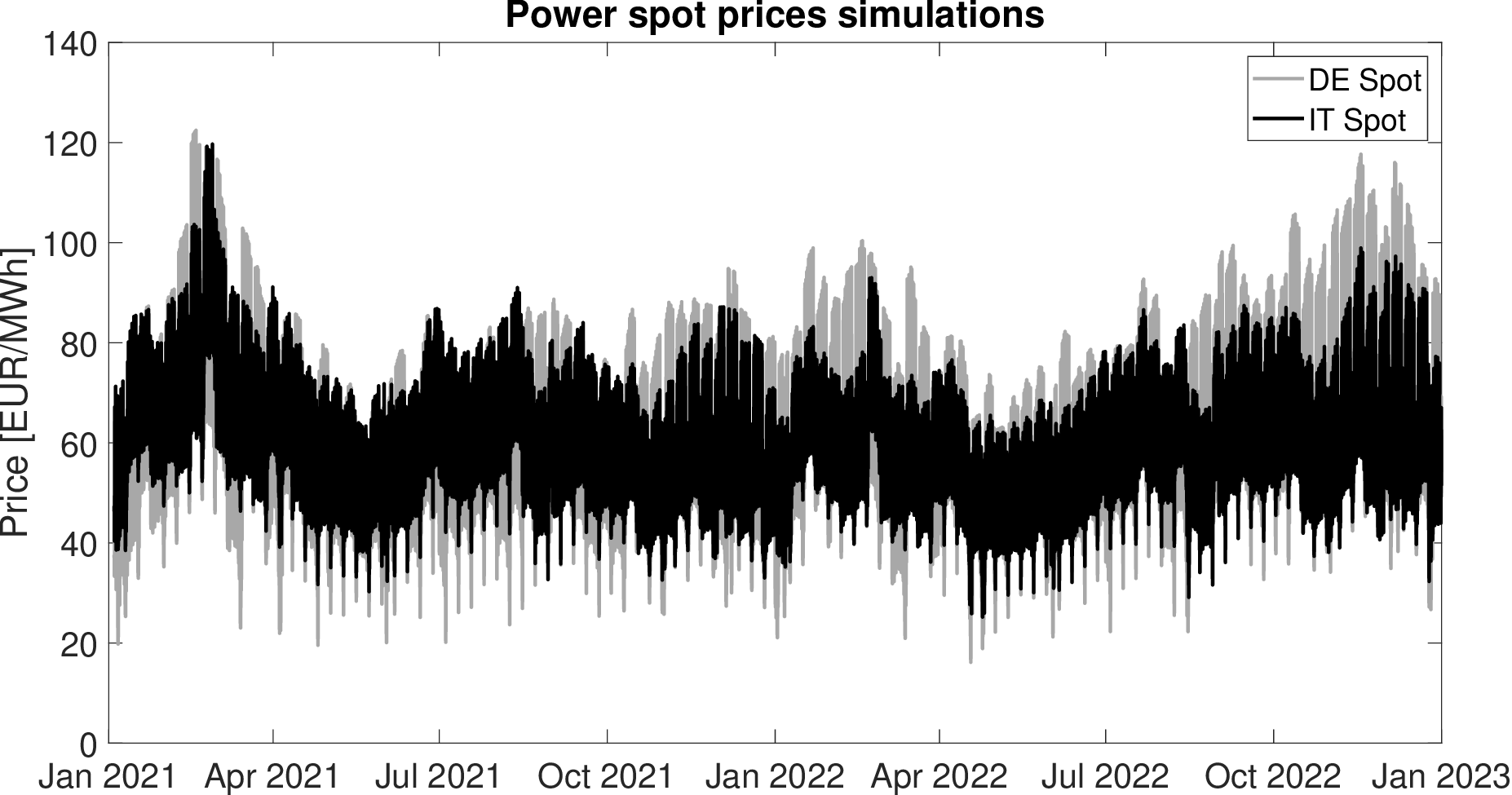}\label{fig:power_spot_simulations_2_years}}
	\subfigure[Zoom spot prices simulations.]{\includegraphics[width=1\textwidth]{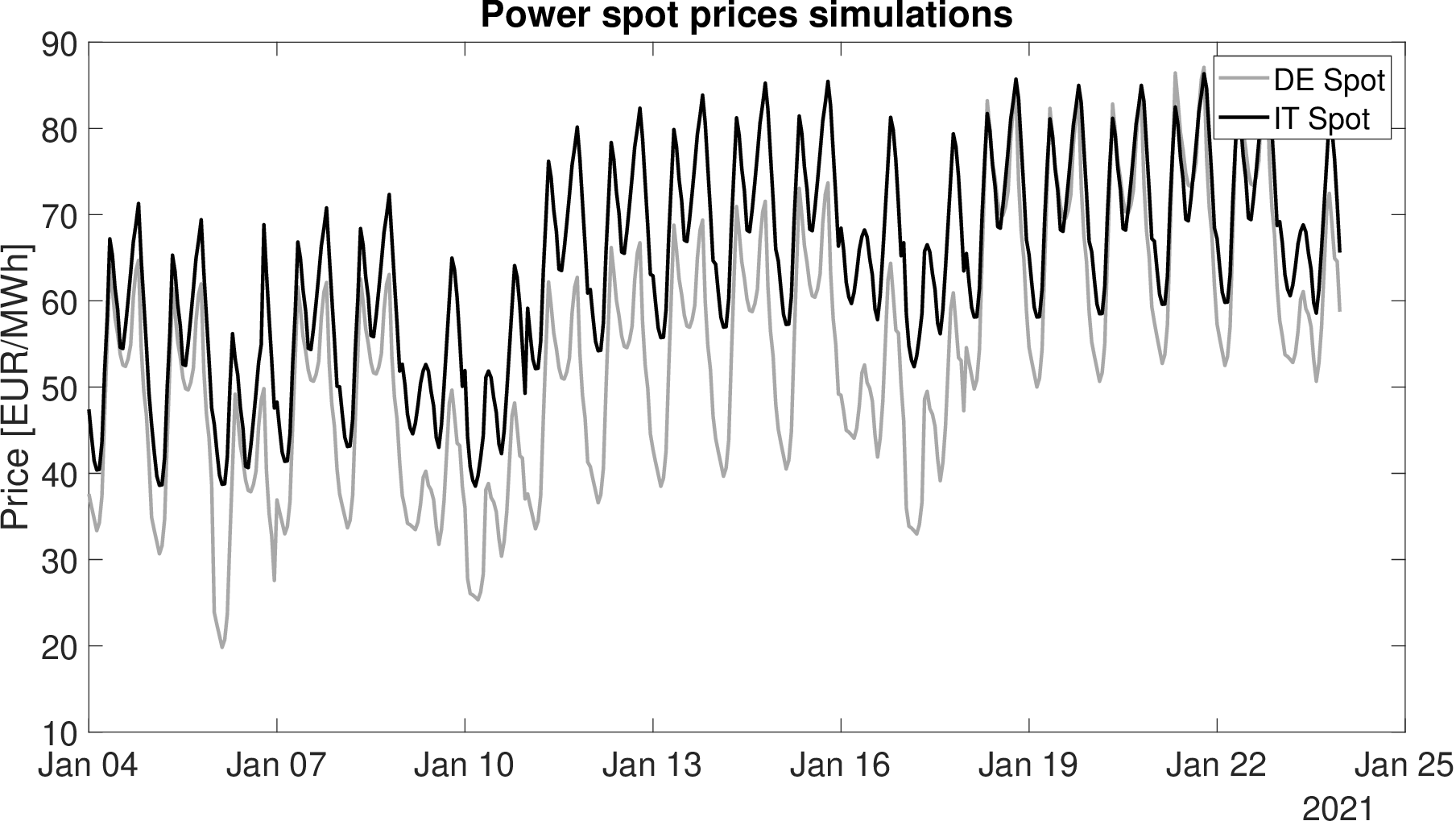}\label{fig:power_spot_simulations_2_years_zoomed}}
	\caption{German and Italian power spot prices simulations.}
	\label{fig:power_spot_simulations_DE_IT}
\end{figure}

\begin{figure}[!h]
	\centering
	\includegraphics[width=1\textwidth]{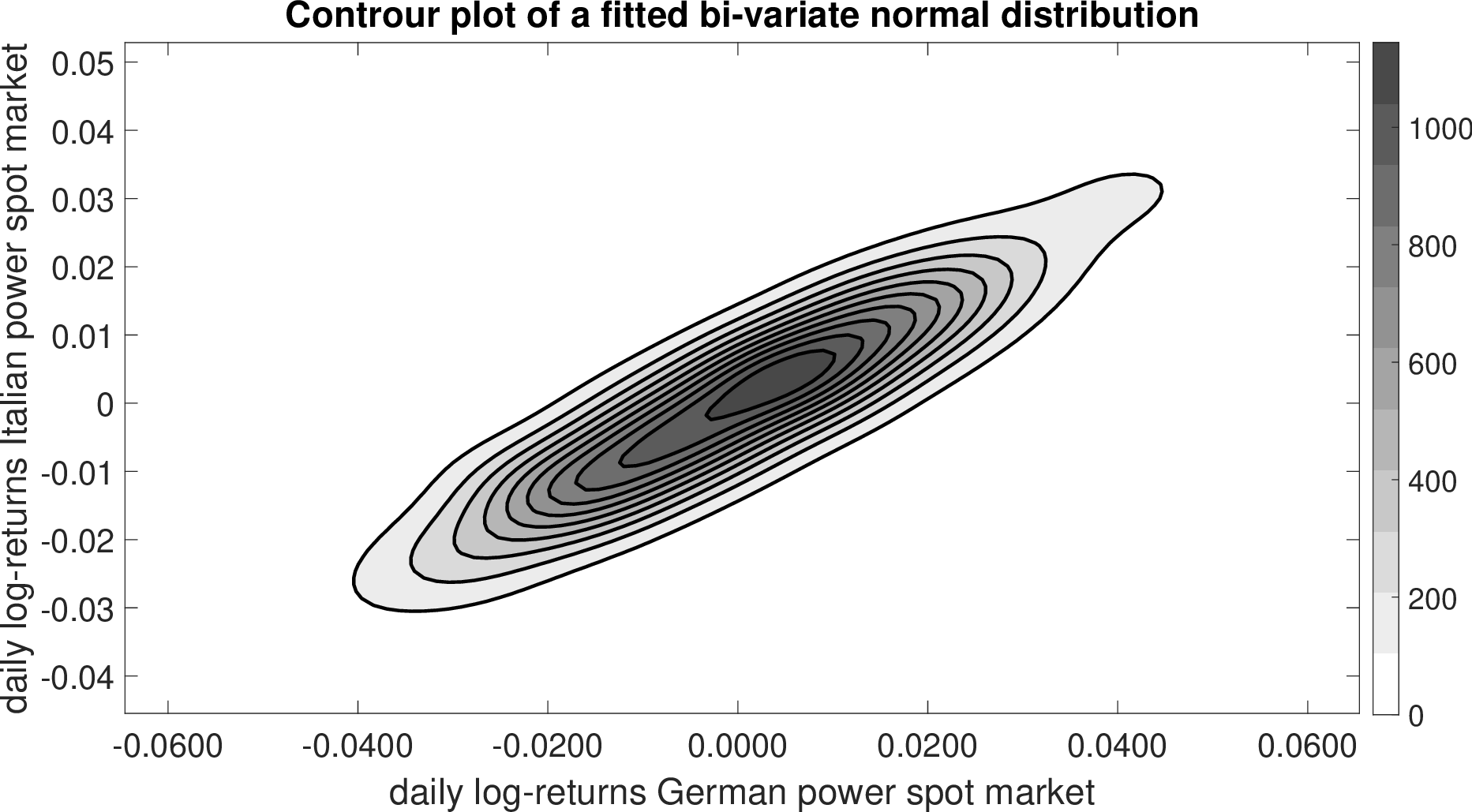}
	\caption{Contour plot of a bi-variate normal distribution fitted on daily log-returns of German and Italian power spot prices simulation. Correlation in log-returns $\rho=0.98$.}
	\label{fig:power_spot_log_returns_distribution}
\end{figure}

As final issue, we investigate the spot prices produced by the HJM framework. In this step, for the sake of conciseness, we focus only of two markets: German and Italian power spot markets. In Figure \ref{fig:power_spot_simulations_DE_IT} we show a single realization of the stochastic process $S^{k}=\left\{S^{k}(t); t \in \left[t_{0},T\right]\right\}$ for the two different commodities, according to the dynamics presented by Equation \eqref{eqn:spot_exact_solution}. As we can observed the spot prices tend to move together since they are driven by the same stochastic factors. In this case the correlation level in daily spot log-returns is high, approximately $\rho=0.98$ and this is confirmed by Figure \ref{fig:power_spot_log_returns_distribution}, where we displayed the contour plot of a bi-variate Gaussian distribution fitted on daily log-returns.
\par It is worth remembering that many European power spot markets are \emph{coupled} together: market coupling optimizes the allocation of cross-border capacities between countries. One of the possible effects on the prices dynamics is that at the same hour the electricity price in two different countries can be the same. The proposed model does not take in account this behavior: in order to include such an effect a fundamental component must introduced, as proposed by \citet{Carmona2014} and \citet{KIESEL2016}. Another limitation of the approach we have developed is that the correlation we introduce in the spot simulations is the one we derive from futures prices which is typically higher than the one we get historically in the spot market.
\par Once that Monte Carlo simulations are available both for power and futures products derivatives, pricing of complex energy financial claims, such as virtual power plants or storage might be easily performed following the algorithms presented by \citet{TB2000} and \citet{Dorr2003} respectively, as will be sketched in the next section.

\begin{figure}[!h]
	\centering
	\includegraphics[width=1\textwidth]{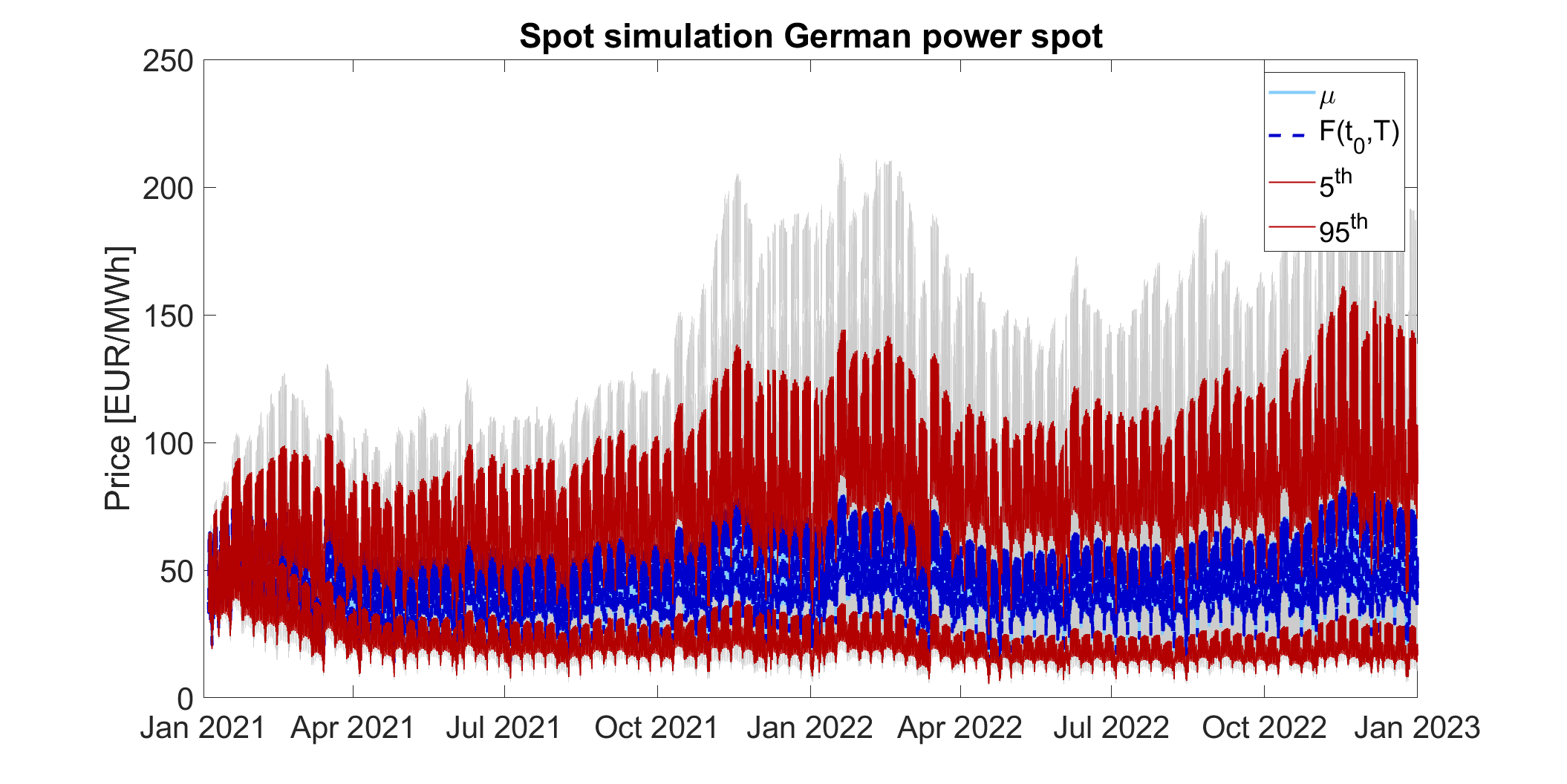}
	\caption{Sample paths for the German power spot price.}
	\label{fig:power_spot_simulations_DE}
\end{figure}

\subsection{\textcolor{black}{Virtual Power Plant pricing}}
In this section, we use Monte Carlo simulation to price a Virtual Power Plant (VPP): the approach we used started from the one presented by \citet{TB2000}. A VPP is a financial contract which mimics the way in which a power plant works: it convert a fuel, usual natural gas, into electricity. In this example we assume that there are hourly markets both for electricity and gas market and we focus on Italian power and gas market.  
\par The heat rate of a plant, $H$, is defined as the conversion rate between electricity and fuel. Assuming $H$ is known, and denoting by $S(t)^{E}$ and $S(t)^{F}$ the electricity and the fuel prices respectively, for every $MWh$ of electricity generation we have the following payoff:
\begin{equation*}
	\Phi(t) = \max\left(S(t)^{E} - H \cdot S(t)^{F},0 \right).
\end{equation*}
We can decide to turn on the plant if $S(t)^{E} > H \cdot S(t)^{F}$. For this reason \citet{HSU1998} propose as a VPP value over a time period $\left[0,T\right]$:
\begin{equation*}
	\sum_{t=1}^{T}\E^{\Q}\left[\max\left(S(t)^{E} - H \cdot S(t)^{F},0 \right)\right],
\end{equation*}
so that the value of the VPP can be estimated by a series of European spark-spread call options. In reality, a power plant has some physical constraints, such as non-zero startup/shutdown costs which usually are included in the virtual power plant contract. Ignoring this physical constraints leads to an overestimation of the VPP contract. For this reason they must be included from a modeling point of view. In this example we include only some constraints: for more details we refer to \citet{TB2000}. In particular we consider:
\begin{itemize}
	\item $t^{on}$: the minimum number of hours the plant must remain on after it was turned on.
	\item $t^{on}$: the minimum number of hours the plant must remain off after it was turned off.
	\item $q(t)$: the amount of electricity generated at time $t$.
	\item $q^{min}$: minimum rated capacity of the unit.
	\item $q^{max}$: maximum rated capacity of the unit.
	\item $S_{u}$: startup cost associated with turning on the unit.
	\item $S_{d}$: shutdown cost associated with turning off the unit.
	\item $H$: it is considered constant and not a function of $q(t)$.
\end{itemize}

The problem to solve has the following form:
\begin{equation*}
	\sup_{q(t)} \E\left[\sum_{t=1}^{T}q(t)\max\left(S(t)^{E} - H \cdot S(t)^{F},0 \right)\right],
\end{equation*}
subject to the constraints $q^{min}\le q(t) \le q^{max},\; t \in \left[0,T\right]$, the power plant must be on (off) for at least $t^{on} (t^{off})$ hours when it is switched on (off) and considering startup/shutdown costs.  
\par
\begin{figure}[!h]
	\centering
	\includegraphics[width=1\textwidth]{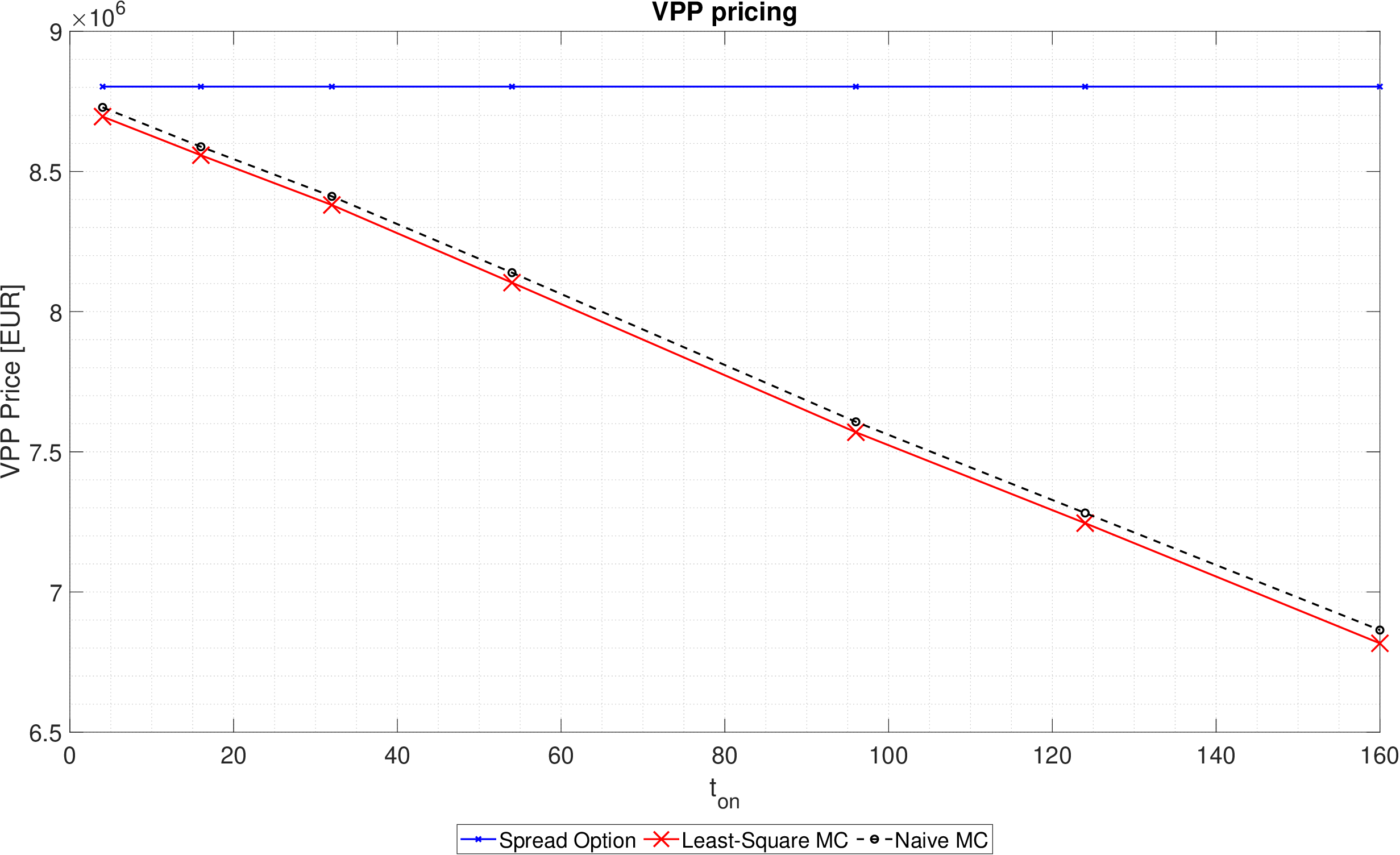}
	\caption{Example of the value of a VPP varying the parameters $t^{on}$ and $t^{off}$.}
	\label{fig:vpp_pricing}
\end{figure}

 To incorporate unit constraints, the problem can be formulated as a multistage stochastic problem which can be solved by using a backward stochastic dynamics programming (BSDP) as the one proposed by \citet{LSW01}. In this case, no closed pricing formula are available. We compute the price of a VPP contract with the set of parameters in Table \ref{tbl:vpp_pricing_parameters}. Observe that we compute the value of the contract for different values of $t^{on}$ and $t^{off}$. We expect that as the the contract loses in flexibility, as it value decreases. Moreover, in order to check that the pricing is coherent, we can compute an upper-bound for the value of the contract. It is given by a strip of European call options on the spark-spread. We also use a \emph{Naive MC} evaluation, namely we compute path-by-path the optimal behavior of the virtual power plant and the VPP value is obtained by taking the mean of all the optimal payoffs. This procedure, should leads to a lower value than the one we obtain by the BSDP approach. Results are reported in Table \ref{fig:vpp_pricing} and in Figure \ref{fig:vpp_pricing} and confirm the intuitions. Computational time increases as $t^{on}/t^{off}$ increases since the state variables, which have been used to implement the algorithm, increase in $t^{on}/t^{off}$ and this cause the slowdown of the solution algorithm.

\begin{table}[h!]
	\begin{center}	
		\begin{tabular}{c|c} 
			\textbf{Parameters} & \textbf{Value} \\
			\hline
			$\tau^{s}$&  1/10/2023 \\
			$\tau^{e}$ & 31/10/2023 \\
			$t^{on}$ & $4, 16,24,54, 96, 124, 160$ \\
			$t^{off}$ & $4, 16,224,54, 96, 124, 160$ \\
			$q^{min}$ & $180$ (MW) \\
			$q^{max}$ & $360$ (MW) \\
			$S_{u}$ & $2000$ (EUR) \\
			$S_{d}$ & $7000$ (EUR) \\
			$H$ & $40\%$
		\end{tabular}
	\end{center}
	\caption{VPP contract parameters for a delivery period $\left[\tau^{s} ,\tau^{e} \right]$ from 1/10/2023 to 31/10/2023.}
	\label{tbl:vpp_pricing_parameters}
\end{table}

\begin{table}[h!]
	\begin{center}	
		\begin{tabular}{c|c|c} 
			$\boldsymbol{t^{on}$ $(t^{off})}$ & \textbf{VPP value (EUR)} & \textbf{Computational Time (s)} \\
			\hline
			1 (1) & $88.02 \cdot 10^{6}$ & 0.21 \\
			4 (4) & $86.95 \cdot 10^{6}$ &  14.93\\
			16 (16) & $85.57 \cdot 10^{6}$ & 22.15  \\
			24 (24) & $84.11 \cdot 10^{6}$ & 33.67 \\
			54 (54) & $81.03 \cdot 10^{6}$ &  51.71 \\
			96 (96) & $75.79 \cdot 10^{6}$ & 96.12 \\
			124 (124) & $72.45 \cdot 10^{6}$ & 132.17 \\
			160 (160) & $68.16 \cdot 10^{6}$ & 186.42  \\
		\end{tabular}
		\caption{VPP evaluation varying $t^{on}$ and $t^{off}$.}
	\end{center}
	\label{tbl:vpp_pricing}
\end{table}

\clearpage 
\subsection{\textcolor{black}{Swing option pricing}}
In this section we valuate a swing option following the algorithm proposed by \citet{Dorr2003}. Consider as underlying commodity the daily spot Dutch natural gas (TTF) $S(t)$ on a given delivery, say from 1/10/2023 to 31/10/2023 and assume that the owner of the swing option can exercise at most $u_{max}$ daily upswings obtaining a payoff of $\left(S(t)-K\right)^{+}$ and at most $d_{max}$ daily downswings obtaining a payoff of $\left(S(t)-K\right)^{+}$, where $K$ is a given strike price. Usually the payoff are multiplied by a given quantity $Q$ which in this example is assumed unitary. We assume that $K$ is the same for the up-downswing but it is immediate to use to different strike prices. In particular this type of contract consists on the sum of multiple exercise American put and call options. 
\par We recall that the value of an American put option $P^{Am}$ with strike $K$ and maturity $T$ can be computed as:
\begin{equation*}
	P^{Am}(K,T) = \sup_{t \le \tau \le T} \E^{\Q}\left[e^{-r(T-t)} \left(K-S(\tau)\right)\right],
\end{equation*}
as shown in \citet[Chapter~21]{Bjork2009}.\\

\par
\begin{figure}[!h]
	\centering
	\includegraphics[width=1\textwidth]{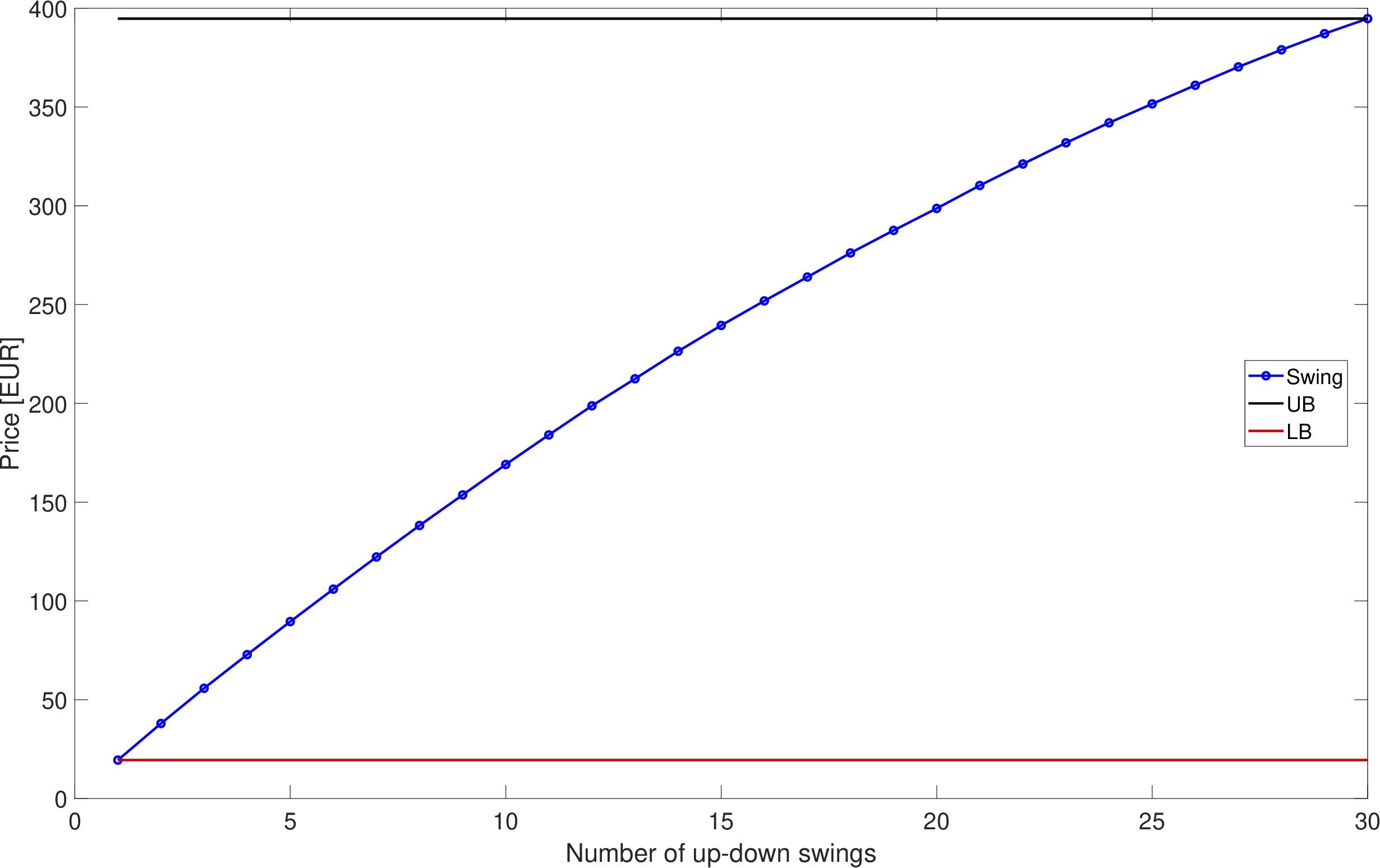}
	\caption{Example of the value of a swing contract varying the number of $u_{max}$ and $d_{max}$.}
	\label{fig:swing_pricing}
\end{figure}

 If we assume that the option can be exercised daily we have that a simple lower-bound for this contract is given by:
\begin{equation*}
	lb_{swing} = C^{Am}(K,T) + P^{Am}(K,T),
\end{equation*} 
where $C^{Am}(K,T)$ and $P^{Am}(K,T)$ are the prices of an American Call and American Put option with strike price $K$ and maturity $T$ equal to the end of the delivery period. On the other hand a natural upper bound for the contract is given by:
\begin{equation*}
	ub_{swing} = \sum_{t=1}^{T} C^{Eu}(K,t) + P^{Eu}(K,t),
\end{equation*} 
where $C^{Eu}(K,t)$ and $P^{Eu}(K,t)$ are the prices of an European Call and European Put option with strike price $K$ and maturity $t$, for all $t$ from now to the end of the delivery period. The value of the swing contract can be obtained, once again, by extending the algorithm by \citet{LSW01}, as proposed by \citet{Dorr2003}. As parameters we use the ones in Table \ref{tbl:swing_pricing_parameters} whereas Table \ref{tbl:swing_pricing} contains the results of the pricing varying the number of maximum up-downswings ($u_{max}$ and $d_{max}$). In Figure \ref{fig:swing_pricing} we show how the value of the swing option increases as $u_{max}$ and $d_{max}$ increase. We observe that its value lies between the two proposed bounds. In particular, when the number of up-downswings is equal to the number of days in the delivery period the level of the upper-bound is reached.

\begin{table}[h!]
	\begin{center}	
		\begin{tabular}{c|c} 
			\textbf{Parameters} & \textbf{Value} \\
			\hline
			$\tau^{s}$&  1/10/2023 \\
			$\tau^{e}$ & 31/10/2023 \\
			$u_{max}$ & $1,\dots, 31$ \\
			$d_{max}$ & $1,\dots, 31$ \\
			$K$ & $39.83$ (EUR/MWh)
		\end{tabular}
	\end{center}
	\caption{Swing contract parameters for a delivery period $\left[\tau^{s} ,\tau^{e} \right]$ from 1/10/2023 to 31/10/2023.}
	\label{tbl:swing_pricing_parameters}
\end{table}

\begin{table}[h!]
	\begin{center}	
		\begin{tabular}{c|c|c} 
			$\boldsymbol{u_{max}$ $(d_{max})}$ & \textbf{Swing value (EUR)} & \textbf{Computational Time (s)} \\
			\hline
			1 (1) & 19.43 & 0.06 \\
			5 (5) & 89.54 & 0.26 \\
			10 (10) & 169.06 & 0.90 \\
			15 (15) & 239.46 & 2.11 \\
			20 (20) & 298.67 & 3.88 \\
			25 (25) & 351.58 & 6.21 \\
			30 (30) & 394.74 & 9.13 \\
		\end{tabular}
	\end{center}
	\caption{Swing evaluation varying $u_{max}$ and $d_{max}$.}
	\label{tbl:swing_pricing}
\end{table}

\clearpage 
\subsection{\textcolor{black}{Gas Storage evaluation}}
More complex energy contracts, as the ones analyzed by \citet{Boeger2009} or \citet{Kluge2006} can be prices by using similar Monte Carlo techniques. In this section we follow the approach proposed by \citet{Boeger2009} and we price a gas storage contract. 
\par By considering the storage contract from the prospective of the holder of the contract we assume it is signed at time $t=0$ and settled at time $t=T+1$. Every day the holder can choose to inject gas, do nothing or withdraw gas, for a volume $\Delta v(t)$ within certain volumetric constraints. $\Delta v$ is considered positive in case of injection In this example, The payoff at time $t$ is assumed to be:
\begin{equation*}
	h(S(t),\Delta v(t)) = \begin{cases} 
	-S(t)\Delta v(t) & \text{inject} \\
	0 & \text{do nothing} \\
	S(t)\Delta v(t) & \text{withdrawn}.
\end{cases}
\end{equation*}
Furthermore, we assume that the volume of the storage at time $t$ satisfies $v_{min} \le v(t) \le v_{max}$ and that the injection or withdrawal is subjected to $\i^{min}\le \Delta v(t) \le i^{max}$. All these constraints can be easily considered time variant. Moreover, a constraint on the terminal value of the volume $v(T+1)$ can be considered. This can be done by considering a \enquote{penalty function} $q(S(T+1),v(T+1))$.
\par The value of the storage contract $U(0,S(0),v(0))$ is the expected value of the accumulated future payoff $h\left(S(t),\Delta v(t)\right)$ under the most optimal strategy (or policy) $\boldsymbol{\pi} = \left\{\pi\left(1,S(1),v(1)\right),\dots,\pi\left(T,S(T),v(T)\right)\right\}$, namely:
\begin{equation}
	U(0,S(0),v(0)) = \sup_{\pi} \E \left[\sum_{t=0}^{T} e^{-r \Delta t} h\left(S(t), \Delta v(t)\right) + e^{-r(T+1)} q\left(S(T+1), v(T+1)\right) \right].
	\label{eqn:storage_contract_equation}
\end{equation}
This problem can be solved adopting standard dynamic programming (SDP) techniques in stochastic setting. Computational efficiency can be improved by using a Least-Squares approach as detailed in \citet{Boeger2009}.   \\

\par In our example, we consider a storage contract written on TTF natural gas with delivery period the whole 2024. The parameters are listed in Table \ref{tbl:gas_storage_pricing_parameters}.

\begin{table}[h!]
	\begin{center}	
		\begin{tabular}{c|c} 
			\textbf{Parameters} & \textbf{Value} \\
			\hline
			$\tau^{s}$&  1/1/2024 \\
			$\tau^{e}$ & 31/12/2024 \\
			$v_{min}$ & 0 (MWh)\\
			$v_{max}$ & 250,000 (MWh)\\
			$v_{0}$ & 100,000 (MWh)\\
			$v_{T+1}$ & 100,000 (MWh)\\
			$i_{min}$ & -8,000 (MWh)\\
			$i_{max}$ &  8,000 (MWh)
		\end{tabular}
	\end{center}
	\caption{Gas storage contract parameters for a delivery period $\left[\tau^{s} ,\tau^{e} \right]$ from 1/1/2024 to 31/12/2024.}
	\label{tbl:gas_storage_pricing_parameters}
\end{table}

\begin{figure}[!h]
	\centering
	\includegraphics[width=1\textwidth]{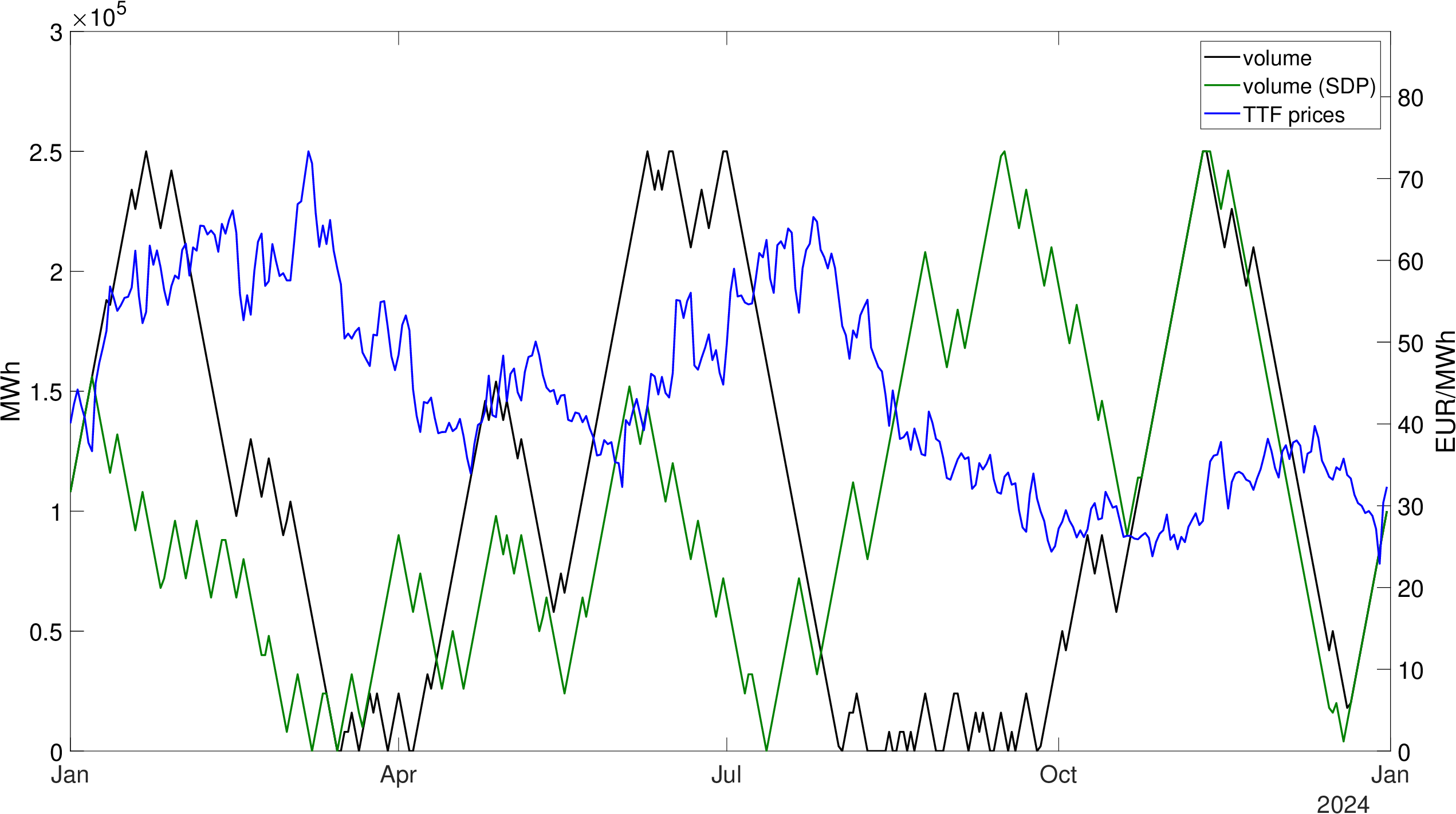}
	\caption{Gas storage volume for a simulation of TTF prices computed using the deterministic approach and the SDP one. Observe how the deterministic approach (black) extract the maximum value from the contract, since the storage is filled in low-prices periods and emptied in high-prices ones.}
	\label{fig:storage_price}
\end{figure}

In Table \ref{tbl:gas_storage_pricing} we report the gas storage price evaluated using 2500 simulations. The \emph{deterministic} approach consists in maximizing path-by-path the payoff and finally taking the average. This approach solves the following problem:
\begin{equation*}
	 \E \left[\sup_{\pi} \sum_{t=0}^{T} e^{-r \Delta t} h\left(S(t), \Delta v(t)\right) + e^{-r(T+1)} q\left(S(T+1), v(T+1)\right) \right]
\end{equation*}
and clearly tends to overprice the contract since it assume a perfect foresight of the future. The correct value of the contract is given by solving \eqref{eqn:storage_contract_equation} using the SDP. We observe that the value of the contract is smaller than in the deterministic case. Furthermore, following \cite{LS2002} we perform an out-of-sample test of the optimal policy $\boldsymbol{\pi}$. We observe that the value of the contract priced on the out-of-sample paths is very close the the one given by the solution of Equation \eqref{eqn:storage_contract_equation}. This lead us to assume that the pricing algorithm is successful since it leads to out-of-sample values that closely approximate the in-sample values for the option. In Figure \ref{fig:storage_price} deterministic and stochastic optimal trajectories of the level of the storage are shown, related the one price process $S(t)$ realization.                                                                                                                                                                                                                                                                                                                                                                                                                                                                                                                                                                                                                                                                                                                                                                                                                                                                                                                                                                                                                                                                                                                               

\begin{table}[h!]
	\begin{center}	
		\begin{tabular}{c|c|c} 
			\textbf{Approach} & \textbf{Swing value (EUR)} & \textbf{Computational Time (s)} \\
			\hline
			Deterministic & 19,826,114 & 25,920.02 \\
			SDP & 7,457,729 & 1,020.77 \\
			SDP (out sample) &  7,462,901 & 23.04
		\end{tabular}
	\end{center}
	\caption{Gas storage evaluation.}
	\label{tbl:gas_storage_pricing}
\end{table}

\clearpage

\section{Conclusions}
\label{sec:conclusions}
In this article we discussed in detail the implementation and a possible application of the Heath-Jarrow-Morton framework to energy markets, focusing, in particular on the European power and natural gas markets. We introduced a risk neutral dynamics synthetic products $F_{M_{h}}(t,T)$ and we derived that of spot $S(t)$ and futures prices $F(t,\tau^{s},\tau^{e})$. By showing that the power and natural gas futures markets are strictly dependent, by the PCA algorithm we selected only few stochastic factors (i.e. independent Brownian motions) which explain a large part of the market variance. Furthermore, following \citet{Boeger2009} we have shown how a closed form solution for European vanilla option is available. Finally, we have applied the model to real market data from European power and natural gas markets. We discussed the daily log-returns correlation structures and we have shown that the models fits the market narrowly. Moreover, we analyzed the futures and spot simulations in output from the model. Finally, we used the Monte Carlo simulation to valuate Virtual Power Plants and swing contracts.
\par HJM model is easy to implement, its calibration almost immediate, due to the hypothesis of normality of log-returns and simulations are fast to perform in an exact way.
\par Clearly, the model presents some drawbacks which must be considered when the output are used for risk-metric computation or for derivative pricing. First of all, log-returns are normally distributed and this is not the case in almost all financial markets: jumps, volatility smiles and clustering are often present and are not considered by the proposed framework. In order to consider such stylized facts one can introduce stochastic volatility, jumps in price dynamics maybe using subordination techniques. On the other hand, even if such models work great in uni-variate setting, they are too hard to calibrate in a multivariate framework. In this latter case, a Gaussian approach could be preferred in term of computational al calibration complexity. On the other hand, if one needs to focus on the pricing of a derivative written o a single underlying asset, more complex models based on \Levy\ processes or on stochastic volatility can be used in order to include many market stylized facts. 
\par Other drawbacks follow from the fact that the calibration has been performed on historical futures prices. First of all, it is not guaranteed that the model replicates the (actually few) options quoted in the market. On the other hand, option power markets are not very liquid and hence, in many situations, an historical calibration is the only possible one. Another drawback of the model is that, since we did not considered spot quotations for the calibration, the correlation and the volatility in simulated spot log-returns directly inherit from the forward one instead of from the spot one.
\par Nevertheless, the proposed framework appears to be one of the easiest one in which dependent futures and spot prices simulation in a multi-commodity setting can be obtained. For this reason, it is widely used by practitioners and, at the moment, it is the benchmark against which testing innovative approaches. For example, it would be worth to include jumps in prices' dynamics or to consider a stochastic volatility approach in the multi-dimensional framework, by preserving both mathematical and numerical tractability. This could be the direction of possible future researches, both in academia and in industry. 



\clearpage

\bibliographystyle{plainnat}
\bibliography{library.bib}

\end{document}